\documentclass{article}
\usepackage[top=1in, bottom=1in, left=1in, right=1in]{geometry}

\usepackage{url} 

\usepackage{bm}\usepackage{soul}
\usepackage{empheq}\usepackage{mathtools}
\usepackage{mathrsfs,color}
\usepackage{amsfonts}\usepackage{multirow}
\usepackage{caption,comment}
\usepackage{amsmath}\usepackage{float}
\usepackage{amssymb,amsthm}
\usepackage{graphicx,afterpage,cancel}

\usepackage[colorlinks = true,
            linkcolor = black,
            urlcolor  = red,
            citecolor = blue,
            anchorcolor = blue]{hyperref}

\usepackage{subfig}
\usepackage{mwe}
\newlength{\tempheight}
\newlength{\tempwidth}

\usepackage{tikz}
\usepackage{tikz-network}

\newcommand{\rowname}[1]
{\rotatebox{90}{\makebox[\tempheight][c]{\textbf{#1}}}}

\newcommand{\columnname}[1]
{\makebox[\tempwidth][c]{\textbf{#1}}}

\definecolor{orange}{RGB}{255,127,0}

\def\d{{\, \rm d}}

\afterpage{\clearpage}

\usepackage{authblk}

\title{Combining Stochastic Parameterized Reduced-Order Models with Machine Learning for Data Assimilation and Uncertainty Quantification with Partial Observations}

\author[1]{Changhong Mou\thanks{cmou3@wisc.edu}}
\author[1,2]{Leslie M. Smith\thanks{lsmith@math.wisc.edu}}
\author[1]{Nan Chen\thanks{chennan@math.wisc.edu}}
\affil[1]{Department of Mathematics, University of Wisconsin-Madison, Madison}
\affil[2]{Department of Engineering Physics, University of Wisconsin-Madison, Madison}

\begin{document}

\maketitle

\begin{abstract}
A hybrid data assimilation algorithm is developed for complex dynamical systems with partial observations. The method starts with applying a spectral decomposition to the entire spatiotemporal fields, followed by creating a machine learning model that builds a nonlinear map between the coefficients of observed and unobserved state variables for each spectral mode. A cheap low-order nonlinear stochastic parameterized extended Kalman filter (SPEKF) model is employed as the forecast model in the ensemble Kalman filter to deal with each mode associated with the observed variables. The resulting ensemble members are then fed into the machine learning model to create an ensemble of the corresponding unobserved variables. In addition to the ensemble spread, the training residual in the machine learning-induced nonlinear map is further incorporated into the state estimation that advances the quantification of the posterior uncertainty. The hybrid data assimilation algorithm is applied to a precipitating quasi-geostrophic (PQG) model, which includes the effects of water vapor, clouds, and rainfall beyond the classical two-level QG model. The complicated nonlinearities in the PQG equations prevent traditional methods from building simple and accurate reduced-order forecast models. In contrast, the SPEKF model is skillful in recovering the intermittent observed states, and the machine learning model effectively estimates the chaotic unobserved signals. Utilizing the calibrated SPEKF and machine learning models under a moderate cloud fraction, the resulting hybrid data assimilation remains reasonably accurate when applied to other geophysical scenarios with nearly clear skies or relatively heavy rainfall, implying the robustness of the algorithm for extrapolation.
\end{abstract}

\section{Introduction}
Complex dynamical systems are ubiquitous in many areas, including geoscience, engineering, neural science, material science, etc. \cite{frisch1995turbulence, majda2006nonlinear, vallis2017atmospheric, salmon1998lectures}. These systems are often high dimensional and strongly nonlinear. Multiscale structures, intermittent instabilities, and non-Gaussian probability density functions (PDFs) are typical features observed in these systems. Modeling and predicting these complex dynamical systems are central scientific problems with significant societal impacts. Since many of these systems are chaotic or turbulent, an accurate estimation of the model states at the initialization stage is vital to facilitate the subsequent forecast. Therefore, developing suitable data assimilation algorithms, which optimally integrate different sources of information to improve the state estimation of a complex dynamical system, becomes an essential prerequisite for reaching skillful forecast results \cite{kalnay2003atmospheric, law2015data, lahoz2010data, majda2012filtering}.
A typical data assimilation cycle contains two steps. The first step involves a statistical prediction using a given forecast model starting from the previously estimated state. The resulting prior distribution is then corrected based on the statistical input of noisy observation in the second step, leading to the so-called posterior distribution. These two steps are known as `forecast' (or prediction) and `analysis' (filtering or correction), respectively.

One of the fundamental challenges in data assimilation is its high computational cost, especially at the forecast step. In fact, due to the intrinsic high-dimensionality and complicated multiscale nonlinear features, simulating a single realization of the underlying dynamics is already computationally expensive, 
let alone running the model forward many times when applying the ensemble-based data assimilation methods. Reduced-order models are thus widely used to mitigate such a computational issue. In general, if the governing equation of the underlying system has simple explicit structures, then direct projections of the original system to a reduced order space with suitable closures can be applied for model reduction \cite{hoteit2002simplified, cstefuanescu2015pod, he2011use, meldi2017reduced, peherstorfer2015dynamic}. Stochastic parameterizations are often incorporated into the resulting reduced-order models to further improve the computational efficiency and accuracy \cite{grooms2015ensemble, mana2014toward, berner2017stochastic, christensen2017stochastic, duan2007stochastic, schneider2021learning, chen2018conditional}. On the other hand, as the primary goal for the time integration of the model is to seek the forecast statistics, nonlinear statistical reduced-order models have been developed to predict the leading
few moments that help reconstruct the forecast PDF \cite{majda2014blended, qi2022machine, chen2022physics, sapsis2013statistically}. In addition, various machine learning forecast approaches have recently been utilized as surrogate forecast models in data assimilation to produce forecast statistics or forecast ensemble members \cite{ruckstuhl2021training, gottwald2021combining, tang2020deep, yang2021machine, tsuyuki2022nonlinear, chattopadhyay2022deep, chattopadhyay2021towards, maulik2022aieada, penny2022integrating, chen2021bamcafe, pawar2020long}.

In this paper, a hybrid data assimilation algorithm is developed for complex turbulent systems, where the observations contain only a subset of the state variables, known as partial observations. The method combines stochastic parameterized reduced-order models with machine learning to facilitate computational efficiency. The idea of the method is first to apply a spectral decomposition to the entire spatiotemporal fields, followed by developing a machine learning model that builds a nonlinear map between the coefficients of the observed and the unobserved state variables for each spectral mode. Afterward, different strategies are adopted to filter these two sets of state variables. Specifically, a cheap low-order nonlinear stochastic parameterized extended Kalman filter (SPEKF) model \cite{gershgorin2010test, gershgorin2010improving} is employed as the forecast model in the ensemble Kalman filter (EnKF) \cite{evensen2009data, burgers1998analysis, houtekamer2005ensemble} to filter the coefficient of each spectral mode associated with the observed state variables. The resulting ensemble members are then fed into the machine learning model to create an ensemble of the unobserved variables associated with the same spectral mode. Notably, due to the turbulent nature and the intrinsic unpredictable components, a training residual typically exists in the optimally calibrated machine learning model, which accounts for the uncertainty in discovering the nonlinear dependence between the observed and unobserved variables \cite{chen2021bamcafe}. Therefore, in addition to the ensemble spread, the training residual in the machine learning-induced nonlinear map is further incorporated into the state estimation that advances the quantification of the posterior uncertainty. The final posterior state is described by a mixture distribution that can capture the underlying non-Gaussian features.

The hybrid data assimilation approach has several unique merits.
First, the spectral decomposition breaks the entire high-dimensional data assimilation problem into many low-dimensional subproblems, each focusing on filtering one primary spectral mode (and a few aliasing modes in the case with sparse observations, if applicable). Coping with such a low-dimensional subproblem facilitates the development of cheap stochastic parameterized models for filtering the observed variables and proper machine learning models for estimating the unobserved states.

Second, the nonlinear SPEKF model is systematically calibrated to reproduce the non-Gaussian statistics of the true signal. Therefore, it can accurately estimate the observed state variables, including recovering intermittency and extreme events, with appropriate uncertainty quantification. The simple but effective stochastic parameterizations in the SPEKF forecast model also play a vital role in compensating for the effect of the complicated nonlinearities in the governing equation of the time series associated with each spectral mode when carrying out the statistical forecast \cite{majda2018model, li2020predictability, harlim2008filtering, kang2012filtering}. Such a statistical approximation allows the filtering of different spectral modes independently, significantly reducing the computational cost. Despite the independence of the forecast models, the estimated states of different spectral modes are naturally correlated after the analysis step when the observations are involved in correcting the forecast errors. In addition, spatial dependence between the state variables at different grid points in physical space is automatically recovered after the spatial reconstruction in light of all the spectral modes. Notably, the simple nonlinear SPEKF model is generally more efficient than complicated machine learning models in terms of both model calibration and statistical forecast. This distinguishes the hybrid strategy from utilizing a sophisticated neural network to fully replace the original dynamical model as the forecast system.

Third, reduced-order models are often essential in ensemble-based data assimilation to accelerate the computations. Since there is a lack of observations to directly rectify the error in forecasting the unobserved states, the skill of filtering these states relies heavily on the accuracy of the reduced order models in characterizing the complex nonlinear statistical dependence between the observed and unobserved variables. However, the structures of many dynamical systems, especially those in fluids and geophysics, are too complicated to allow the development of skillful reduced-order models with simple analytic expressions using traditional approaches.
Nevertheless, since the primary goal of these reduced-order models is to provide statistical forecast results, it is often unnecessary to explicitly take into account the exact physics in developing these models. Thus, machine learning is an effective surrogate of the traditional parametric reduced-order models in discovering complicated nonlinear dependence between different state variables. Particularly, as the focus of the machine learning is only on a low-dimensional subspace after the spectral decomposition, designing and training the machine learning model becomes much more accessible.
Finally, beyond the deterministic forecast, the uncertainty in the machine learning model is incorporated into the estimation of the posterior distribution for these turbulent models. This advances a more accurate quantification of the uncertainty in data assimilation.

The hybrid data assimilation strategy is applied to a two-level, precipitating quasi-geostrophic (PQG) model \cite{smith2017precipitating, edwards2020spectra, hu2021initial}. Different from the classical
QG model \cite{salmon1980baroclinic, qi2016low, vallis2017atmospheric} (hereinafter referred to as dry QG), the PQG model takes into account the effects of water vapor, clouds, and rainfall. Therefore, it can more realistically describe atmospheric dynamics.
It also has the potential to elucidate aspects of the hydrological cycle and the effects of latent heat release on synoptic-scale midlatitude dynamics. The PQG equations include Heaviside nonlinearities due to phase changes, which can potentially have a significant influence on QG turbulence. However, these Heaviside nonlinearities, together with solving a (nonlinear) elliptic equation to recover the streamfunction
at each step, prevent using traditional methods for building effective reduced-order models with a simple explicit expression. The machine learning model then becomes crucial to discover the nonlinear dependence between different state variables in facilitating data assimilation.

The rest of the paper is organized as follows. The hybrid data assimilation strategy is presented in Section \ref{Sec:Strategy}. The PQG model, which serves as the test model for the hybrid data assimilation strategy, is summarized in Section \ref{Sec:PQG}. The data assimilation results are shown in Section \ref{Sec:DA}. Section \ref{Sec:Conclusion} includes additional discussions and a conclusion.

\section{Hybrid Data Assimilation Strategy}\label{Sec:Strategy}
Consider the following general form of complex dynamical systems:
\begin{subequations}\label{Starting_System}
\begin{align}
  \frac{\partial \mathbf{u}(\mathbf{x},t)}{\partial t} &= \mathcal{F}_1\big(\mathbf{u}(\mathbf{x},t),\mathbf{h}(\mathbf{x},t)\big),\\
  \frac{\partial \mathbf{h}(\mathbf{x},t)}{\partial t} &= \mathcal{F}_2\big(\mathbf{u}(\mathbf{x},t),\mathbf{h}(\mathbf{x},t)\big),
\end{align}
\end{subequations}
where the vectors $\mathbf{u}(\mathbf{x},t)$ and $\mathbf{h}(\mathbf{x},t)$ are the state variables with dimensions $S_1$ and $S_2$, respectively, $\mathbf{x}$ is the spatial coordinate vector, and $t$ is the time. On the right-hand side of \eqref{Starting_System}, $\mathcal{F}_1$ and $\mathcal{F}_2$ are complicated nonlinear functions containing
differential operators, which may not
necessarily have explicit expressions. For the convenience of presenting the hybrid data assimilation framework, periodic boundary conditions are imposed for all the state variables in \eqref{Starting_System}, which facilitates the use of Fourier basis functions in the spectral decomposition. Other spectral decomposition methods, such as the proper orthogonal decomposition (POD) \cite{berkooz1993proper} and the autoencoder networks \cite{kramer1991nonlinear}, can be adopted for more general situations.

In many practical problems, only a subset of the state variables is observed, known as partial observations. To this end, assume that $\mathbf{u}$ contains the observed state variables, while there are no observations for $\mathbf{h}$. Further, assume that the observational locations are sparse in space at regularly-spaced grid points.  More generally, sparse irregularly-spaced observations can be interpolated to the regularly-spaced model grid points.
See \cite{majda2012filtering, harlim2011interpolating} for the detailed procedure and the mathematical theory of the error estimates.

Denote by $u^{\{s_1\}}$ and $h^{\{s_2\}}$ a scalar component of $\mathbf{u}$ and $\mathbf{h}$, respectively, where $s_1\in\{1,\ldots, S_1\}$ and $s_2\in\{ 1,\ldots, S_2\}$.
Apply a spectral decomposition to $u^{\{s_1\}}(\mathbf{x},t)$ and $h^{\{s_2\}}(\mathbf{x},t)$ with Fourier bases,
\begin{equation}\label{Fourier}
\begin{split}
  u^{\{s_1\}}(\mathbf{x},t) &= \sum_{\mathbf{k}\in\mathcal{K}}\hat{u}^{\{s_1\}}_\mathbf{k}(t)\exp(i\mathbf{k}\mathbf{x}),\\
  h^{\{s_2\}}(\mathbf{x},t) &= \sum_{\mathbf{k}\in\mathcal{K}}\hat{h}^{\{s_2\}}_\mathbf{k}(t)\exp(i\mathbf{k}\mathbf{x}),
\end{split}
\end{equation}
where $\mathcal{K}$ is a finite index set containing wavevectors. 
Let $\hat{\mathbf{u}}_\mathbf{k}(t) = (\hat{u}^{\{1\}}_\mathbf{k}(t),\ldots, \hat{u}^{\{S_1\}}_\mathbf{k}(t))$ and $\hat{\mathbf{h}}_\mathbf{k}(t) = (\hat{h}^{\{1\}}_\mathbf{k}(t),\ldots, \hat{u}^{\{S_2\}}_\mathbf{k}(t))$ be the collections of the Fourier coefficients for a fixed wavevector $\mathbf{k}$.

\begin{figure}[H]
\centering
\includegraphics[width=.8\textwidth]{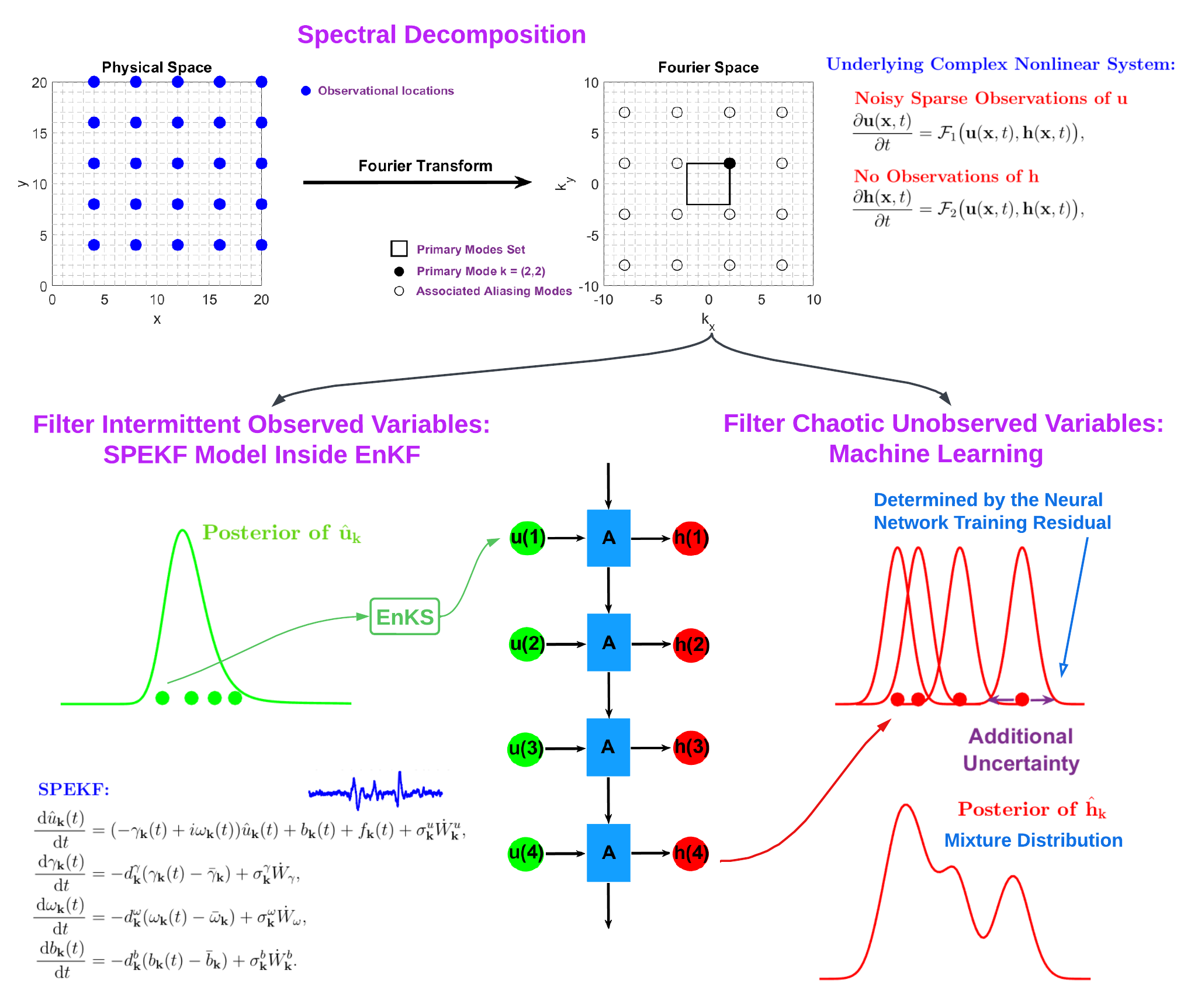}
\caption{A schematic illustration of the hybrid data assimilation method.}\label{Fig:Schematic}
\end{figure}

A schematic illustration of the algorithm is shown in Figure \ref{Fig:Schematic} with the details being discussed in the following subsections.

\subsection{Filtering the Observed Variables Using Stochastic Parameterized Forecast Models}
The exact governing equation of $\hat{u}^{\{s_j\}}_\mathbf{k}(t)$ involves complicated nonlinear interactions with not only all the other components in $\hat{\mathbf{u}}_\mathbf{k}(t)$ and $\hat{\mathbf{h}}_\mathbf{k}(t)$ but also those in $\hat{\mathbf{u}}_{\mathbf{k}^\prime}(t)$ and $\hat{\mathbf{h}}_{\mathbf{k}^\prime}(t)$ for $\mathbf{k}^\prime\neq\mathbf{k}$. Utilizing such an equation as the forecast model of $\hat{u}^{\{s_j\}}_\mathbf{k}(t)$ in data assimilation is computationally expensive for at least two reasons. First, the entire high-dimensional system of $\hat{\mathbf{u}}_\mathbf{k}(t)$ and $\hat{\mathbf{h}}_\mathbf{k}(t)$ for all $\mathbf{k}\in\mathcal{K}$ has to be integrated forward at the same time to reach the forecast state of even a single component $\hat{u}^{\{s_j\}}_\mathbf{k}(t)$. 
{Second, the right-hand side of the governing equation of each $\hat{u}^{\{s_j\}}_\mathbf{k}(t)$ contains complicated nonlinear terms, whose calculation at each time step requires a significant amount of computational time.}

To this end, a stochastic parameterized extended Kalman filter (SPEKF) model \cite{gershgorin2010test, gershgorin2010improving} is utilized as the forecast model of each $\hat{u}^{\{s_j\}}_\mathbf{k}(t)$ to reduce the computational cost. The idea of the SPEKF model is to build a statistical approximate forecast model that utilizes simple stochastic parameterizations to compensate for the complicated nonlinear interactions between $\hat{u}^{\{s_j\}}_\mathbf{k}(t)$ and the coefficients of the other spectral modes, such that the statistical forecast of $\hat{u}^{\{s_j\}}_\mathbf{k}(t)$ stays close to that using the exact governing equation. The resulting equation of  $\hat{u}^{\{s_j\}}_\mathbf{k}(t)$ depends only explicitly on itself but not the coefficients of other spectral modes. This not only reduces the complexity on the right-hand side of $\hat{u}^{\{s_j\}}_\mathbf{k}(t)$ by a significant amount, but also allows the filtering of different spectral modes independently.

Below, the framework of the SPEKF is presented for a single scalar component of the Fourier coefficient of the observed state variables.
For notational simplicity, the superscript in $\hat{u}^{\{s_j\}}_\mathbf{k}(t)$ is omitted, and the scalar Fourier coefficient is simply denoted by $\hat{u}_\mathbf{k}(t)$.

\subsubsection{The SPEKF Model}

The forecast model of $\hat{u}_\mathbf{k}(t)$ is approximated by the following SPEKF model:
\begin{subequations}\label{SPEKF}
\begin{align}
\frac{\d \hat{u}_\mathbf{k}(t)}{\d t} &= (-\gamma_\mathbf{k}(t) + i\omega_\mathbf{k}(t) ) \hat{u}_\mathbf{k}(t) + b_\mathbf{k}(t) + f_\mathbf{k}(t)+\sigma^u_\mathbf{k}\dot{W}^u_\mathbf{k},\label{SPEKF_u}\\
\frac{\d \gamma_\mathbf{k}(t)}{\d t} &= -d^\gamma_\mathbf{k}(\gamma_\mathbf{k}(t) -\bar\gamma_\mathbf{k}) + \sigma^\gamma_\mathbf{k}\dot{W}^\gamma_\mathbf{k},\label{SPEKF_gamma}\\
\frac{\d \omega_\mathbf{k}(t)}{\d t} &= -d^\omega_\mathbf{k}(\omega_\mathbf{k}(t) -\bar\omega_\mathbf{k}) + \sigma^\omega_\mathbf{k}\dot{W}^\omega_\mathbf{k},\label{SPEKF_omega}\\
\frac{\d b_\mathbf{k}(t)}{\d t} &= -d^b_\mathbf{k}(b_\mathbf{k}(t) -\bar{b}_\mathbf{k}) + \sigma^b_\mathbf{k}\dot{W}^b_\mathbf{k},\label{SPEKF_b}
\end{align}
\end{subequations}
where three additional stochastic processes $\gamma_\mathbf{k}(t)$, $\omega_\mathbf{k}(t)$ and $b_\mathbf{k}(t)$ are coupled to the governing equation of $\hat{u}_\mathbf{k}(t)$. They represent the stochastic damping, stochastic phase and stochastic forcing, respectively.
In \eqref{SPEKF}, $\hat{u}_\mathbf{k}(t)$ and $b_\mathbf{k}(t)$ are complex variables while $\gamma_\mathbf{k}(t)$ and $\omega_\mathbf{k}(t)$ are real-valued.
The function $f_\mathbf{k}(t)$ is a known time-periodic function that often represents seasonal effects.  The other parameters $\sigma^u_\mathbf{k}$, $d^\gamma_\mathbf{k}$, $\bar\gamma_\mathbf{k}$, $\sigma^\gamma_\mathbf{k}$, $d^\omega_\mathbf{k}$, $\bar\omega_\mathbf{k}$, $\sigma^\omega_\mathbf{k}$, $d^b_\mathbf{k}$, $\bar{b}_\mathbf{k}$, $\sigma^{b}_\mathbf{k}$ are all constants, where the three damping coefficients $d^\gamma_\mathbf{k}$, $d^\omega_\mathbf{k}$ and $d^b_\mathbf{k}$ are positive. In addition, $\dot{W}^u_\mathbf{k}$, $\dot{W}^\gamma_\mathbf{k}$, $\dot{W}^\omega_\mathbf{k}$ and $\dot{W}^b_\mathbf{k}$ are independent white noises.

It is important to note that if the damping $\gamma_\mathbf{k}$, the frequency $\omega_\mathbf{k}$ and the additional forcing $b_\mathbf{k}$ in \eqref{SPEKF_u} are all constants, then $\hat{u}_\mathbf{k}$ becomes an Ornstein–Uhlenbeck (OU) process \cite{gardiner1985handbook}, known as the mean stochastic model (MSM), which has linear dynamics and Gaussian statistics. However, intermittency and non-Gaussian features with extreme events and fat tails are often observed in complex turbulent systems. Therefore, $\gamma_\mathbf{k}$, $\omega_\mathbf{k}$ and $b_\mathbf{k}$ are parameterized by stochastic processes in \eqref{SPEKF}, which facilitate extra temporal variabilities in $u_\mathbf{k}$ that lead to rich nonlinear dynamical features and non-Gaussian statistics. Specifically, the alternating between positive and negative values in $\gamma_\mathbf{k}$ corresponds to the damping and the anti-damping of $\hat{u}_\mathbf{k}$, where the latter is crucial in triggering intermittency, extreme events and non-Gaussian distributions.
Similarly, the random evolution of the phase $\omega_\mathbf{k}$ advances $\hat{u}_\mathbf{k}$ 
{to have a wide range of the spectrum that is consistent with the typical characteristics of turbulent signals.} Since the primary role of $\gamma_\mathbf{k}$, $\omega_\mathbf{k}$ and $b_\mathbf{k}$ is to trigger intermittency and turbulent features of $\hat{u}_\mathbf{k}$, simple structures of these stochastic parameterizations are preferred that allow an efficient numerical integration. Thus, the OU processes are adopted to parameterize the time evolution of these three variables. Note that the interactions between $\gamma_\mathbf{k}$ and $\hat{u}_\mathbf{k}$ as well as $\omega_\mathbf{k}$ and $\hat{u}_\mathbf{k}$ are both quadratic nonlinear. Therefore the full SPEKF system \eqref{SPEKF} is nonlinear and $\hat{u}_\mathbf{k}$ can have non-Gaussian statistics.
Finally, it is worthwhile to remark that despite the nonlinearity in \eqref{SPEKF}, the time evolution of the moments can all be written down using closed analytic formulae, which provide an efficient and accurate statistical forecast of the SPEKF system \eqref{SPEKF} in data assimilation. Nevertheless, since the SPEKF model \eqref{SPEKF} only involves four state variables, a direct Monte Carlo simulation with a moderate ensemble size can also be utilized for the forecast, which will not significantly increase the computational cost.

It has been shown in \cite{branicki2018accuracy} that the SPEKF can compete with or outperform an optimally tuned 3DVAR algorithm in filtering turbulent signals, and it can overcome competing sources of error in a range of dynamical scenarios.  It has also been shown that the SPEKF model has much higher skill than the classical Kalman filter 
{using the MSM} in filtering and predicting signals with intermittency and extreme events \cite{majda2012filtering, majda2018model, chen2018conditional, branicki2012quantifying, branicki2013non}.

In the following, the SPEKF model \eqref{SPEKF} is adopted as the forecast model in the EnKF to filter each spectral mode of the observed state variables, where in total $J$ ensemble members are generated, denoted by $\{\hat{u}^{(j)}_\mathbf{k}(t), j=1,\ldots,J\}$. These ensemble members will also be utilized to create a set of inputs for the machine learning to seek the ensemble of the unobserved state variables.

\subsubsection{Estimating the Parameters in the SPEKF Model}\label{Subsec:SPEKF_ParameterEstimation}
Determining the model parameters is a prerequisite for applying the SPEKF model as the forecast model in data assimilation. As the large-scale $f(t)$ is often prescribed and $\sigma_\mathbf{k}^u$ can be estimated from computing the quadratic variation, the focus here is mainly on estimating the $9$ parameters in the three stochastically parameterized processes \eqref{SPEKF_gamma}--\eqref{SPEKF_b}.

In light of the special structure of the SPEKF model \eqref{SPEKF}, a simple iterative algorithm is developed to advance the estimation of the parameters in \eqref{SPEKF_gamma}--\eqref{SPEKF_b}. Given a time series of $\hat{u}_\mathbf{k}$ and the initial guess of the parameters in \eqref{SPEKF}, an efficient conditional sampling approach with a closed analytic formula can be applied to obtain a time series of $\gamma_\mathbf{k}$, $\omega_\mathbf{k}$ and $b_\mathbf{k}$ (see the Appendix for details). Next, the mean, variance, and decorrelation time (integration of the autocorrelation function) of  $\gamma_\mathbf{k}$ are computed from this time series, which are denoted by $m^\gamma_\mathbf{k}$, $E^\gamma_\mathbf{k}$ and $\tau^\gamma_\mathbf{k}$, respectively (similarly for those associated with $\omega_\mathbf{k}$ and $b_\mathbf{k}$). The three parameters in the OU process, i.e., $d^\gamma_\mathbf{k}$, $\bar{\gamma}_\mathbf{k}$ and $\sigma^\gamma_\mathbf{k}$, are uniquely determined by these three statistics,
\begin{equation}\label{Parameter_Estimation}
  d^\gamma_\mathbf{k} =1/\tau^\gamma_\mathbf{k},\qquad  \bar{\gamma}_\mathbf{k}=m^\gamma_\mathbf{k},\qquad\mbox{and}\qquad
  \sigma^\gamma_\mathbf{k}= \sqrt{2E^\gamma_\mathbf{k}/\tau^\gamma_\mathbf{k}}.
\end{equation}
The updated parameters from \eqref{Parameter_Estimation} are then combined with the conditional sampling method to improve the sampled trajectory. Repeating such a procedure a few times leads to the convergence of the estimated parameters.

\subsubsection{Filtering Sparse Regularly Spaced Observations with Aliasing}
Aliasing comes from the nature of sparse observations in space. In the presence of sparse observations, the noisy observation after transferring to spectral space is given by
\begin{equation}\label{Aliasing}
  \hat{v}_\mathbf{k} = g\hat{u}_\mathbf{k} + g\sum_{\mathbf{k'}\neq\mathbf{k},\mathbf{k'}\in\mathcal{A}_\mathbf{k}} \hat{u}_\mathbf{k'} + \sigma^o_\mathbf{k}
\end{equation}
where $\mathcal{A}_\mathbf{k}$ is the index set involving the spectral modes that belong to the same aliasing set as mode $\mathbf{k}$,  $\sigma^o_\mathbf{k}$ is the observational noise projected to spectral space, and $g$ is the observational operator. There are many ways of dealing with aliasing, which can be chosen depending on the properties of the underlying dynamics. See \cite{majda2012filtering} for a summary of these strategies.

The specific strategy utilized to filter the PQG model is the so-called reduced Fourier domain Kalman filter (RFDKF) \cite{grote2006stable}. The RFDKF approximation is based on the intuitive idea that for sufficiently rapid decay in the spectrum of the true turbulent signal, the primary mode contains the most energy, so only this mode should be actively filtered. Thus, RFDKF always trusts the dynamics for all the aliased modes yielding a Kalman gain vector with only the first component (corresponding to the primary mode) being non-zero.
Utilizing the RFDKF, the second term on the right-hand side of \eqref{Aliasing} is replaced by the known equilibrium mean of the aliased modes in the analysis step. Thus, a simple relationship between the observation $\hat{v}_\mathbf{k}$ and the primary mode $\hat{u}_\mathbf{k}$ appears.

\subsection{Filtering the Unobserved Variables Using a Machine Learning Model}
One of the main difficulties in filtering complex systems with only partial observations is the state estimation of the unobserved variables. Unlike the observed variables, where observations play an important role in directly mitigating the model error in the forecast, the observational information has only an indirect impact on rectifying the forecast error of the unobserved variables. Therefore, the skill of filtering these states heavily relies on the accuracy of the forecast model in characterizing the complex nonlinear dependence between the observed and unobserved variables.
\subsubsection{Discovering the Nonlinear Dependence Between Observed and Unobserved Variables Using Machine Learning}
To advance the characterization of such a  dependence without involving complicated parametric approximate models,  machine learning is adopted to develop an effective surrogate model. Given the filtered ensembles of the observed variables for each spectral mode resulting from the EnKF, machine learning aims to find the corresponding posterior ensembles of the unobserved state variables.

The machine learning model can be regarded as a generalization of the linear analysis of linear partial differential equations (PDEs) with at least two sets of state variables $\mathbf{u}(\mathbf{x},t)$ and $\mathbf{h}(\mathbf{x},t)$, where the full solution can be written as a superposition of different eigenmodes,
\begin{equation}\label{EigenDecomposition}
  \left(
    \begin{array}{c}
      \mathbf{u}(\mathbf{x},t) \\
      \mathbf{h}(\mathbf{x},t) \\
    \end{array}
  \right) = \sum_{\mathbf{k}\in\mathcal{K},\alpha\in\mathcal{C}}\hat{v}_{\mathbf{k},\alpha}(t)\exp(i\mathbf{k}\mathbf{x})\mathbf{r}_{\mathbf{k},\alpha}.
\end{equation}
In \eqref{EigenDecomposition}, $\mathcal{C}$ is the set that contains the indices of different eigenmodes, the Fourier coefficient $\hat{v}_\mathbf{k}(t)$ is a scalar, and different state variables are linked via the eigenvectors $\mathbf{r}_{\mathbf{k},\alpha} = (\mathbf{r}_{\mathbf{k},\alpha}^\mathbf{u}, \mathbf{r}_{\mathbf{k},\alpha}^\mathbf{h})^\mathtt{T}$. For a fixed wavevector
, the spectral coefficients of the two state variables are given by the following linear relationship
\begin{equation}
  \left(
    \begin{array}{c}
      \hat{\mathbf{u}}_\mathbf{k}(t) \\
      \hat{\mathbf{h}}_\mathbf{k}(t) \\
    \end{array}
  \right) = \sum_{\alpha\in\mathcal{C}}\hat{v}_{\mathbf{k},\alpha}(t)\mathbf{r}_{\mathbf{k},\alpha}.
\end{equation}
Therefore, $\hat{\mathbf{h}}_\mathbf{k}(t)$ is uniquely determined by $\hat{\mathbf{u}}_\mathbf{k}(t)$ and $\mathbf{r}_{\mathbf{k},\alpha}$.

For complex nonlinear dynamical systems, the above simple relationship is generally not held. Nevertheless, machine learning becomes a natural choice to build a nonlinear relationship between $\hat{\mathbf{u}}_\mathbf{k}(t)$ and $\hat{\mathbf{h}}_\mathbf{k}(t)$ for each fixed $\mathbf{k}$. Then, in light of the filtered solution $\hat{\mathbf{u}}_\mathbf{k}(t)$ from the EnKF, the machine learning model provides an ensemble of $\hat{\mathbf{h}}_\mathbf{k}(t)$.

Below, a recurrent neural network (RNN) \cite{dupond2019thorough} is adopted as the machine learning model to build the link between  $\hat{\mathbf{u}}_\mathbf{k}$ and $\hat{\mathbf{h}}_\mathbf{k}$:
\begin{equation}\label{NN_Model}
  \hat{\mathbf{h}}^{(j)}_\mathbf{k}(t) = \mbox{RNN}(\hat{\mathbf{u}}^{(j)}_\mathbf{k}(t-m:t)),
\end{equation}
where the input is one ensemble member of $\hat{\mathbf{u}}_\mathbf{k}(t-m:t)$ and the output is that of $\hat{\mathbf{h}}_\mathbf{k}(t)$. Running the machine learning model \eqref{NN_Model} $J$ times, an ensemble of the unobserved state variable $\hat{\mathbf{h}}_\mathbf{k}(t)$ can be created, which is denoted by $\{\hat{\mathbf{h}}^{(j)}_\mathbf{k}(t), j = 1,\ldots, J\}$.

It is essential to highlight that a time series of the observed variable $\hat{\mathbf{u}}_\mathbf{k}$ from $t-m$ to $t$ is utilized to find the corresponding value of the unobserved state variable $\hat{\mathbf{h}}_\mathbf{k}$ at time $t$. The length of the input can be chosen as one decorrelation time of the true signal. Such a memory effect in the input of the neural network allows us to take into account the dynamics of the observed variable in recovering the unobserved state variable, which is a crucial part that involves more information from physics.

\subsubsection{Creating the Input in Machine Learning Model}
The RNN \eqref{NN_Model} aims to capture the nonlinear dependence between $\hat{\mathbf{u}}_\mathbf{k}$ and $\hat{\mathbf{h}}_\mathbf{k}$. Therefore, during the training period, the trajectories of the mode $\mathbf{k}$ from the underlying dynamics \eqref{Starting_System} are naturally utilized as the input and the output of the RNN.
As the input in the RNN \eqref{NN_Model} requires a trajectory of the observed state variable $\hat{\mathbf{u}}_\mathbf{k}$ up to the current time instant, the ensemble members of $\hat{\mathbf{u}}_\mathbf{k}$ resulting from the EnKF using the SPEKF can be used as the input during the testing period. Yet, since filtering utilizes only the information from the past, such input may not fully capture the underlying dynamics. To create ensembles that are more dynamically consistent with the truth, an ensemble Kalman smoother (EnKS) \cite{evensen2000ensemble} is adopted to correct the error in the ensemble members from the EnKF. The resulting ensemble members are then fed into the RNN \eqref{NN_Model}.

\subsubsection{Incorporating the Machine Learning Training Residual into the Uncertainty of the Posterior Distribution}
Due to the turbulent nature and the intrinsic unpredictable components, the training residual from the machine learning model needs to be taken into account as an additional source of the uncertainty in the state estimation of the unobserved variable $\hat{\mathbf{h}}_\mathbf{k}(t)$. The deterministic forecast from the RNN can be regarded as an analog of the mean part of the solution associated with a knowledge-based stochastic model in the content of the classical mean-fluctuation decomposition \cite{muller2006equations}. In contrast, the training residual mimics the fluctuation that the RNN cannot fully characterize. A separate validation period is included in the training to prevent the over-fitting of the RNN, which will bring about the underestimation of the uncertainty in the machine learning model. An appropriate RNN leads to a comparable level of the residual in such an independent validation period as in the training period for calibrating the RNN.

To compute this additional uncertainty, each ensemble member of the unobserved state variable $\hat{\mathbf{h}}^{(j)}_\mathbf{k}(t)$ from \eqref{NN_Model} is modified by adding a non-Gaussian distribution $\epsilon$ to the point-wise value,
\begin{equation}\label{eq: mean}
	p(\hat{\mathbf{h}}^{(j)}_\mathbf{k}(t)) = \hat{\mathbf{h}}^{(j)}_\mathbf{k}(t) + \epsilon,
\end{equation}
where $\epsilon$ is the distribution of residual in the machine learning model \eqref{NN_Model} during the training period. One simple way to obtain the uncertainty is to compute
\begin{equation}\label{eq: residual}
	\epsilon \sim \textrm{PDF of } \Big(\hat{\mathbf{h}}_\mathbf{k}(t^\prime) - \mbox{RNN}(\hat{\mathbf{u}}_\mathbf{k}(t^\prime-m:t^\prime)),\quad\mbox{for}~ m < t \leq T \Big),
\end{equation}
where $T$ is the total length of the training period \cite{chen2021bamcafe}.
The posterior state estimate of the unobserved variable $\hat{\mathbf{h}}_\mathbf{k}(t)$ is then represented by a non-Gaussian distribution. It is a mixture distribution, where each mixture component is another non-Gaussian distribution associated with one ensemble member computed from \eqref{eq: mean}--\eqref{eq: residual}. Such uncertainty quantification is utilized in the numerical tests below. Note that \eqref{eq: residual} assumes that the 
residual $\epsilon$ is a constant over time, which is a crude but simple approximation. A more refined approach to determine the uncertainty is to let $\epsilon$ depend on the input $\hat{\mathbf{u}}_\mathbf{k}(t^\prime-m:t^\prime)$. This can be achieved by first partitioning the input $\hat{\mathbf{u}}_\mathbf{k}(t^\prime-m:t^\prime)$ into several clusters in the training period and then computing the distribution of $\hat{\mathbf{h}}_\mathbf{k}(t^\prime) - \mbox{RNN}(\hat{\mathbf{u}}_\mathbf{k}(t^\prime-m:t^\prime)$ within each cluster to form a specific $\epsilon$. In the testing period, find the cluster that the input belongs to and then add the corresponding value of $\epsilon$ to $\hat{\mathbf{h}}^{(j)}_\mathbf{k}(t)$.

As a final remark, the residual or the uncertainty associated with the RNN depends on the skill of the RNN in discovering the nonlinear relationship between $\hat{\mathbf{u}}_\mathbf{k}$ and $\hat{\mathbf{h}}_\mathbf{k}$. A more skillful RNN results in a smaller uncertainty. If the RNN can perfectly recover the underlying dynamics, then such uncertainty vanishes.

\section{Precipitating Quasigeostrophic (PQG) Equations}\label{Sec:PQG}
\subsection{The model}

The PQG model is a recently developed moist version of the quasigeostrophic (QG) model \cite{smith2017precipitating, edwards2020spectra, hu2021initial}
to describe synoptic-scale ($\approx 1000$ km) dynamics at mid-latitudes.  Beyond the classical dry QG dynamics, PQG includes additional physics and dynamics associated with water vapor, clouds, phase changes between vapor and liquid, and rainfall. Thus, the PQG model can more realistically describe large-scale, mid-latitude weather patterns.

In the dry QG model, there is a single prognostic variable called the potential vorticity $(PV)$ \cite{salmon1980baroclinic, qi2016low, vallis2017atmospheric}. All other variables in the systems, namely the horizontal winds and potential temperature, can be diagnostically recovered from $PV$ by inverting a linear elliptic operator in so-called $PV$-inversion.  With the addition of water, the PQG model requires an additional prognostic variable $M$, and all other variables (horizontal winds, potential temperature and water) are found from $PV$-and-$M$ inversion of a nonlinear elliptic operator.   The nonlinear terms in the elliptic operator arise from the presence of phase interfaces separating unsaturated and saturated regions of the flow.
These phase boundaries lead to discontinuous coefficients in the elliptic operator, denoted by Heaviside nonlinear functions.

\begin{figure}[H]
\begin{tikzpicture}[scale=.6,every node/.style={minimum size=1cm},on grid]
    \begin{scope}[
     yshift=-170,every node/.append style={
            yslant=0.5,xslant=-1},yslant=0.5,xslant=-1
            ]
                \fill[white,fill opacity=.5] (0,0) rectangle (5,5);
        \draw[black, thick] (0,0) rectangle (5,5);
        \draw [dashed,red] (0,1.5) parabola bend (2.5,2.5) (5,1.5) ;
    \draw [dashed,red](0,.5) parabola bend (2,1) (5,1)  ;
        \draw [dashed,red] (0,2) parabola bend (2.7,2.7) (5,2)  ;
        \draw [dashed,red] (0,2.5) parabola bend (3.5,3.5) (5,2.5)  ;
        \draw [dashed,red] (0,3.5)  parabola bend (2.75,4.5) (5,3.5);
        \draw [dashed,red] (0,4)  parabola bend (2.75,4.8) (5,4);
        \draw [dashed,red] (0,3)  parabola bend (2.75,3.8) (5,3);
    \end{scope}
    	
    \begin{scope}[
    	yshift=0,every node/.append style={
    	    yslant=0.5,xslant=-1},yslant=0.5,xslant=-1
    	             ]

        \fill[white,fill opacity=.5] (0,0) rectangle (5,5);
        \draw[black, thick] (0,0) rectangle (5,5);
                \draw [dashed,red] (0,1.5) parabola bend (2.5,2.5) (5,1.5) ;
    	        \draw [dashed,red](0,.5) parabola bend (2,1) (5,1)  ;
        \draw [dashed,red] (0,2) parabola bend (2.7,2.7) (5,2)  ;
        \draw [dashed,red] (0,2.5) parabola bend (3.5,3.5) (5,2.5)  ;
        \draw [dashed,red] (0,3.5)  parabola bend (2.75,4.5) (5,3.5);
        \draw [dashed,red] (0,4)  parabola bend (2.75,4.8) (5,4);
        \draw [dashed,red] (0,3)  parabola bend (2.75,3.8) (5,3);
    \end{scope}
    \begin{scope}[
               yshift=-83,every node/.append style={
        yslant=0.5,xslant=-1},yslant=0.5,xslant=-1
                  ]
        \draw[black,very thick] (0,0) rectangle (5,5);
        \draw[black, thick] (0,0) rectangle (5,5);
        \draw[step=5mm, black] (0,0) grid (5,5);
        \draw [fill=lime](0,0) circle (.1) ;
        \draw [fill=lime](0,5) circle (.1);
        \draw [fill=lime](5,0) circle (.1);
        \draw [fill=lime](5,5) circle (.1);
    \end{scope} 

    \begin{scope}[
    yshift=80,every node/.append style={
    yslant=0.5,xslant=-1},yslant=0.5,xslant=-1]
        \fill[white,fill opacity=.4] (0,0) rectangle (5,5);
        \draw[black,dashed] (0,0) rectangle (5,5);
    \end{scope}
    \begin{scope}[
    yshift=-240,every node/.append style={
    yslant=0.5,xslant=-1},yslant=0.5,xslant=-1]
        \fill[white,fill opacity=.4] (0,0) rectangle (5,5);
        \draw[black,dashed] (0,0) rectangle (5,5);
    \end{scope}
     \draw[-latex,dashed] (6.2,6) node[right]{Top boundary}
         to[out=180,in=90] (4,5.5);
    \draw[-latex,dashed] (6.2,-7) node[right]{Bottom boundary}
         to[out=0,in=270] (4,-6.5);

    \draw[-latex,thick] (6.2,2) node[right]{$\mathsf{Level\,2}: PV_2, \psi_2,\theta_{e,2},{\bf u}_2$}
         to[out=180,in=90] (4,2);

    \draw[-latex,thick](5.8,-.5)node[right]{$\mathsf{Middle}: M_m,\theta_{e,m},q_{vs,m},q_{r,m},H_u,H_s$}
        to[out=-180,in=-90] (3.9,-1);
    \draw[-latex,thick](5.5,-4.5) node[right]{$\mathsf{Level\,1}: PV_1, \psi_1,\theta_{e,1},{\bf u}_1$}
        to[out=180,in=90] (2,-5);	
\end{tikzpicture}
    \caption{Illustration of the two-level setup of the PQG system.}
    \label{fig:two-level-setup}
\end{figure}
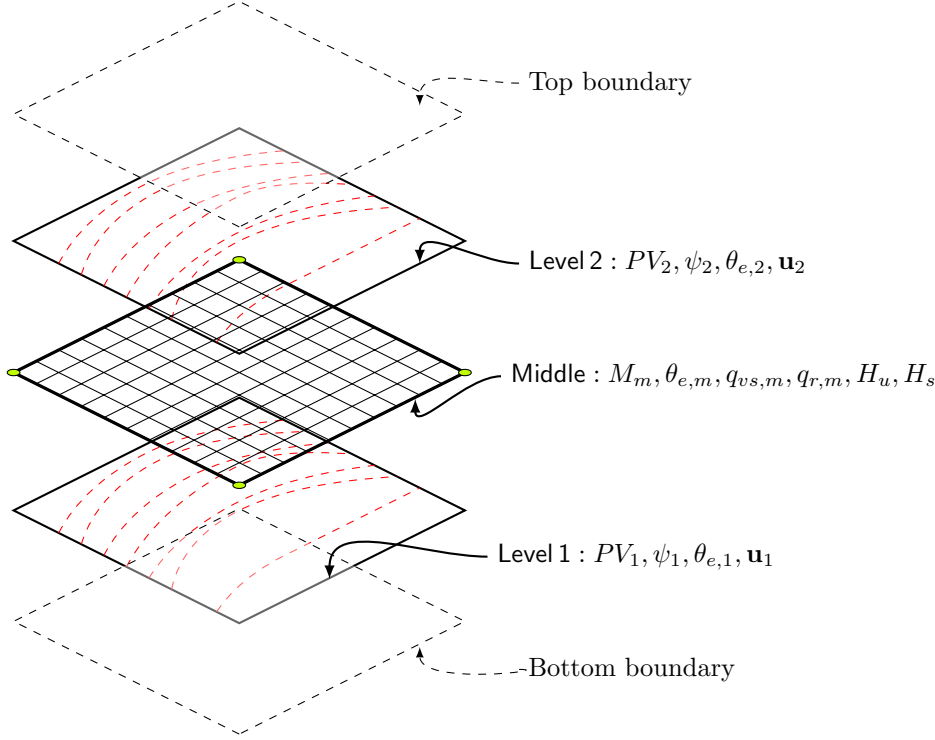
A two-level version of the PQG equations can be derived using a staggered grid in the vertical direction \cite{hu2021initial}, as illustrated in Figure~\ref{fig:two-level-setup}. The subscript $(\cdot)_{j}, j=1,2$ indicates the variables defined at level 1 or level 2, and the subscript $(\cdot)_{m}$ indicates those at the middle level.
The governing equations are given by
\begin{align}
    &\begin{aligned}
     \frac{\partial  { PV_{1}}}{\partial t}
     +
     J(\psi_1,{ PV_{1}})
   -
   U\frac{\partial { PV_{1}}}{\partial x}
   +
   &\beta v_1
   +
      v_1\frac{\partial { PV_{1,bg}}}{\partial y}
    =
    \\
     -
    &{
    \frac{L_{du}}{L_{ds}}\frac{L}{L_{ds}}\frac{\partial\textbf{u}_h}{\partial z}\cdot\nabla_h\theta_{e,1}
    }
    -\kappa\Delta_h \psi_1
   -\nu\Delta_h^{4} {PV_{1}},
\end{aligned}
    \label{eqn:layer-pqg-1}
    \\
    &\begin{aligned}
     \frac{\partial { PV_{2}}}{\partial t}
     +
   J(\psi_2, {PV_{2}})
    +
    U\frac{\partial  { PV_{2}}}{\partial x}
    +
    &\beta v_2
    +      v_2\frac{\partial { PV_{2,bg}}}{\partial y}
    =
    \\
    -
    &{
    \frac{L_{du}}{L_{ds}}\frac{L}{L_{ds}}\frac{\partial\textbf{u}_h}{\partial z}\cdot\nabla_h\theta_{e,2}
    }
    -\nu\Delta_h^{4}  { PV_{2}},
\end{aligned}
    \label{eqn:layer-pqg-2}
\\
    &\begin{aligned}
    {
     \frac{D_m M_m}{Dt} +
     v_m\frac{\partial M_{bg}}{\partial y}
    =-\frac{V_r}{\Delta z}q_{r,m}-\nu\Delta_h^{4} M_m +E
    },
    \label{eqn:layer-pqg-3}
    \end{aligned}
\end{align}
where $PV_{1,2}$ and $M_m$ are related to a streamfunction $\psi_{1,2}$ by nonlinear, elliptic operators (see (\ref{eqn:pv-m-inversion-1}) and (\ref{eqn:pv-m-inversion-2}) below).  From the streamfunction $\psi_{j}$, the horizontal winds ${\bf u}_{h,j} = (u_j,v_j)$ are obtained from the relations
$u_j = -\partial \psi_j/\partial y$, $v_j = \partial \psi_j/\partial x$ with mid-level values $u_m = (u_1+u_2)/\Delta z$, $v_m = (v_1+v_2)/\Delta z$, where $\nabla_h$ is the horizontal part of the gradient operator and $\Delta z$ is the distance between levels 1 and 2. The
equivalent potential temperature $\theta_e$ (a linear combination of potential temperature and water vapor) is found from the streamfunction $\psi_j$ and $M_m$ using the relation
\begin{equation}
\theta_{e,m} = H_s\left(\frac{L}{L_{du}}\frac{\psi_2-\psi_1}{\Delta z}+q_{vs,m}\right)+H_u\left(\frac{1}{1+G_M}\frac{L}{L_{du}}\frac{\psi_2-\psi_1}{\Delta z}+\frac{1}{1+G_M}M_m\right),
\label{eqn:pv-m-inversion-3}
\end{equation}
where $G_m$ is an $O(1)$ parameter related to the background profiles, $L$ is the reference length scale (1000 km), and $L_{du}$ is the Rossby radius of deformation associated with the unsaturated background state.

The total water mixing ratio is $q_{t,m} = q_{v,m} + q_{r,m}$, where $q_{v,m}, q_{r,m}$ are the vapor and liquid components, respectively.
In terms of $M_m$ and $\theta_{e,m}$, $q_{t,m} = M_m - G_M \theta_{e,m}$.
A saturation profile $q_{vs,m}$ separates unsaturated regions (water vapor only) from saturated regions (vapor and rain water).  In the current two-level model, $q_{vs,m}$ may be written in  terms of the streamfunction as
\begin{align}
&q_{vs,m} = q_{vs}^0+q_{vs}^1 \frac{\psi_2-\psi_1}{\Delta z },
\label{def:qvs-2level}
\end{align}
where the parameters $q_{vs}^0$ and $q_{vs}^1$ are non-negative constants.  The absence or presence of rain $q_{r,m}$ is diagnosed from $M_m, \theta_{e,m}$ and $q_{vs,m}$ according to the relation
$q_{r,m} = \max(0,M-G_M \theta_{e,m}-q_{vs,m})$.
The cloud indicator (Heaviside) functions are then
\begin{align}
    &H_s =\begin{cases}
        1 & \text{if } q_r>0\\
        0 & \text{if } q_r=0
    \end{cases},  &
        &H_s =\begin{cases}
        0 & \text{if } q_r>0\\
        1 & \text{if } q_r=0
    \end{cases}.
    \label{def:cloudindicator}
\end{align}
At the top and bottom, $PV_{1,2}, M_m$ and all associated unknowns obey periodic boundary conditions in the horizontal directions $(x,y)$.  The top and bottom boundary conditions are derived from a rigid-lid condition (no flow through the boundaries), material invariance of $\theta_e$, and the governing transport equation for total water $q_{t,m}$ including a rainfall term.  A simplified version replaces the latter two conditions with $q_{t,m} = \theta_{e,m} = 0$.

As alluded to above, the model assumes a prescribed background state with
characteristic length-scale $L = 1000$ km. This
environmental state is partly described by the parameter $G_m$, which is the ratio of vertical variations in potential temperature and total water profiles (both linear).
In addition, $\beta$ characterizes the meridional variation of the Coriolis parameter in a narrow mid-latitude band, $U$ is the vertical shear of flow in the zonal direction, and the length scales $L_{du}, L_{ds}$ are the Rossby deformation radii associated with unsaturated and saturated flow regions, respectively.  Given parameters $U, G_m, L, L_{du}$ and $q_{vs}^1$, the background profiles $PV_{j,bg}$ and $M_{m,bg}$ are
\begin{align}
    &PV_{j,bg} = (-1)^j(1+q_{vs}^1)\frac{1}{(\Delta z)^2}\frac{L^2}{L^2_{du}} (2Uy)  ,
    \\
    &\begin{aligned}
    M_{bg}
    & = -(q_{vs}^1+G_M(1+q_{vs}^1))\frac{1}{\Delta z} \frac{L}{L_{du}} (2Uy),
    \end{aligned}
\end{align}
where $q_{vs}^1$ appears in (\ref{def:qvs-2level}) as the threshold for phase changes. The remaining (constant) parameters in (\ref{eqn:layer-pqg-1})-(\ref{eqn:layer-pqg-3}) are the evaporation source term $E$ which regulates the cloud fraction in this setup, as well as dissipation terms parameterized by $\kappa$ (friction at the bottom level) and $\nu$ (dissipation of small-scale turbulence).

To close the PQG system, it is necessary to perform $PV$-and-$M$ inversion to find the streamfunction $\psi$ and associated quantities ${\bf u}_h, \theta_e, q_t$, etc.  In the two-level setup, the nonlinear elliptic equations relating $PV$ and $M$ to the streamfunction $\psi$ are given by
\begin{align}
&\begin{aligned}
    PV_{1} &= \nabla_h^2 \psi_1+
    H_s\left(
    \left(\frac{L}{L_{ds}}\frac{1}{\Delta z}\right)^2(\psi_2-\psi_1)+\frac{L_{du}}{L_{ds}}\frac{L}{L_{ds}}\frac{1}{\Delta z} q_{vs,m}
    \right)+
    \\
    &H_u\left(
    \left(\frac{L}{L_{du}}\frac{1}{\Delta z}\right)^2(\psi_2-\psi_1)+\frac{L}{L_{du}}\frac{1}{\Delta z} M_{m}
    \right)
    \label{eqn:pv-m-inversion-1}
\end{aligned}
\\
&\begin{aligned}
    PV_{2} &= \nabla_h^2 \psi_2+
    H_s\left(
    \left(\frac{L}{L_{ds}}\frac{1}{\Delta z}\right)^2(\psi_1-\psi_2)-\frac{L_{du}}{L_{ds}}\frac{L}{L_{ds}}\frac{1}{\Delta z} q_{vs,m}
    \right)+
    \\
    &H_u\left(
    \left(\frac{L}{L_{du}}\frac{1}{\Delta z}\right)^2(\psi_1-\psi_2)-\frac{L}{L_{du}}\frac{1}{\Delta z} M_{m}
    \right).
    \label{eqn:pv-m-inversion-2}
\end{aligned}
\end{align}
{As discussed in Section \ref{subsec:numericalsolverPQG}, $PV$-and-$M$ inversion is the most expensive part of PQG numerical solution, and thus a strong motivation for exploring inexpensive and efficient surrogate models as described herein. }

\subsection{Model properties}

The PQG model was derived as a distinguished asymptotic limit of the rotating Boussinesq equations with idealized cloud microphysics \cite{smith2017precipitating}. As such, the model retains the limiting (small-Rossby number) dynamics of a moist atmosphere under the conditions of strong moist stratification and rapid rotation.
Energetics of the continuous PQG model and its two-level version have been analyzed in \cite{smith2017precipitating,hu2021initial}.
In particular, the moist potential energy may be decomposed into three parts representing the familiar unsaturated and saturated potential energies, as well as a moist latent energy that is released upon change of phase.

Initial studies of the two-level PQG system have tested the model's ability to capture canonical large-scale phenomena associated with important weather events.  In
 \cite{edwards2020rivers}, purely saturated computations exhibited the formation of atmospheric rivers (ARs). In nature, ARs are long narrow corridors, typically associated with extratropical cyclones, and accounting for most of the poleward transport of water from the tropics to the mid-latitudes.   Standard AR-identification algorithms found ARs in the saturated PQG two-level system in the presence of sufficiently strong meridional moisture gradient and precipitation. Variability of the mid-latitute zonal jet was studied in \cite{hu2021initial}, including phase changes of water with small-to-moderate cloud fractions in the range [0,0.2].  Different behaviors were observed depending on cloud fraction, including poleward propagation of the latitude of the jet.

\subsection{Numerical solver of the PQG equations}
\label{subsec:numericalsolverPQG}

Here we give a brief overview of the numerical methods used to solve the system (\ref{eqn:layer-pqg-1})-(\ref{eqn:pv-m-inversion-2}),
following \cite{hu2021initial}.
The evolution equations (3.1)-(3.3) are discretized using a pseudospetral method.  A third-order Runge Kutta scheme is used for the temporal evolution with time step $\Delta t = 0.01$ satisfying the CFL stability criterion.  During this portion of the algorithm, computation of the nonlinear terms is the most expensive component, and is achieved using FFTs.

The $PV$-and-$M$ inversion (\ref{eqn:pv-m-inversion-1})-(\ref{eqn:pv-m-inversion-2}) is treated in physical space.
In particular, a standard centered difference method is coupled with the conjugate gradient method to iteratively solve for the updated streamfunction $\psi$ and updated cloud indicator function $H_s$, the latter which depends on the updated streamfunction itself.  The iterative procedure starts with $H_s$ evaluated at the previous time step.  Given updated $PV_{1,2}$, $M$ and previous $H_s$, the discretized nonlinear elliptic equations (\ref{eqn:pv-m-inversion-1})-(\ref{eqn:pv-m-inversion-2}) are inverted to find $\psi$ using conjugate gradient.  Then (\ref{eqn:pv-m-inversion-3})-(\ref{def:qvs-2level}), the relation $q_r = \max(0,M-G_M \theta_{e,m}-q_{vs,m})$ and (\ref{def:cloudindicator}) are used to check consistency of the previous and updated $H_s$. If they are not the same, the inversion is repeated starting from $PV_{1,2}$, $M$ and the updated $H_s$. The procedure is iterated until convergence of $H_s$. 
{This iterative algorithm is the most computationally expensive element of the solution algorithm as a whole, which becomes prohibitive for forecast modeling.}

\section{Data Assimilation of PQG with Partial Observations}\label{Sec:DA}
\subsection{Setup}

\subsubsection{Model setup}

We consider a standard mid-latitude setting with parameter values that are consistent with previous two-level computations in dry and moist settings \cite{qi2016low,hu2021initial} (see
Table~\ref{table:param-pqg}).
All parameters are fixed except for the evaporation rate $E$ in
(\ref{eqn:layer-pqg-3}), which acts as a source of moisture.  The value of $E$ determines a statistical equilibrium in which the evaporation source $E$ balances the moisture sink from rainfall, on average.
Thus, a larger value of $E$ leads to larger cloud fraction
and more rainfall.
In the following data assimilation tests, the main focus will be on the case $E=0.2$, corresponding to $17\%$ cloud fraction on average over the entire domain (see Section~\ref{ss:result}). This moderate cloud fraction is consistent with the realistic mid-latitude atmosphere. The data assimilation skill will also be tested on experiments with more extreme cases: $E=0.02$ and $E=0.35$ (see Section~\ref{ss-sen}).

\begin{table}[H]
    \centering
    \begin{tabular}{c|c|c|c|c|c|c|c|c|c|c|c|c}
    \hline
     Parameters& $N$  &$\beta$ & $G_M$ &$q_{vs}^0$  &$q_{vs}^1$ &$L$ &$L_{ds}$&$L_{du}$ &$\Delta z$ & $V_m$ & $\kappa$ & $U$\\
      \hline
     Values &$128$ &$2.5$ & $1.0$& $1.0$& $1.0$ & $1.0$ &$1/\sqrt{2}$&$1/2$ & $0.5$ &$1.0$ &$0.05$ &0.1
    \\
      \hline
    \end{tabular}
    \caption{Parameters in the PQG equations.}
    \label{table:param-pqg}
\end{table}

\subsubsection{Data assimilation method}
The ensemble transform Kalman filter (ETKF) \cite{bishop2001adaptive} is adopted as the EnKF method
for filtering the observed component of each of the spectral modes (i.e., the time series of the Fourier coefficient), where the SPEKF model is utilized as the forecast model. The data assimilation is carried out in Fourier space.

\subsubsection{Partial and sparse observations}
The lower layer PV, namely $PV_1$, is chosen to be the observed state variable. In contrast, there are no direct observations for $PV_2$, $M$, and other state variables. Although the model resolution in the horizontal direction is $128\times 128$, only sparse observations at $16 \times 16$ regularly spaced grid points are available in the following data assimilation tests, mimicking realistic situations of atmospheric observations, usually on coarse-grained mesh grids. The reduced Fourier domain Kalman filter (RFDKF) is utilized to deal with the aliasing issue \cite{grote2006stable} inside the ETKF, as the energy decays sufficiently fast outside the range of the resolved modes (see Panel (a) of Figure \ref{fig1:spectrum}) . The primary goal is to recover the leading  $16 \times 16$ Fourier modes via data assimilation.

\subsubsection{Ensemble size, observational time step and observational noise}
\begin{figure}[H]
     \centering
         \includegraphics[width=.48\textwidth]{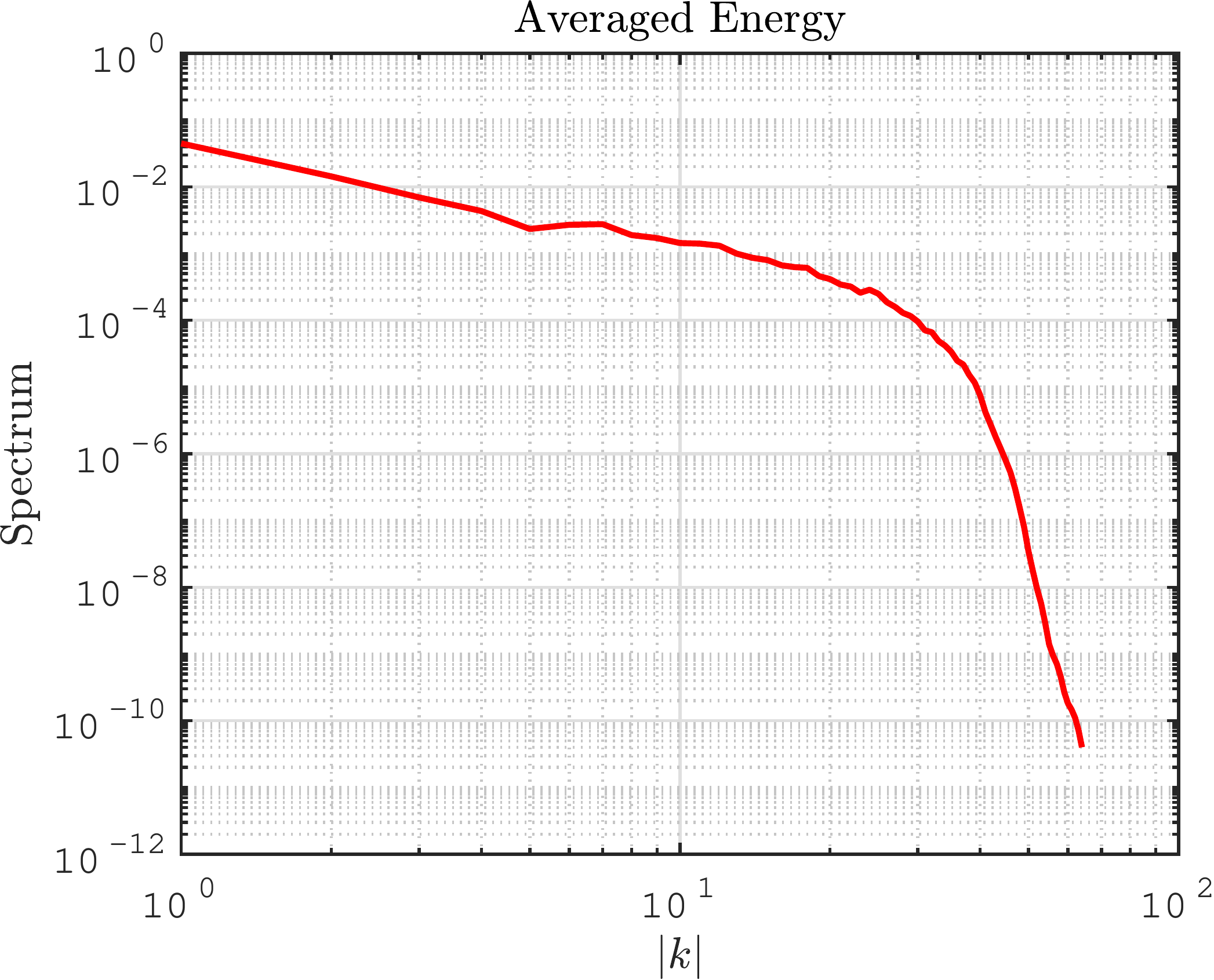}
         \includegraphics[width=.46\textwidth]{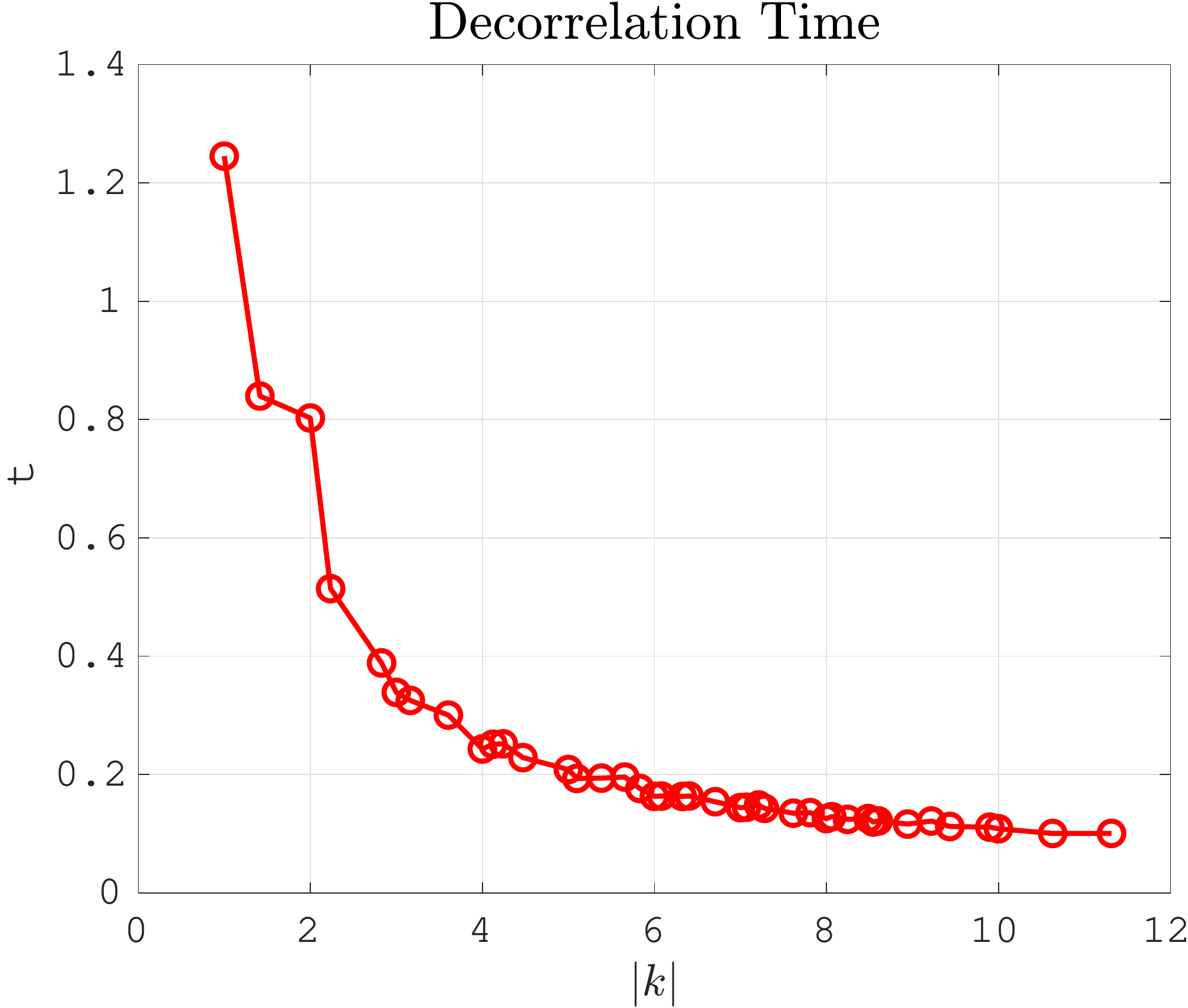}
         \caption{Left panel:
         spectrum of $PV_1$; right panel: decorrelation time of $PV_1$. The energy is computed by taking the average over the modes with the same $|\mathbf{k}|$, which is the variable for the x-axis.
         }
         \label{fig1:spectrum}
\end{figure}

The size of the ensemble number is $50$, the observational time step is chosen to be $\Delta t^{obs}= 0.3$, and the observational noise is $25\%$. 
Figure \ref{fig1:spectrum} shows the spectrum and the decorrelation time of the observed variable, $PV_1$, which are shown as a function of the absolute value of the wavenumber $|k| =\sqrt{k_1^2+k_2^2}$. Note that the observational time step $\Delta t^{obs}= 0.3$ is smaller than the decorrelation time of modes with $|k|\leq 3$ while the decorrelation times of remaining modes in the resolved spatial scale are between $0.3$ and $0.1$. The data assimilation results with different observational time steps and observational noise will be included in Section~\ref{ss-sen}.

\subsubsection{Neural network setup}

The long short-term memory (LSTM) neural network \cite{hochreiter1997long} is adopted here as the RNN. To train the LSTM for each Fourier mode (namely the time series of the Fourier coefficients), $400$ time unit data ($4\times 10^4$ data points) is adopted in the training stage, and $100$ time unit data ($10^4$ data points) is utilized in the validation stage to prevent overfitting, which retains the balance between the accuracy and computational cost. An empirical calibration is used here to systematically determine the LSTM's hidden units of each Fourier mode $(k_i,k_j)$. The number of hidden units $= (\tau^{decor}_{i,j}/\Delta t)\cdot (|k_i| +|k_j|)$, where $\tau^{decor}_{i,j}$ is the decorrelation time of the Fourier mode $(k_i,k_j)$ and $\Delta t$ is step size of PQG simulation. This criterion is motivated by the following two facts. First, the Fourier mode with a larger wave number, i.e., larger $k_i,k_j$ values, usually yields more intermittency. Second, a smaller decorrelation time $\tau^{decor}_{i,j}$ implies a shorter memory of the time series~\cite{kurikawa2020repeated}.

\subsubsection{Comparison with data assimilation using a traditional reduced order forecast model}
The hybrid data assimilation algorithm is compared with a traditional reduced-order model-based data assimilation method (hereafter ``traditional ROM''). In this traditional ROM method, a spectral representation of the original PQG system with a lower resolution of $16\times 16$ grids is developed. This is achieved by first projecting the system onto such a reduced-order basis, where the higher frequencies are truncated. Then the hyper-viscosity is increased to guarantee that the solution is stabilized on such a coarse-grained mesh. Additional stochastic noise is added to the equation of each spectral mode to match the equilibrium variance with the truth. Furthermore, the complicated $PV$-and-$M$ inversion is replaced by the linear inversion corresponding to dry QG. This reduces the computational cost and prevents the truncation error in the input from being significantly amplified by the Heaviside functions. This reduced-order model is then incorporated into the localized ETKF (LETKF) \cite{hunt2007efficient} in physical space for data assimilation. See the Appendix for more details.

This traditional ROM method is fundamentally different from the hybrid strategy. In the traditional ROM, the governing equation of each spectral mode contains the nonlinear interactions between different Fourier modes. In contrast, the SPEKF model in the hybrid strategy utilizes cheap stochastic parameterization to effectively compensate for these complicated nonlinearities. 
Furthermore, the traditional ROM uses reduced order parametric form to build the connection between the observed and unobserved state variables while the hybrid strategy exploits machine learning to achieve this goal.

\subsection{Skill scores in assessing data assimilation skill}
The following skill scores are utilized to quantify the data assimilation skill. They are the normalized root-mean-square error (RMSE) and pattern correlation (Corr) between the truth and the posterior mean time series. They are defined as follows:
\begin{equation}\label{SkillScores}
\begin{split}
  \mbox{Corr} &= \frac{\sum_{i=1}^n(u^M_i-\bar{u}^M)(u_i^{ref}-\bar{u}^{ref})}{\sqrt{\sum_{i=1}^n(u^M_i-\bar{u}^M)^2}\sqrt{\sum_{i=1}^n(u^{ref}_i-\bar{u}^{ref})^2}},\\
  \mbox{RMSE} &=  \frac{1}{\mbox{std}(u^{ref})}\sqrt{\frac{\sum_{i=1}^n(u^M_i-u^{ref}_i)^2}{n}},\end{split}
\end{equation}
where $u^M_i$ and $u^{ref}_i$ are the posterior mean and the truth, respectively, at time $t=t_i$, and $n$ is the total number of the data points to compute these statistics.  The time averages of the assimilated and the true time series are denoted by $\bar{u}^M$, and $\bar{u}^{ref}$, respectively, and $\mbox{std}(u^{ref})$ is the standard deviation of the true time series $u^{ref}$. The normalization used in the RMSE ensures that the skill score is a non-dimensional quantity. A smaller RMSE and a larger pattern correlation imply a more accurate solution. The filtering result is considered unskillful if the normalized RMSE exceeds $1$ or the pattern correlation is below a certain threshold, usually $0.5$.

In addition to the time series, the pattern correlation (Corr) between the truth and assimilated spatial fields can be calculated at a fixed time. It is defined as
\begin{equation}\label{SkillScores}
\begin{split}
  \mbox{Corr} &= \frac{\sum_{i=1}^{n_x}\sum_{j=1}^{n_y}(u^M_k (x_i,y_j)-\bar{u}_k^M)(u^{ref}_k(x_i,y_j)-\bar{u}_k^{ref})}{\sqrt{\sum_{i=1}^{n_x}\sum_{j=1}^{n_y}(u^M_k(x_i,y_j)-\bar{u}_k^M)^2}\sqrt{\sum_{i=1}^{n_x}\sum_{j=1}^{n_y}(u_k^{ref}(x_i,y_j)-\bar{u}_k^{ref})^2}},
  \end{split}
\end{equation}
where $u^M_k(x_i,y_j)$ and $u^{ref}_k(x_i,y_j)$ are the posterior mean and the truth at time $t=t_k$, respectively, at spatial grid $(x_i,y_j)$, and $\bar{u}^M$ and $\bar{u}^{ref}$ are the spatial averages of the assimilated data and the truth at time $t=t_k$.

\subsection{Main data assimilation results\label{ss:result}}

\subsubsection{Data assimilation skill of different Fourier modes}
\begin{figure}[H]
  \centering
  \begin{tabular}{@{}c@{}}
         \includegraphics[width=.95\textwidth]{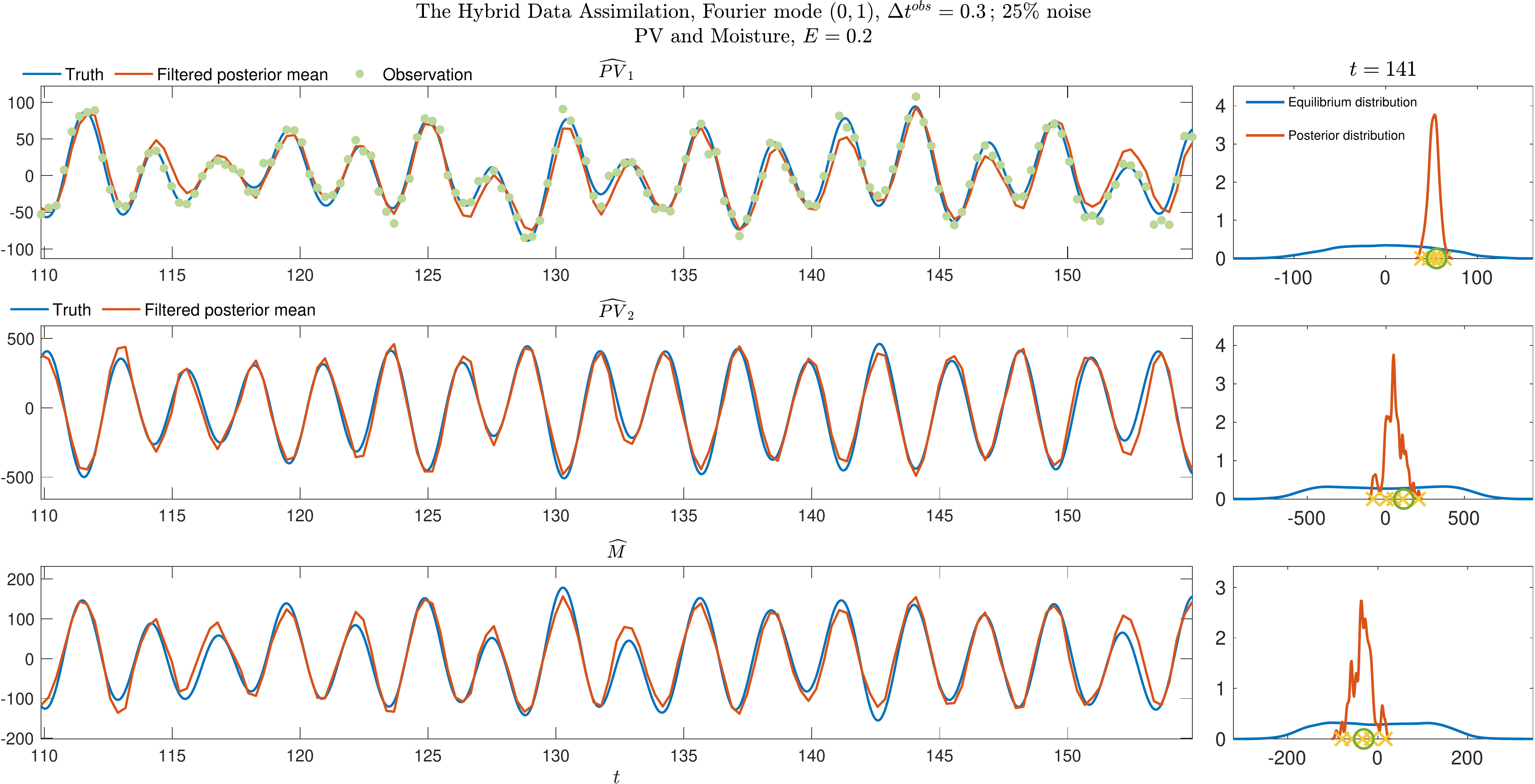}
 \\[\abovecaptionskip]
    \small (a) Fourier mode $(0,1)$
  \end{tabular}

  \vspace{\floatsep}

  \begin{tabular}{@{}c@{}}
         \includegraphics[width=.95\textwidth]{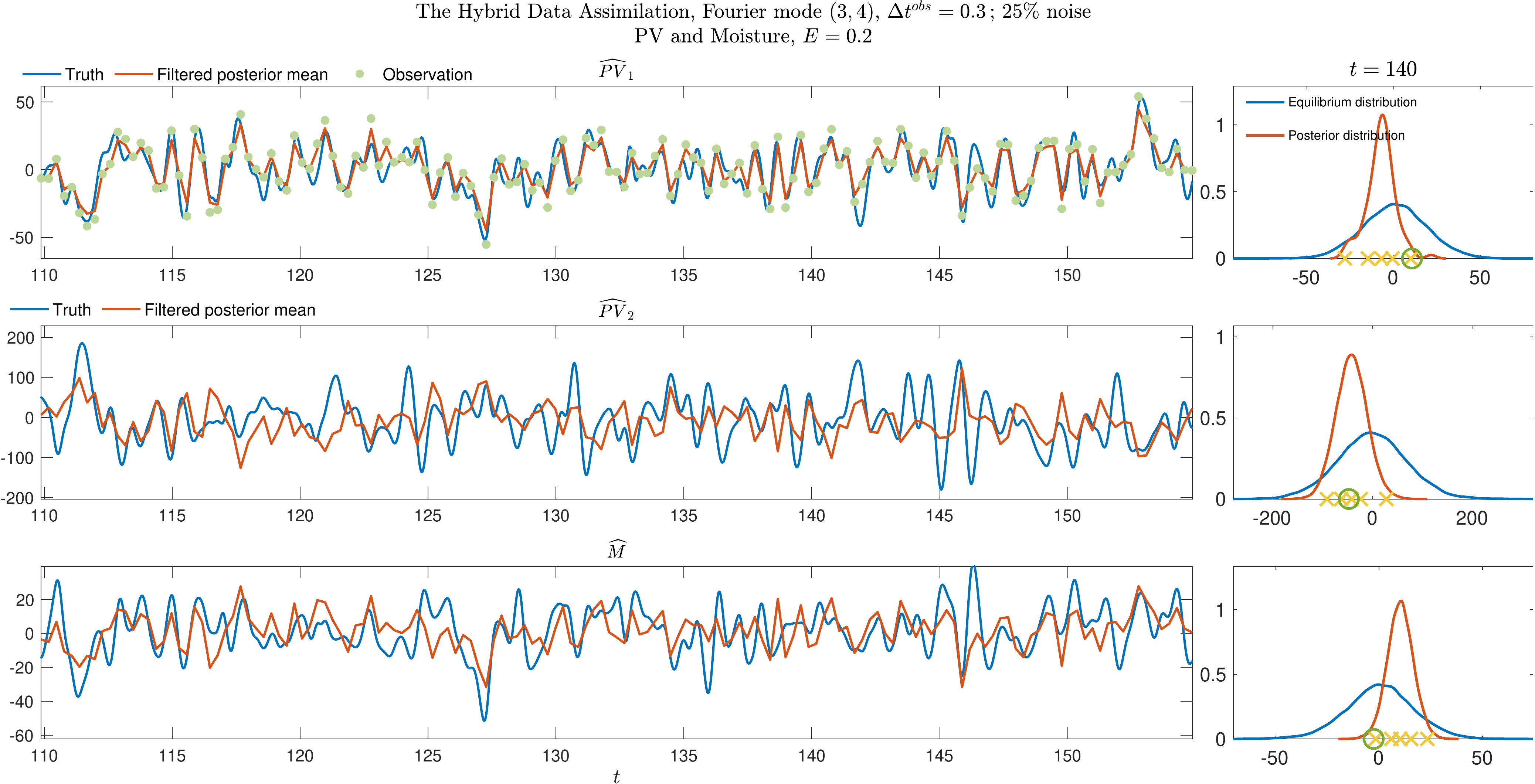}
 \\[\abovecaptionskip]
    \small (b) Fourier mode $(3,4)$
  \end{tabular}
  \caption{Time series and mixture posterior distribution of observed variable $PV_1$ and unobserved variables $PV_2$ and moisture $M$ at Fourier modes $(0,1)$ and $(3,4)$ with observation time $\Delta t^{obs} = 0.3$ and $25\%$ observation noise. The observations (green dots) are the projected values from physical space to each Fourier mode. Left panel: times series of truth (blue), filter (red), of $PV_1,PV_2$ and moisture $M$, and observation (green) for $PV_1$; right panel: equilibrium distribution and uncertainty distribution at $t=141$ for each variable, the green circle denotes the truth value and the yellow crosses denote $1$st, $16$th, $50$th, $84$th, and $99$th percentile of the point-wise ensemble realization values. 
  }
         \label{fig2-time-series}
\end{figure}

Figure \ref{fig2-time-series} shows the posterior mean time series resulting from the hybrid data assimilation algorithm (red), which is compared with the truth generated from the PQG model \eqref{eqn:layer-pqg-1}--\eqref{eqn:layer-pqg-3} (blue). For the largest scale mode $(0,1)$, the posterior mean time series captures the quasi-regular oscillation patterns for both the observed and unobserved state variables. In addition, the posterior uncertainty is quite small, indicating the confidence of the posterior mean estimate.  With the increase of $|k|$ (corresponding to smaller scales), the data assimilation skill deteriorates, as expected. Nevertheless, although the decorrelation time of mode $(3,4)$ approaches the observational time step, the recovery of the observed state variable $PV_1$ remains quite skillful. This is because observations play a direct role in correcting the forecast error. It is worth highlighting that most of the extreme events are captured accurately by the data assimilation solution since the SPEKF forecast model has the mechanism for predicting intermittency with the help of stochastic damping. The SPEKF forecast model is thus more skillful in providing an accurate prior distribution than a simple linear stochastic model (namely, the mean stochastic model) \cite{branicki2018accuracy}. On the other hand, although the recovered unobserved state variables $PV_2$ and $M$ for mode $(3,4)$ are not as accurate as their observed counterpart, the posterior mean time series captures the overall tendency of the truth. Note that the posterior uncertainty of $PV_2$ and $M$ for mode $(3,4)$ is still much smaller than the equilibrium variance, which indicates the benefits of state estimation from data assimilation. Finally, if the machine learning training residual is not included, then the posterior uncertainty comes from only the spread of the posterior ensemble members. As is seen from the right column of Figure \ref{fig2-time-series}, the 1st and the 99th quantile of the ensemble members do not reach the tail of the posterior PDF, which means the posterior uncertainty will be underestimated without considering the training residual. Such a result validates that the mixture distribution advances the uncertainty quantification of the posterior distribution. Nevertheless, the training residual is not significant, which indicates the skill of machine learning in finding the posterior ensemble members of the unobserved state variables.

\subsubsection{Data assimilation skill of the spatial reconstructed fields for the prognostic variables}

\begin{figure}[H]
\setlength{\tempwidth}{.3\linewidth}
\settoheight{\tempheight}{\includegraphics[width=\tempwidth]{example-image-a}}%
\centering
\rowname{Observation}
\subfloat{\includegraphics[width=\tempwidth]{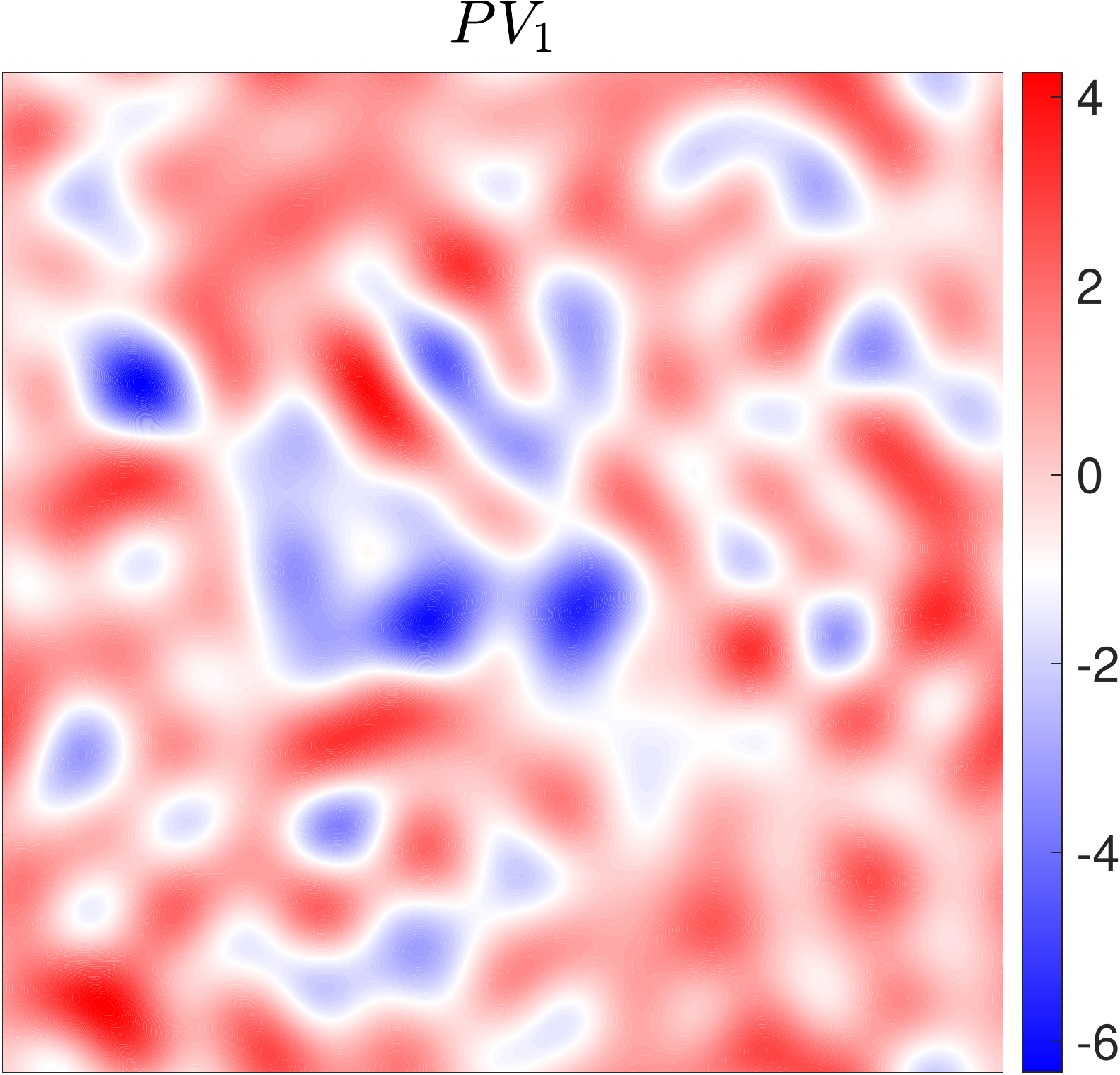}
}\hfil
\subfloat{\includegraphics[width=2\tempwidth]
{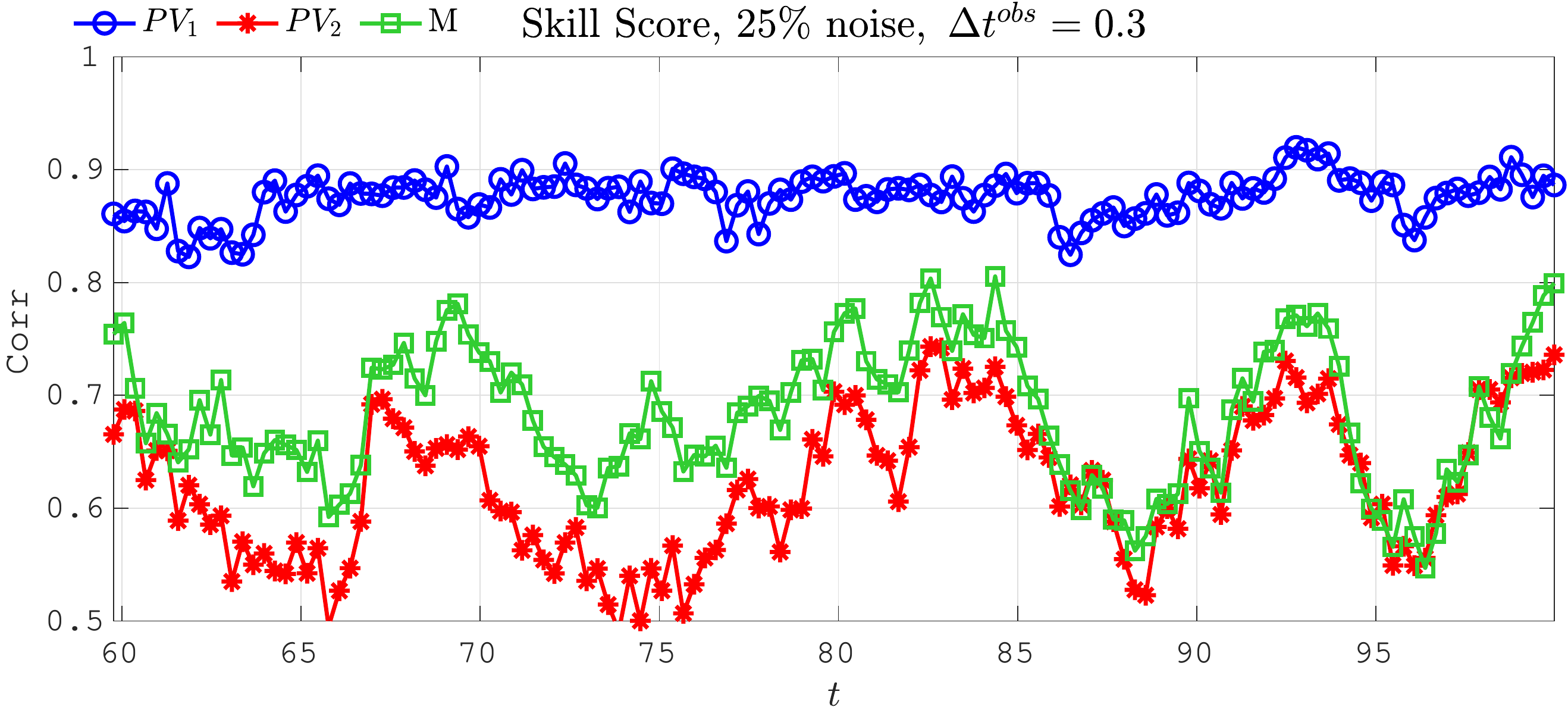}
}\\
\rowname{Truth}
\subfloat{\includegraphics[width=\tempwidth]{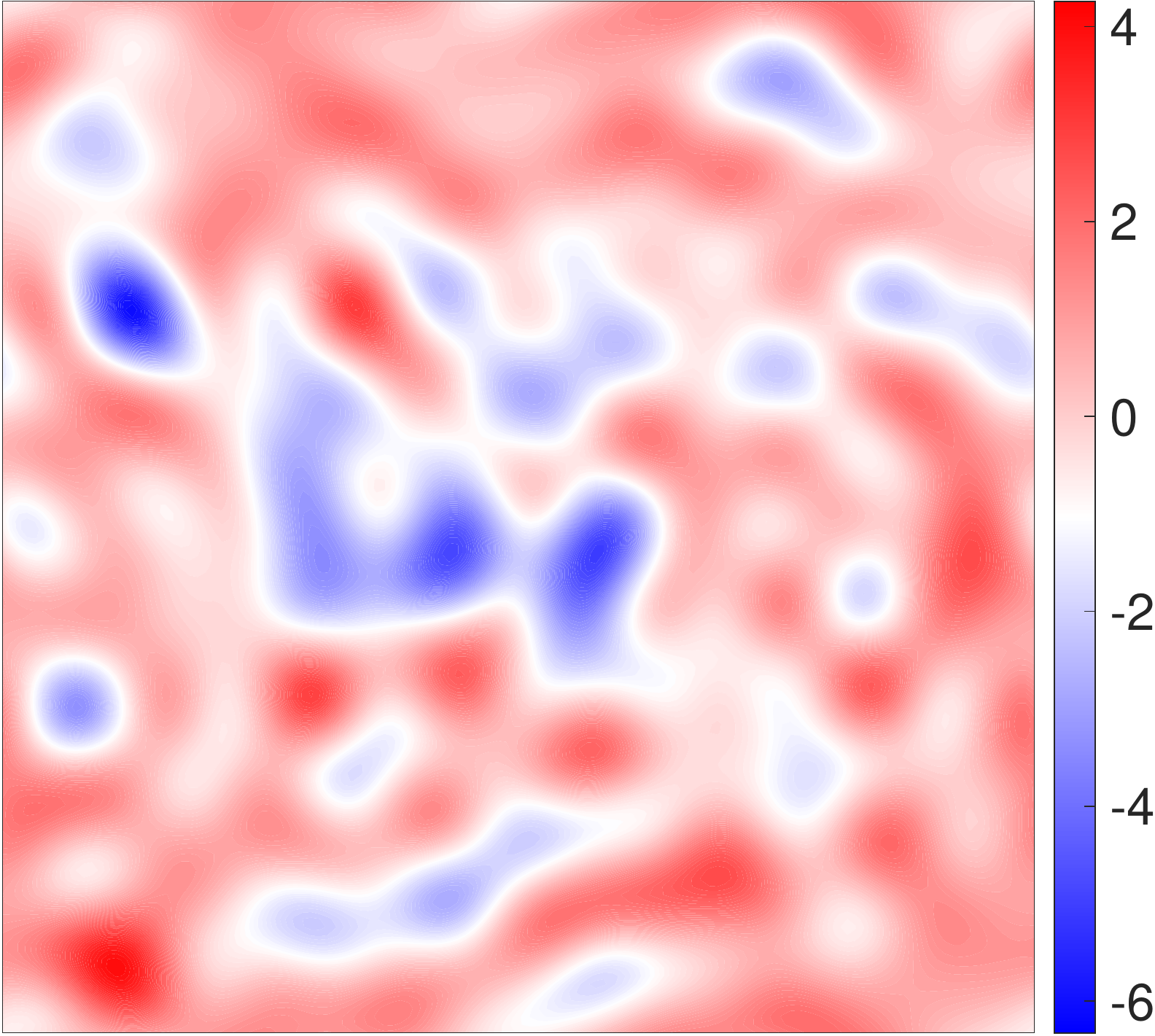}}\hfil
\subfloat{\includegraphics[width=1.02\tempwidth]{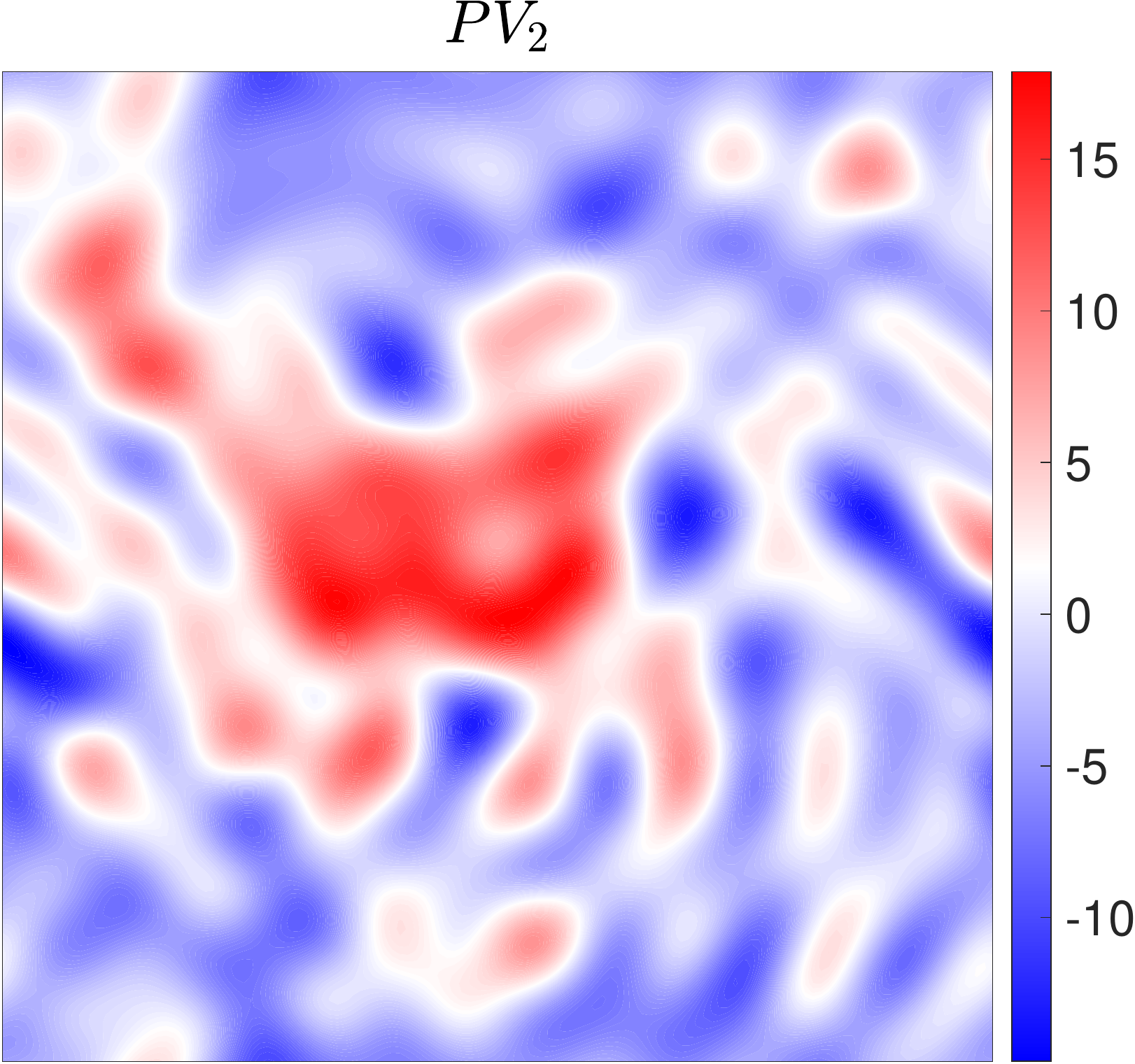}}\hfil
\subfloat{\includegraphics[width=\tempwidth]{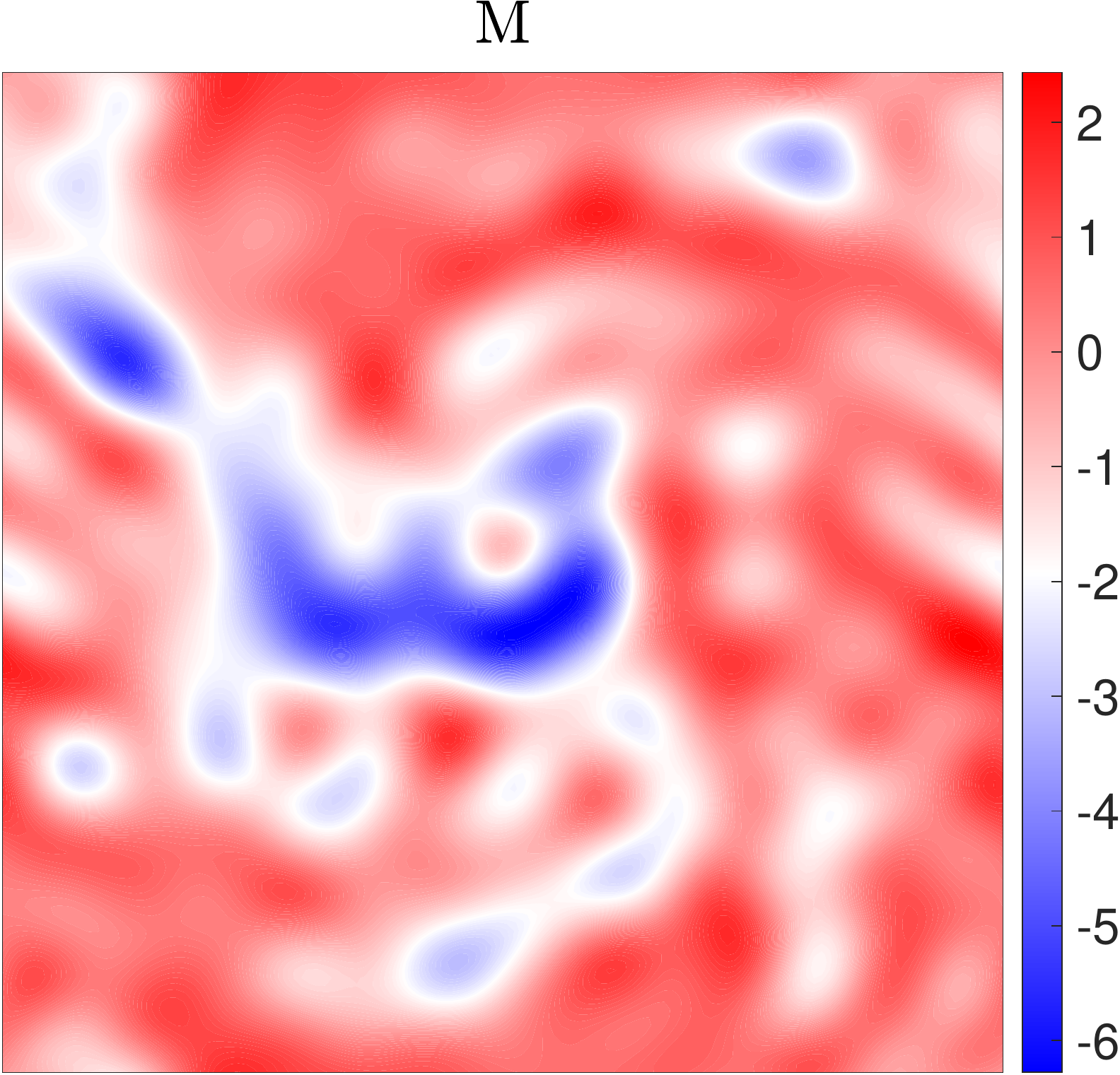}}
\\
\rowname{Filter}
\subfloat{\includegraphics[width=\tempwidth]{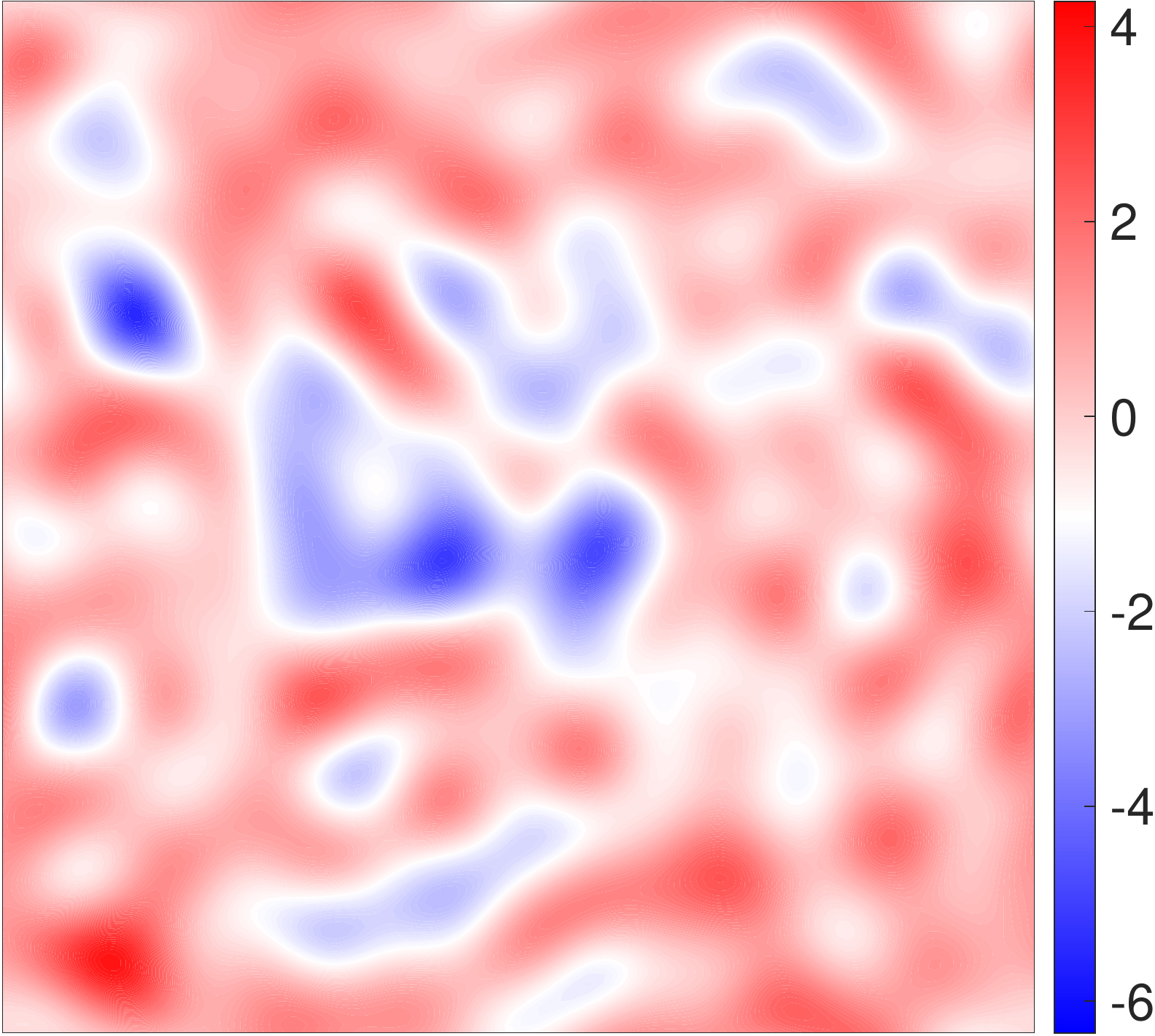}}\hfil
\subfloat{\includegraphics[width=1.02\tempwidth]{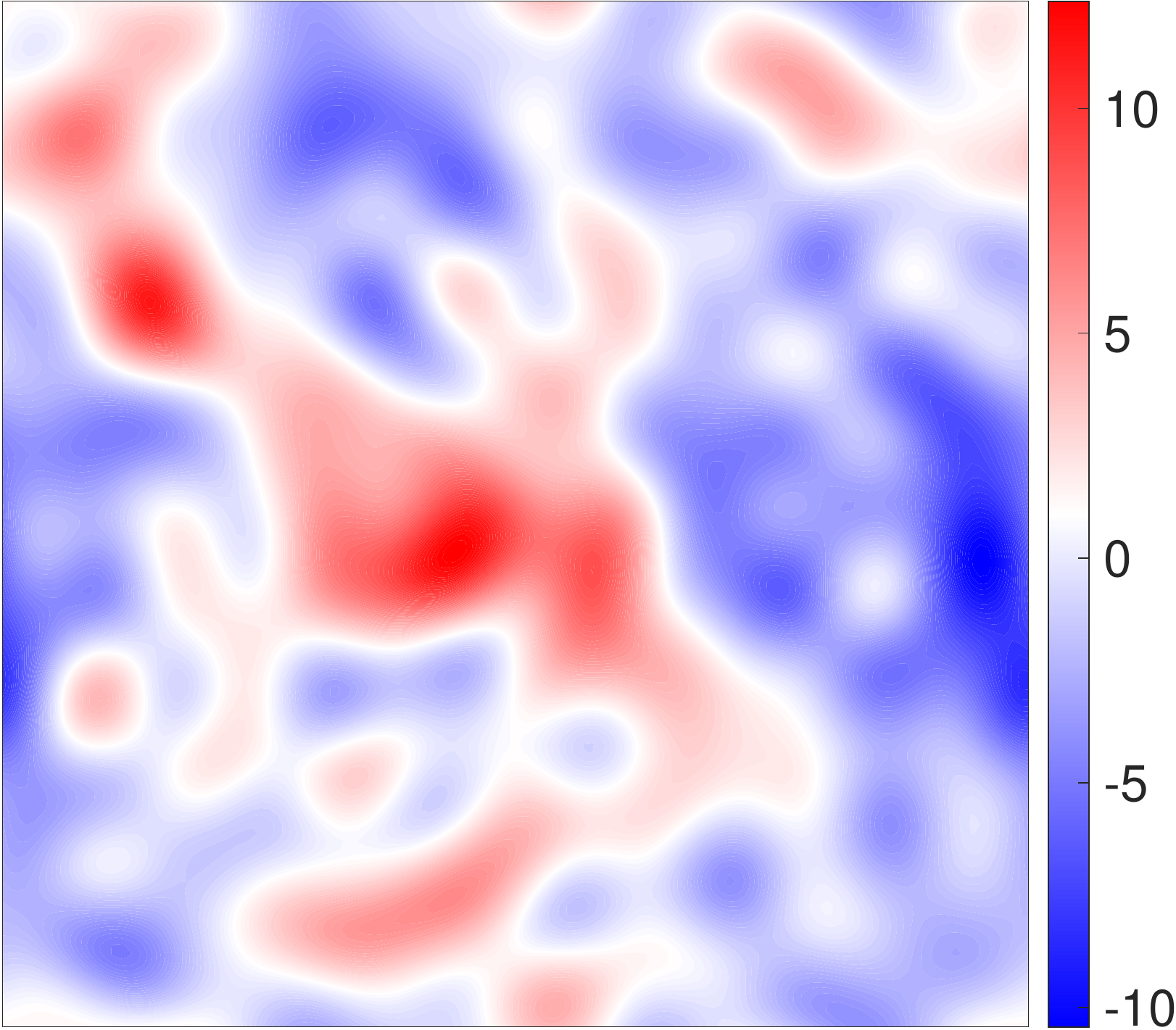}}\hfil
\subfloat{\includegraphics[width=\tempwidth]{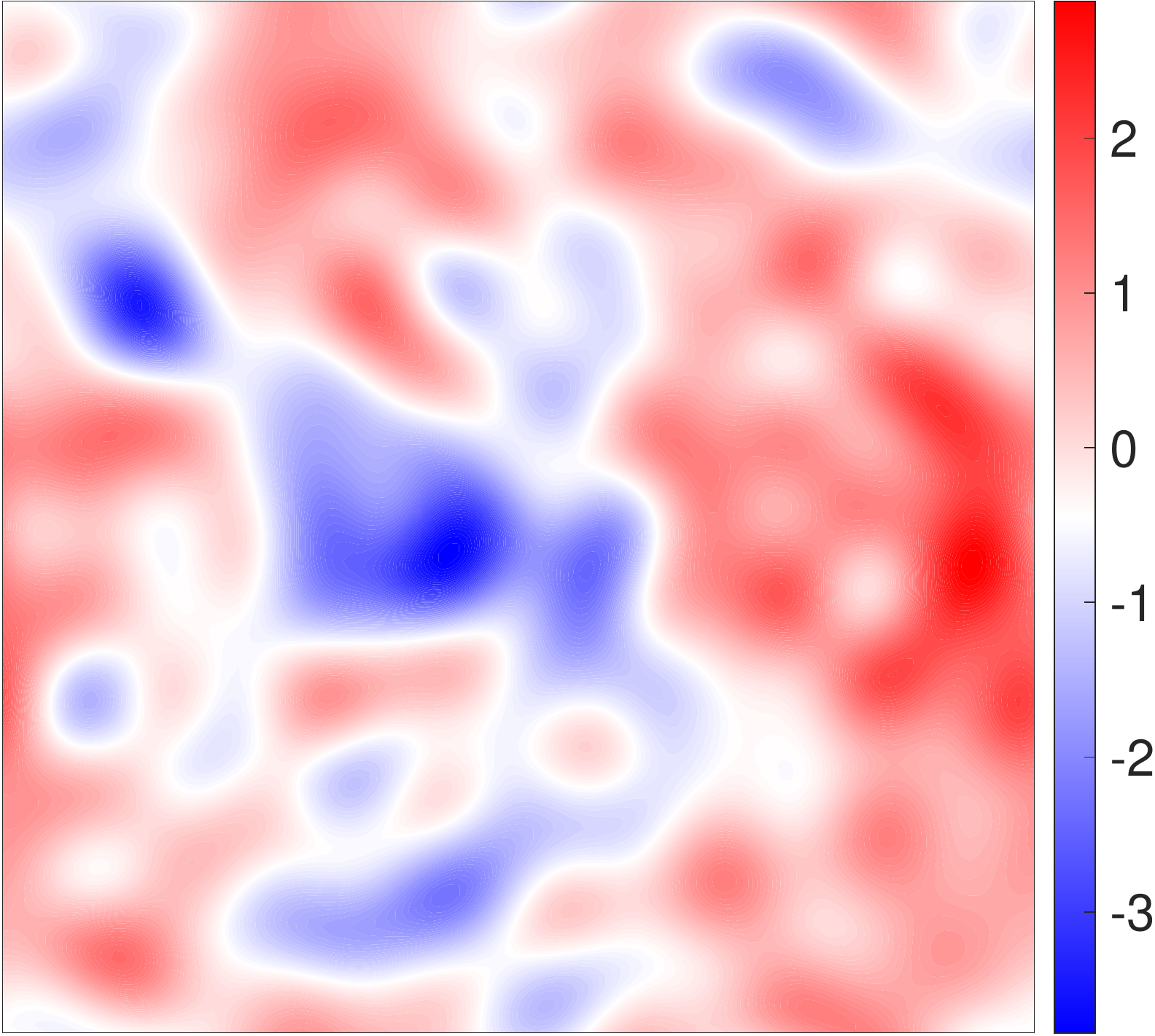}}
\\
\rowname{Trad ROM}
\subfloat{\includegraphics[width=\tempwidth]{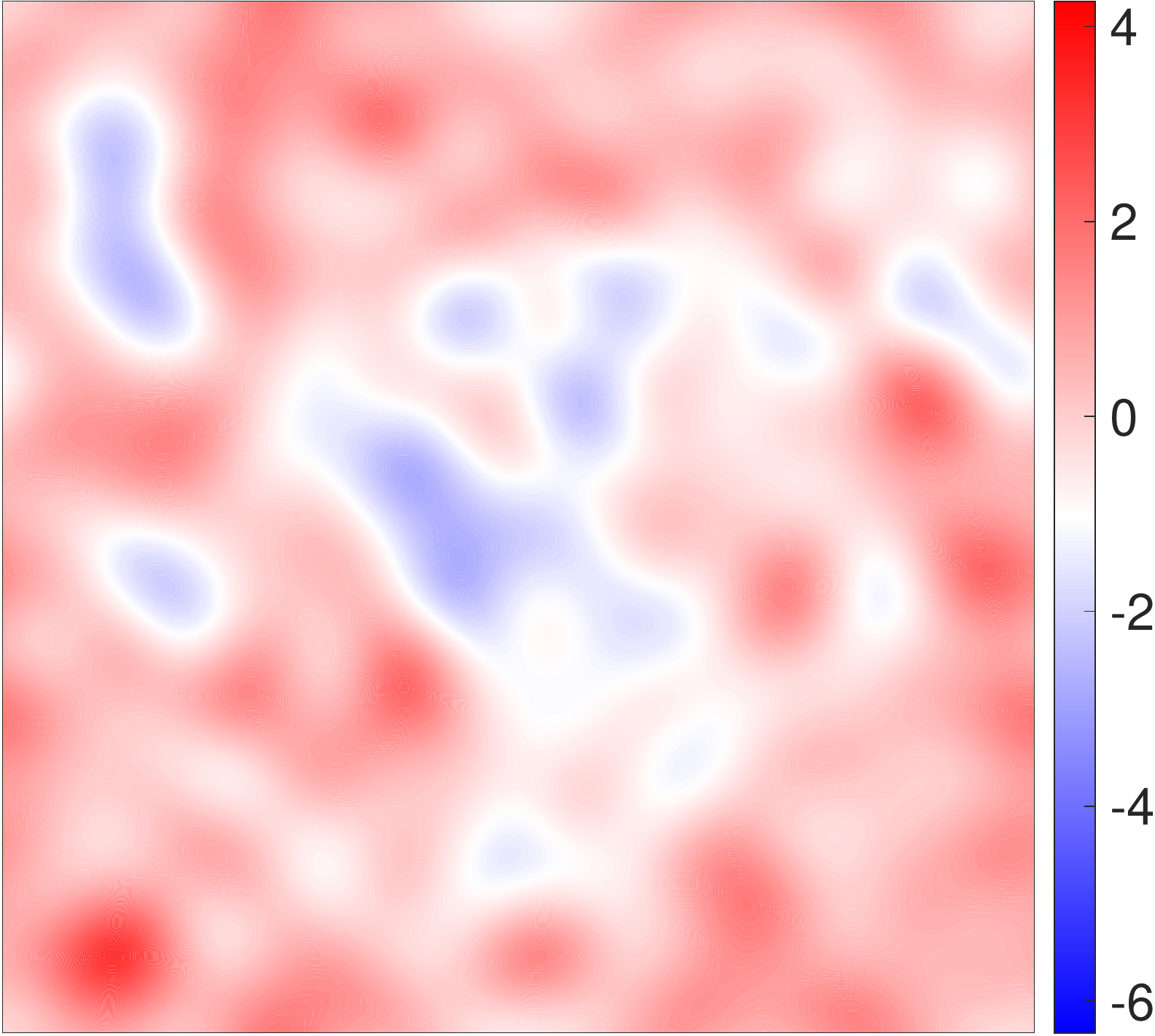}}\hfil
\subfloat{\includegraphics[width=1.02\tempwidth]{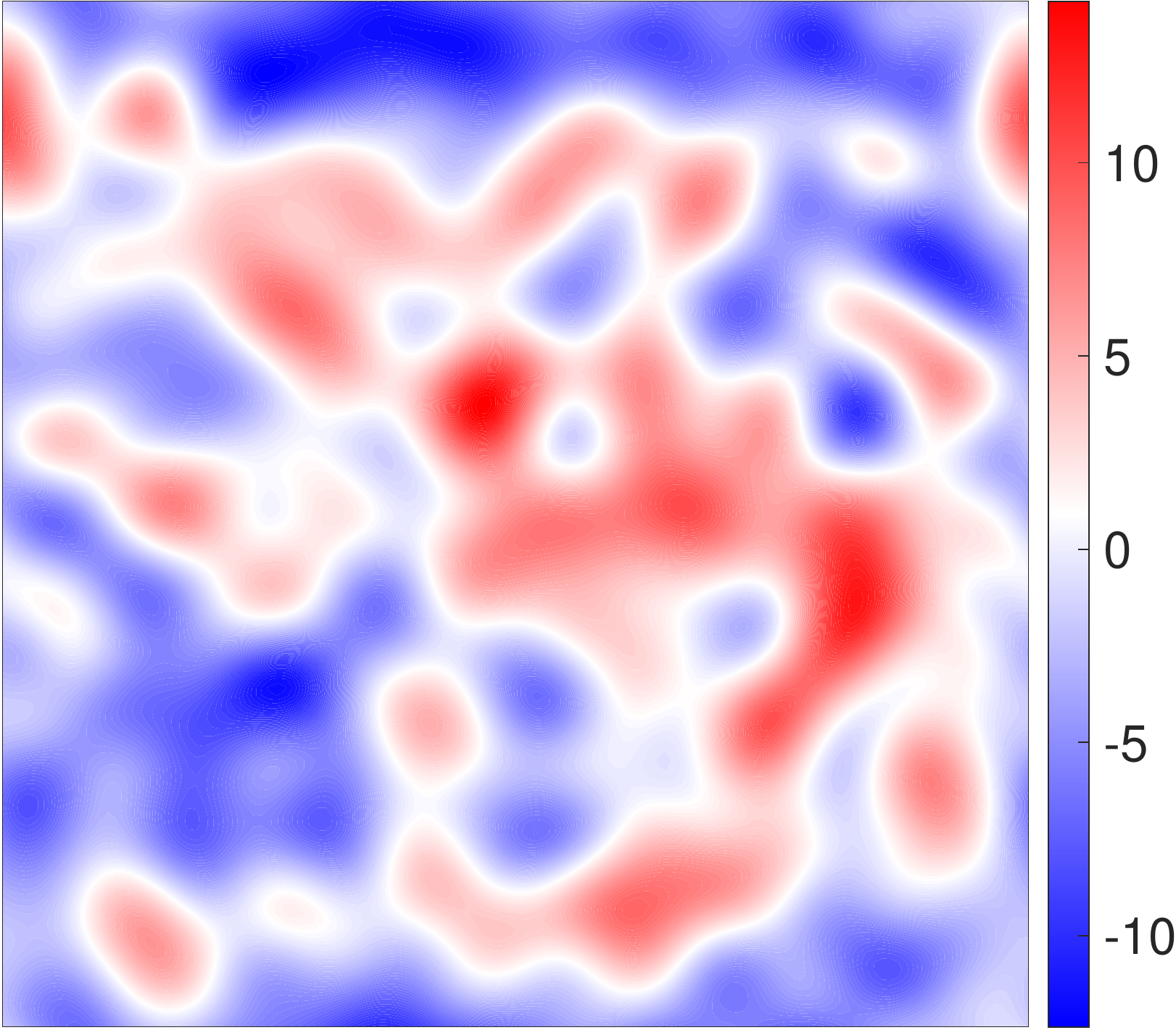}}\hfil
\subfloat{\includegraphics[width=1\tempwidth]{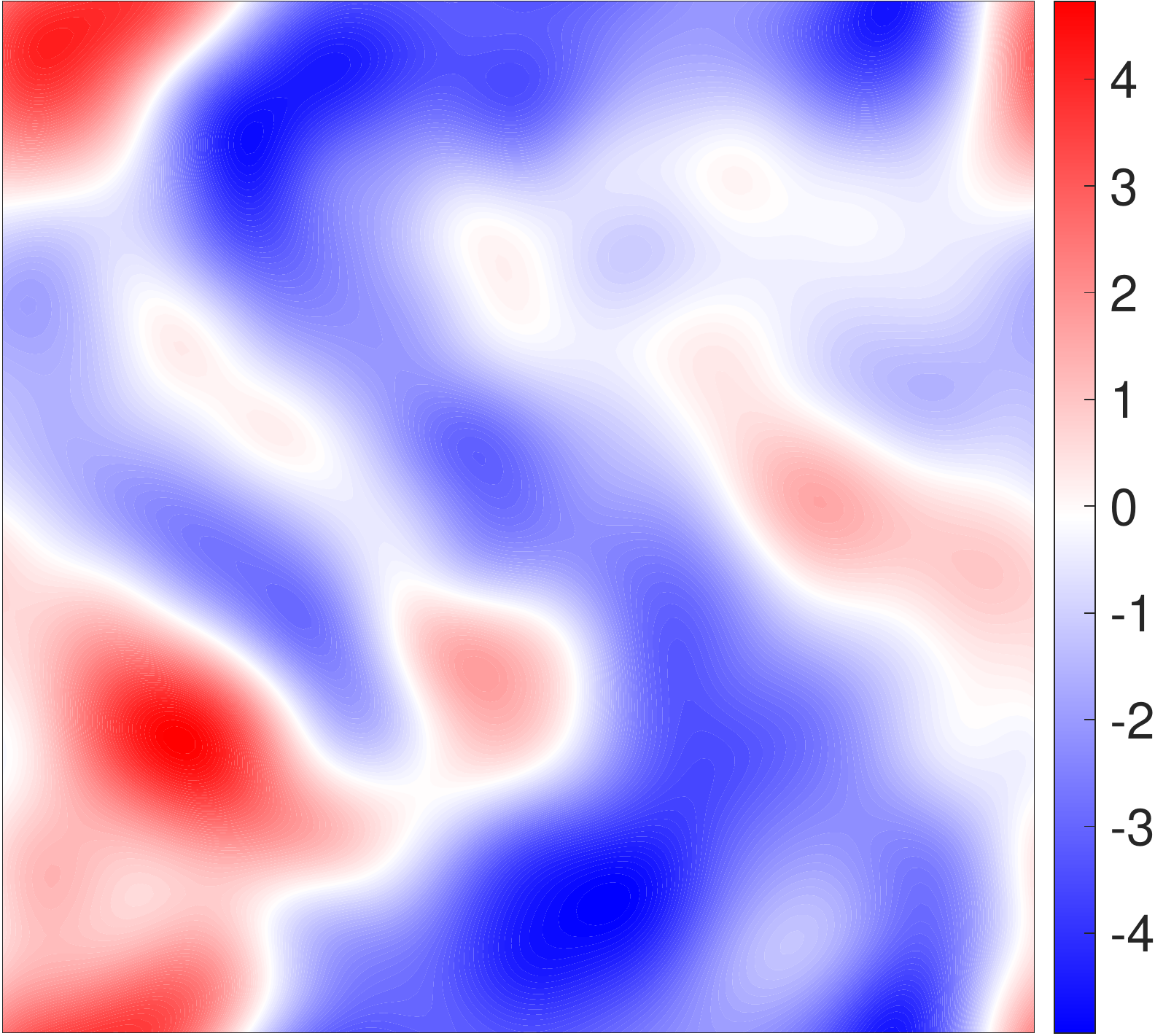}}
\caption{Reconstructed spatial fields and pattern correlation skill scores (upper right panel) of observed variable $PV_1$ (first column) and unobserved variables $PV_2$ (second column) and moisture $M$ (third column) at $t = 63.17$, with observation time $\Delta t^{obs} =0.3$.  }
\label{figure4-spacial-temporal-field}
\end{figure}

Figure~\ref{figure4-spacial-temporal-field} shows the reconstructed observed variable $PV_1$ and the unobserved variables $PV_2$ and $M$ in physical space. In addition to the hybrid data assimilation strategy, the results from the traditional ROM are also presented for comparison.

The first column shows the observation, the truth, and the recovered field of $PV_1$ using the two methods at a specific time instant $t=63.17$. The hybrid method almost perfectly reproduces the truth and contains less errors than the observation. The traditional ROM also works reasonably well by accurately recovering the overall patterns with the help of noisy observations. But certain biases in the intermediate scales are found in the results from the traditional ROM. On the other hand, as is shown in the second and the third columns, despite some small-scale errors, the hybrid strategy remains overall skillful in recovering the spatial pattern of the two unobserved prognostic variables. In contrast, the recovered fields using the traditional ROM lead to much larger errors due to the inaccurate coupling relationship between the observed and unobserved variables in the ROM with truncation. Finally, the pattern correlations between the truth and each recovered field using the hybrid strategy at different time instants are shown at the top right corner of Figure \ref{figure4-spacial-temporal-field}. The pattern correlation remains robust across time. In particular, the pattern correlation is above $0.8$ within the period for the observed variable $PV_1$. The pattern correlation is, on average, 0.65 for the two unobserved variables, namely $PV_2$ and $M$.

\subsubsection{Data assimilation skill of the spatial reconstructed fields for the diagnostic variables}

Recall that one of the critical features of the PQG system is the phase change of water and rainfall over the domain, which is characterized by the quantity of 
rain water $q_r$. In addition, the PQG system models the mid-latitude jet stream with 
bands of zonal wind related to velocity field $u$. Therefore, in addition to the three prognostic variables $PV_1$, $PV_2$, and $M$, it is crucial to study the accuracy of the recovered fields of these two quantities from the hybrid data assimilation.

\begin{figure}[H]
\setlength{\tempwidth}{.22\linewidth}
\settoheight{\tempheight}{\includegraphics[width=\tempwidth]{example-image-a}}%
\centering
\rowname{Truth, $q_r$}
\subfloat{\includegraphics[width=\tempwidth]{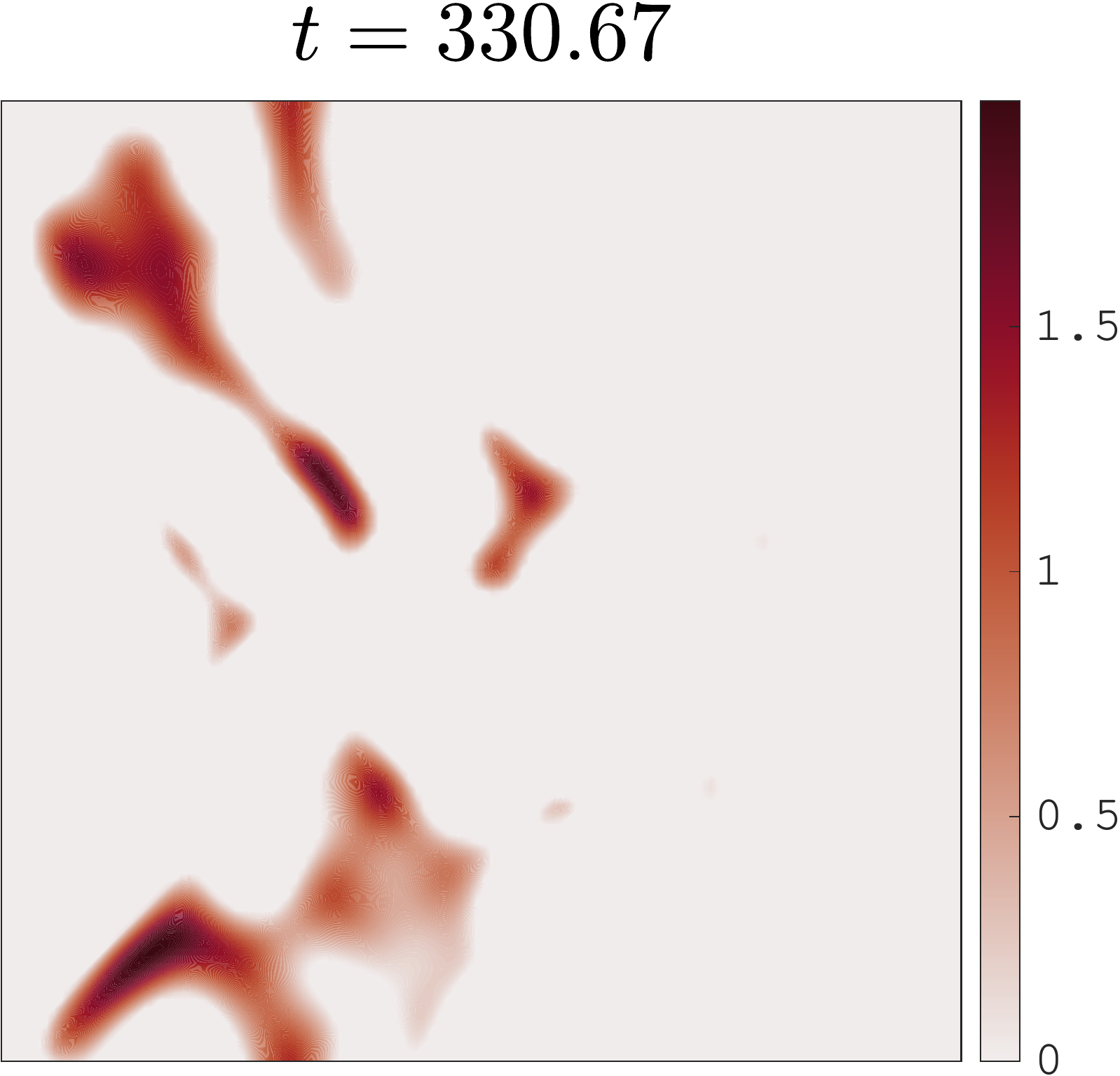}}\hfil
\subfloat{\includegraphics[width=1.02\tempwidth]{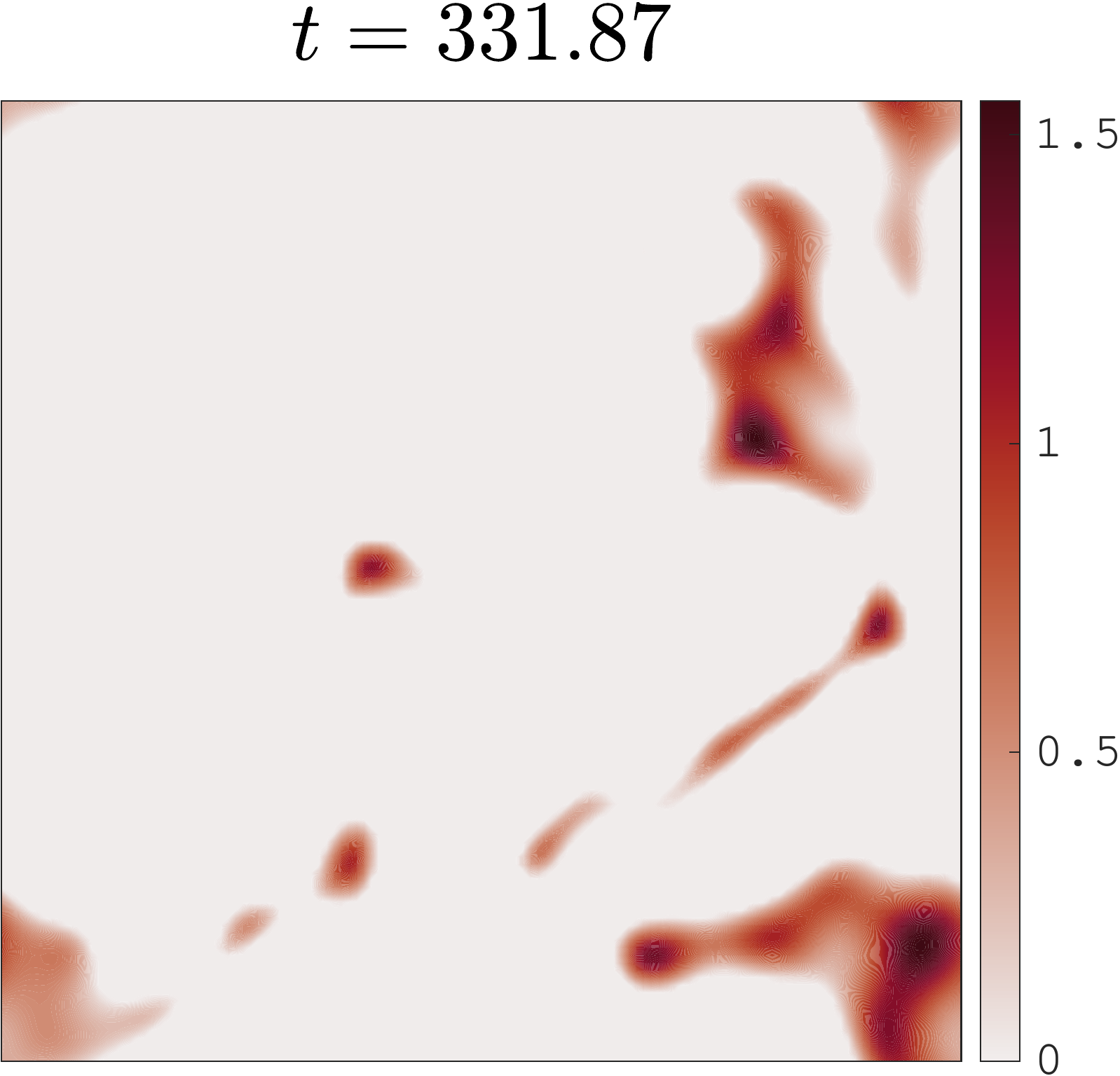}}\hfil
\subfloat{\includegraphics[width=1.02\tempwidth]{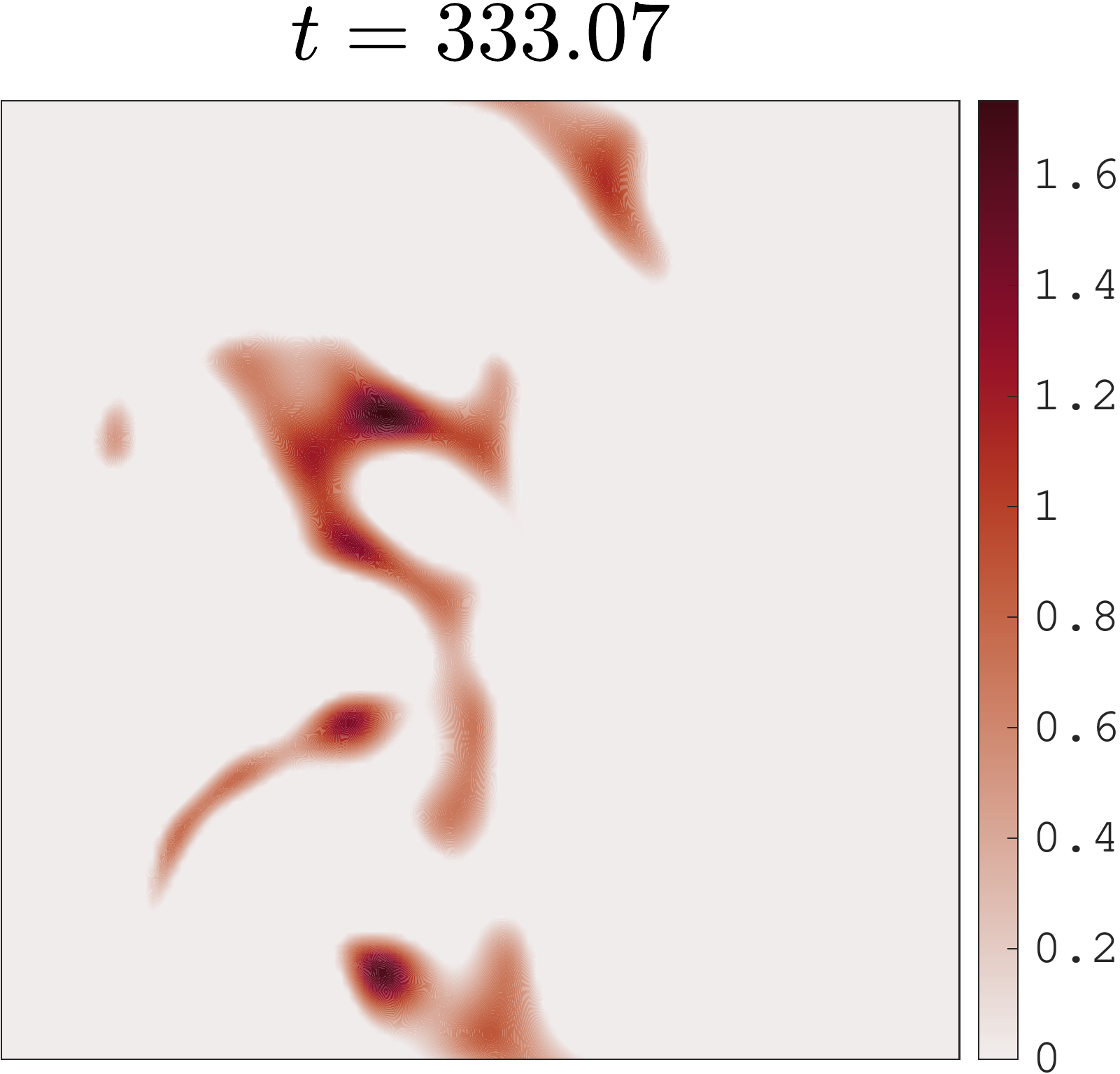}}\hfil
\subfloat{\includegraphics[width=\tempwidth]{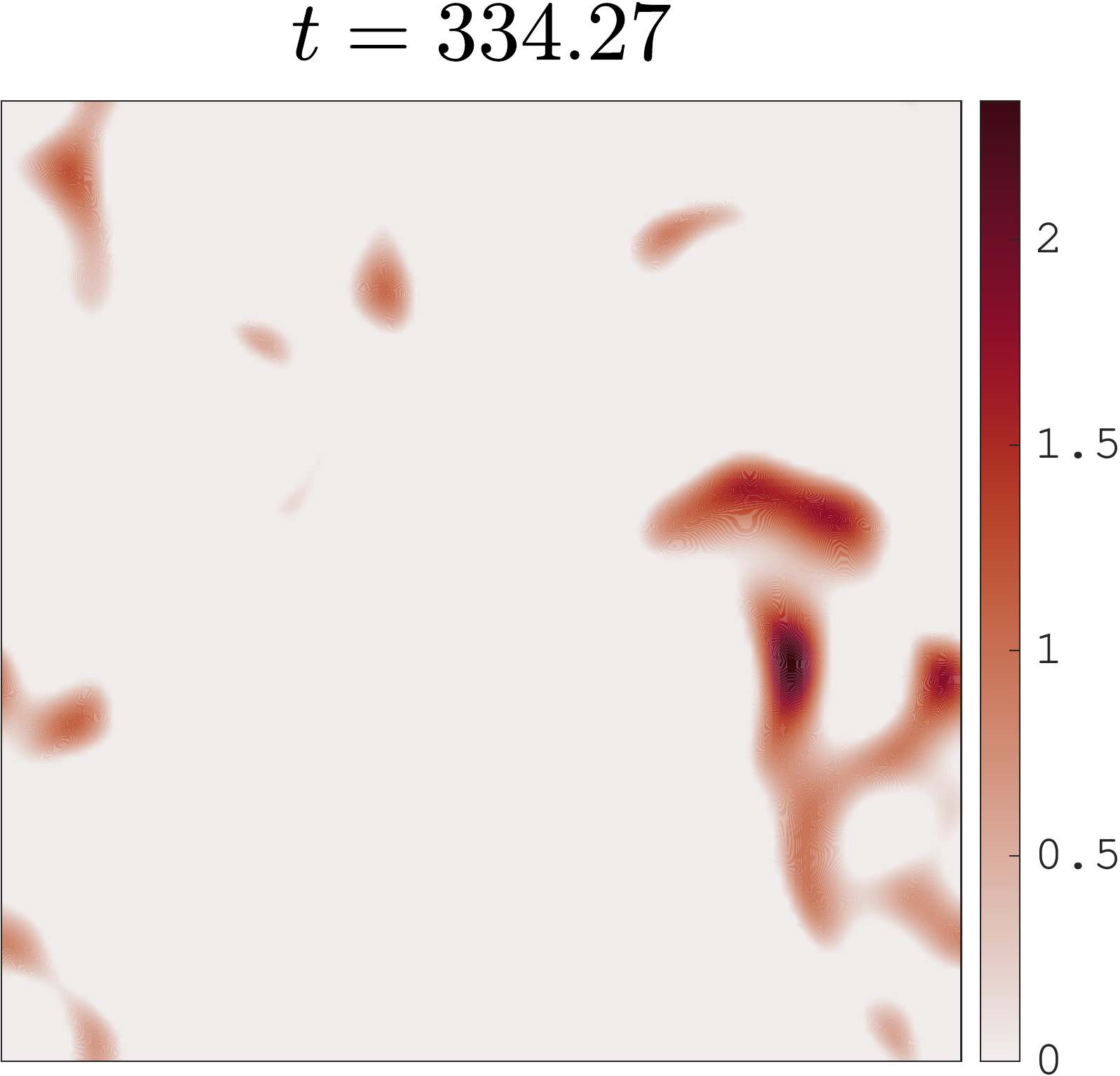}}
\\
\rowname{Filter, $q_r$}
\subfloat{\includegraphics[width=\tempwidth]{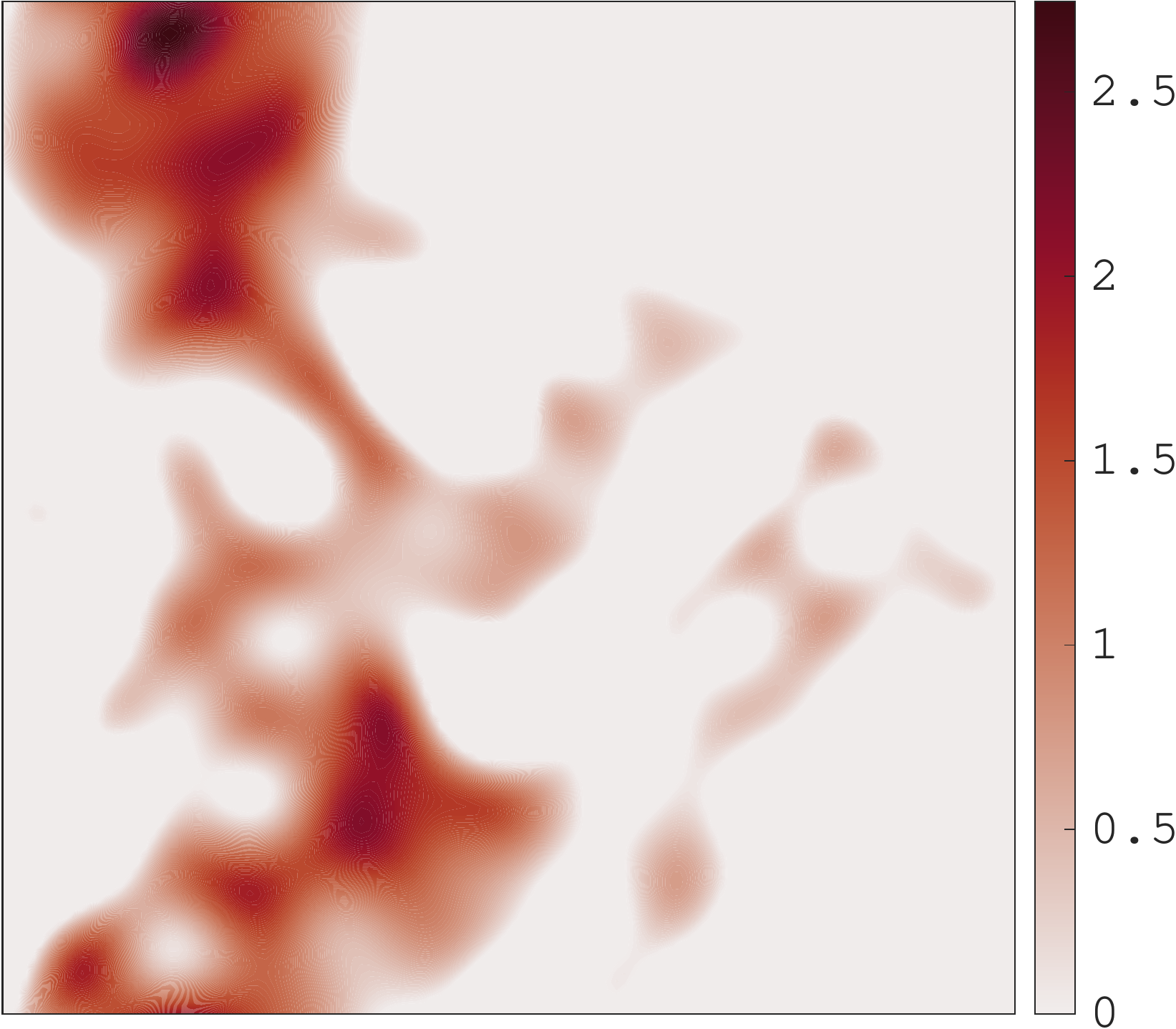}}\hfil
\subfloat{\includegraphics[width=1.02\tempwidth]{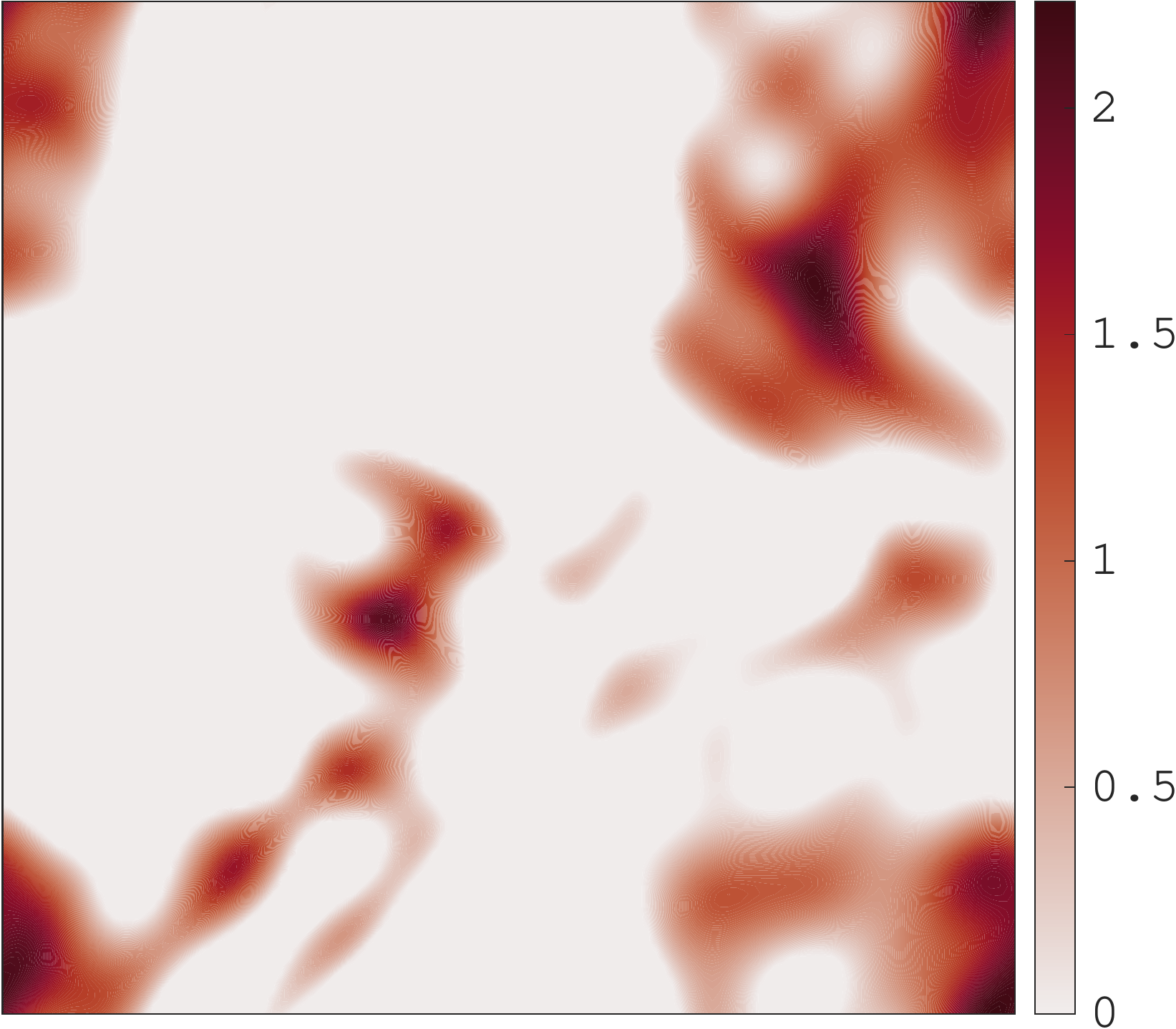}}\hfil
\subfloat{\includegraphics[width=1.02\tempwidth]{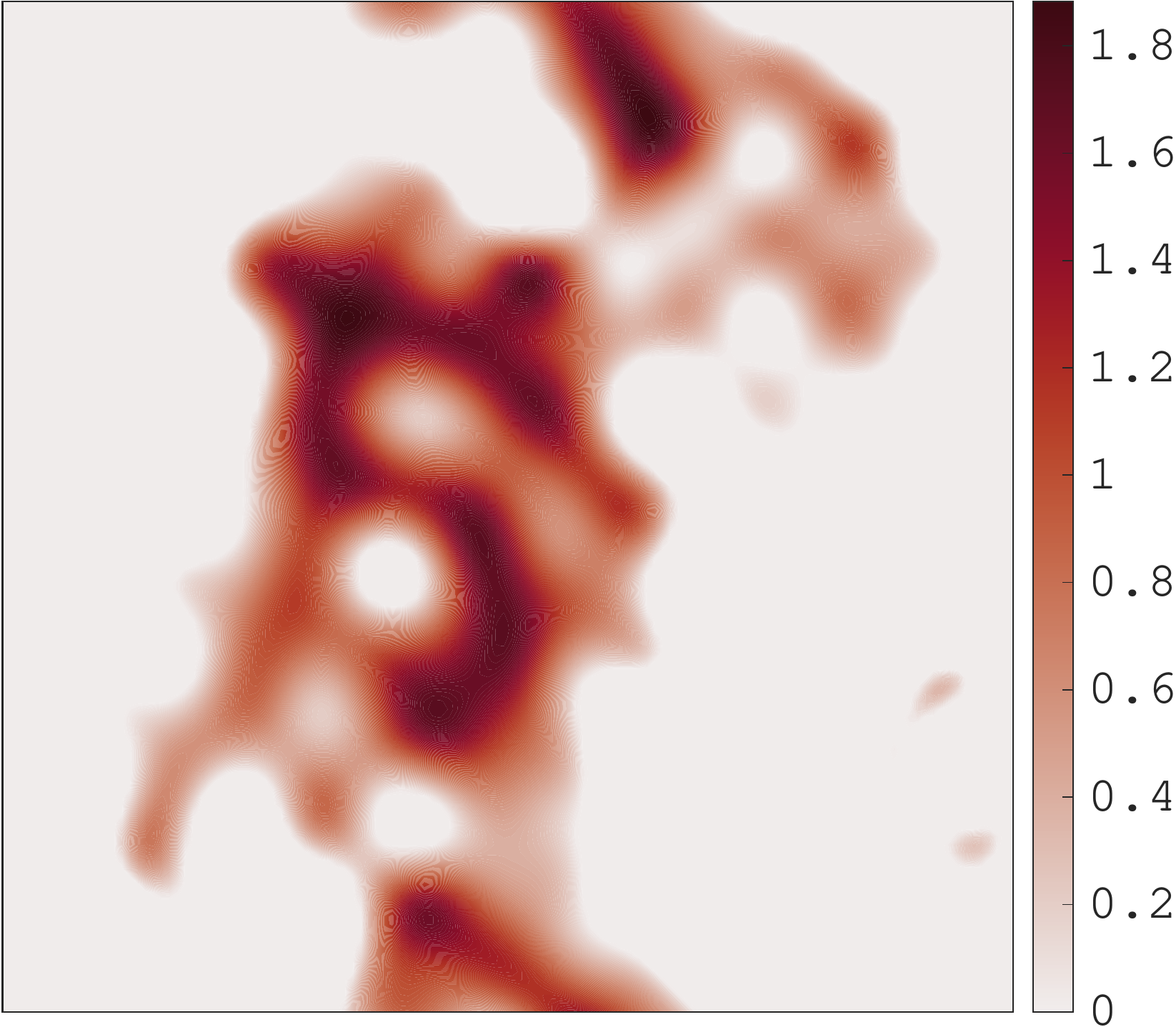}}\hfil
\subfloat{\includegraphics[width=\tempwidth]{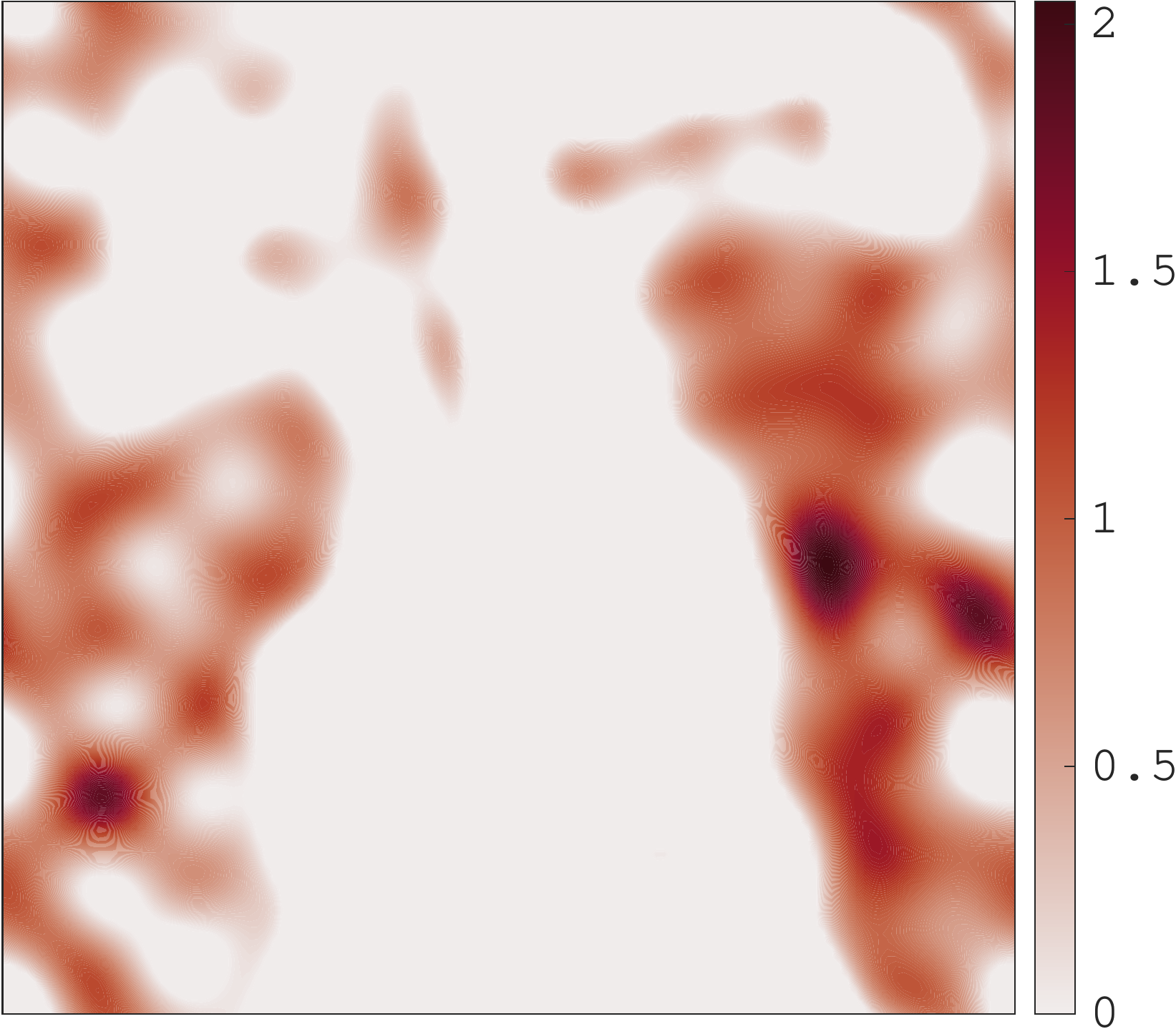}}
\\
\rowname{Truth, $u_2$}
\subfloat{\includegraphics[width=\tempwidth]{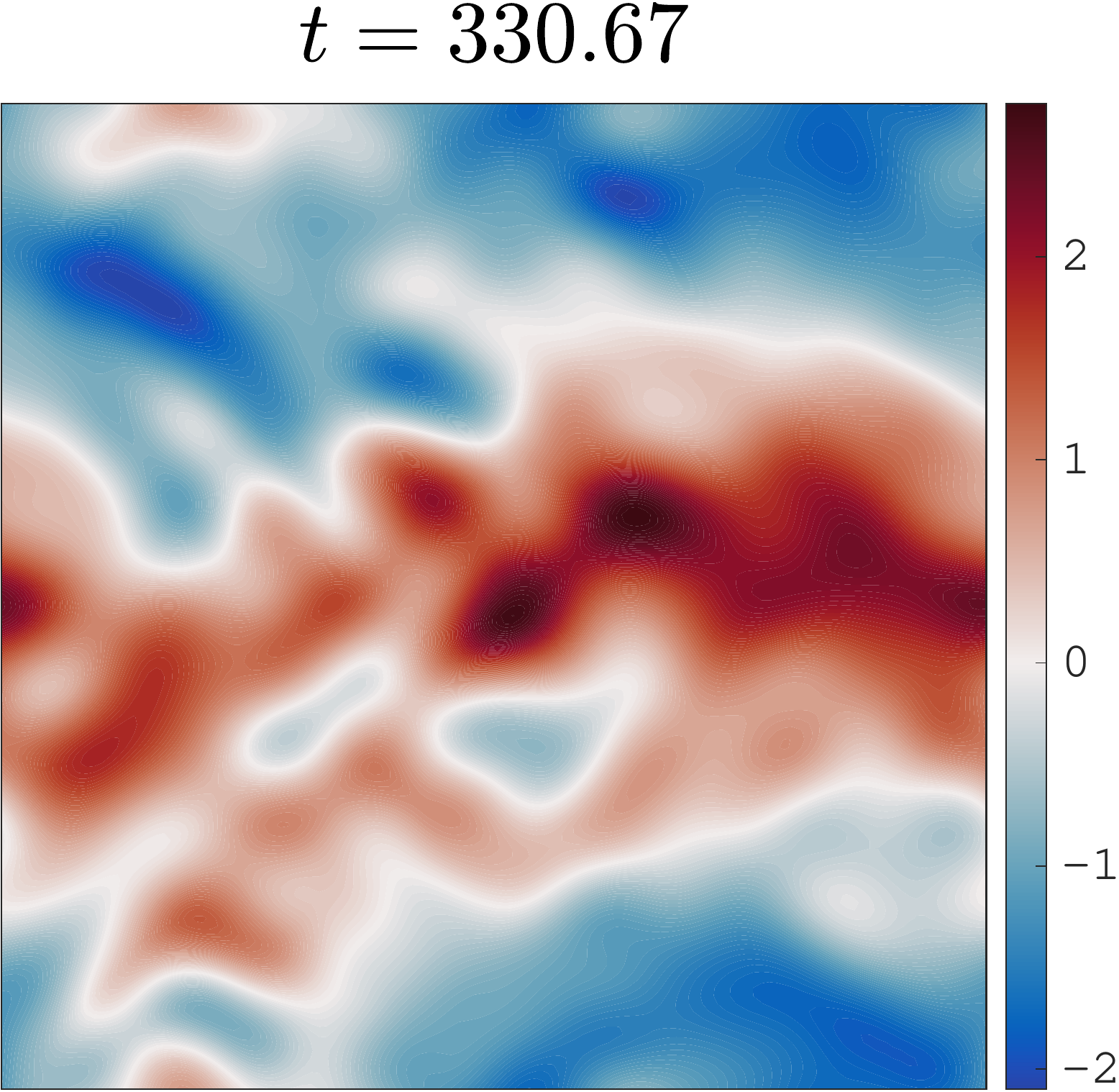}}\hfil
\subfloat{\includegraphics[width=1.02\tempwidth]{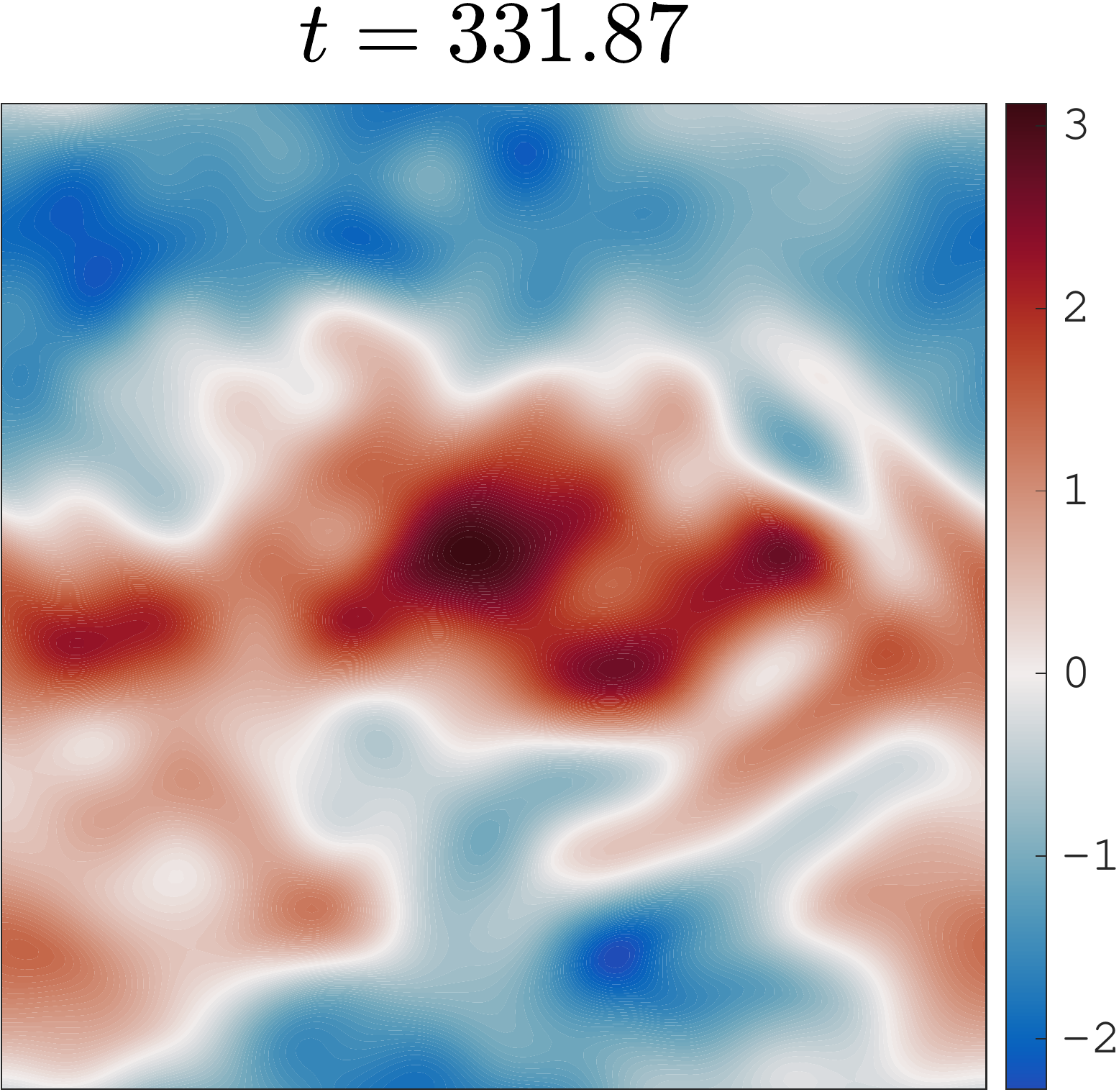}}\hfil
\subfloat{\includegraphics[width=1.08\tempwidth]{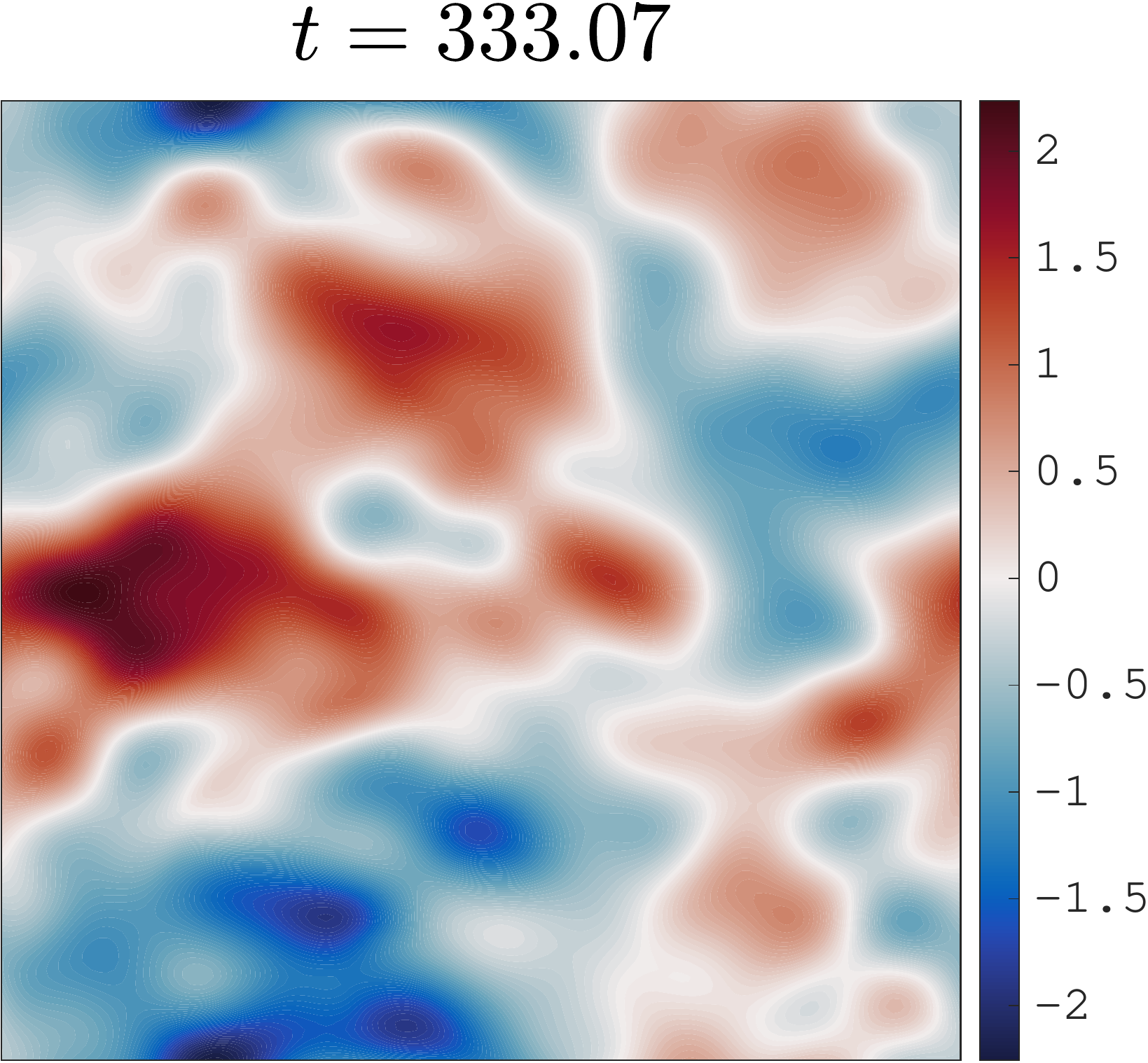}}\hfil
\subfloat{\includegraphics[width=1.09\tempwidth]{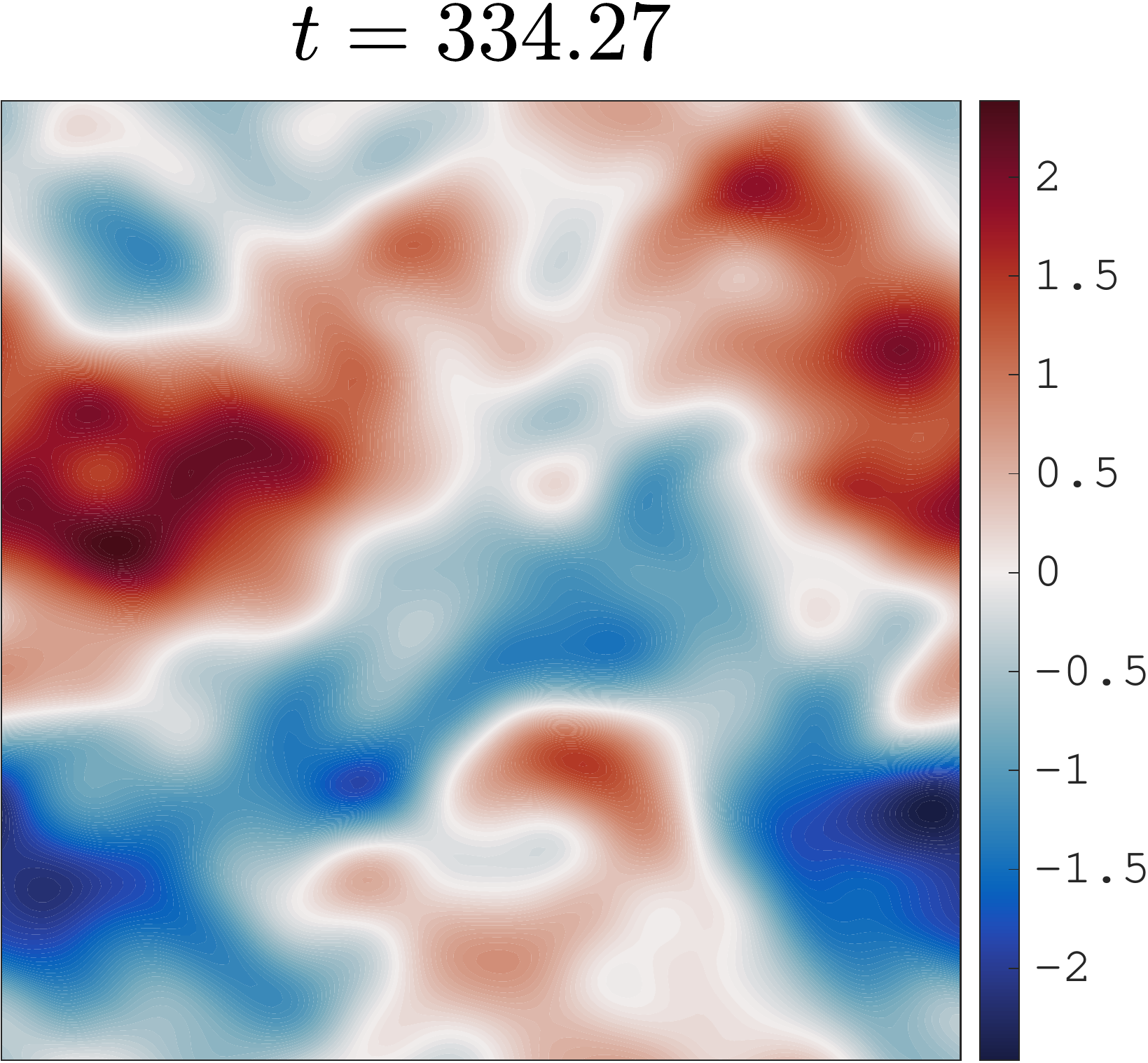}}\hfil
\\
\rowname{Filter, $u_2$}
\subfloat{\includegraphics[width=\tempwidth]{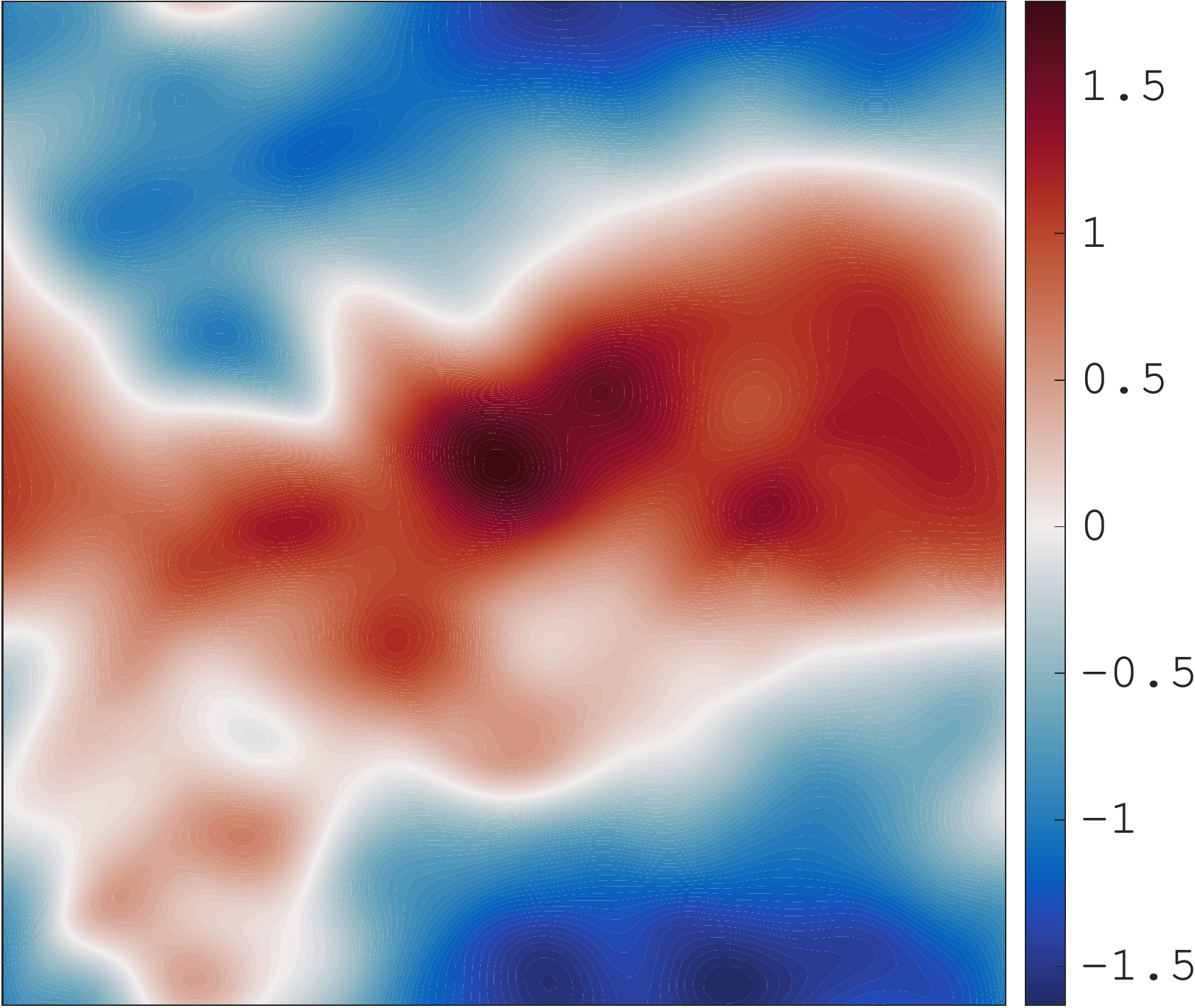}}\hfil
\subfloat{\includegraphics[width=1.02\tempwidth]{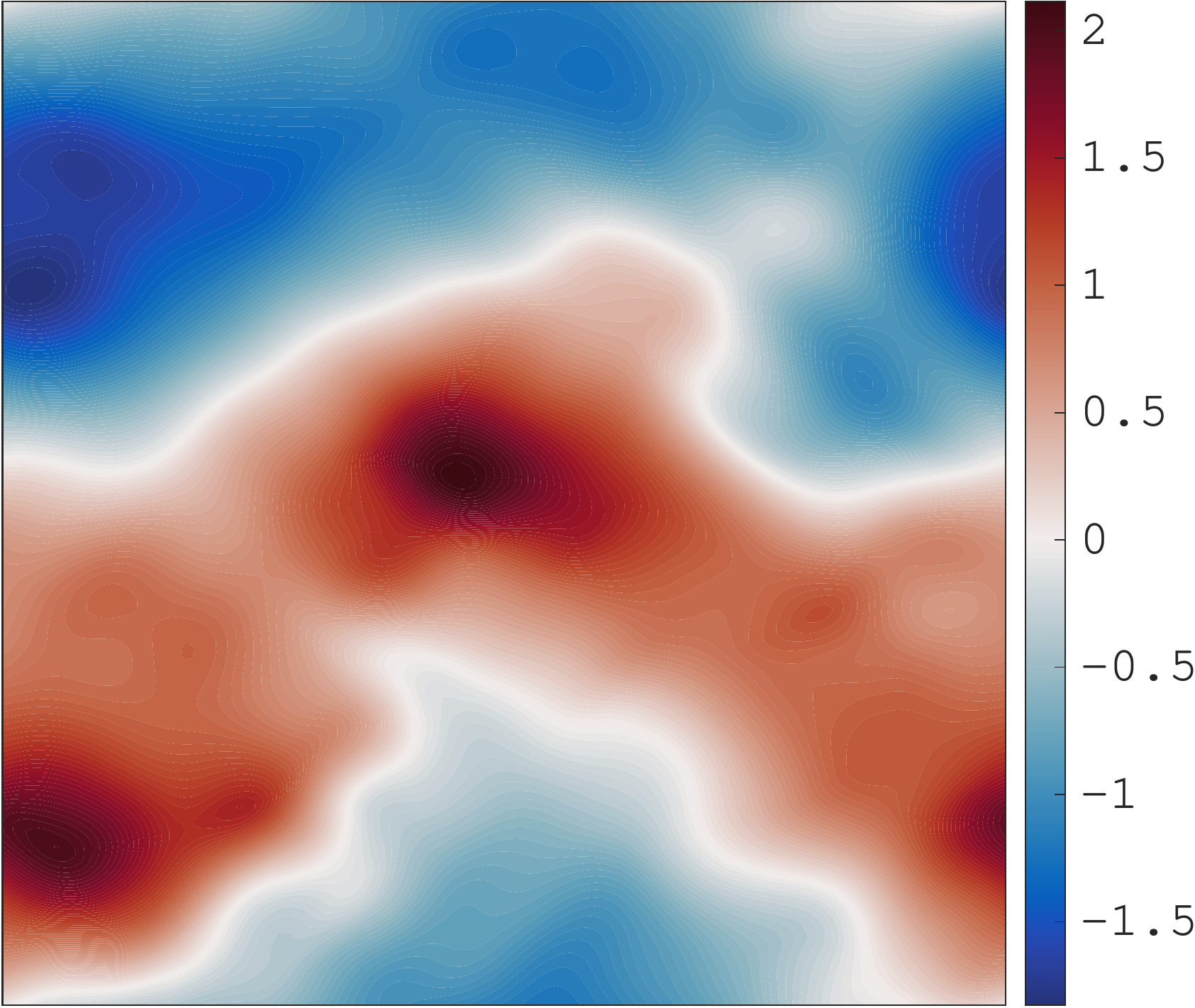}}\hfil
\subfloat{\includegraphics[width=1.02\tempwidth]{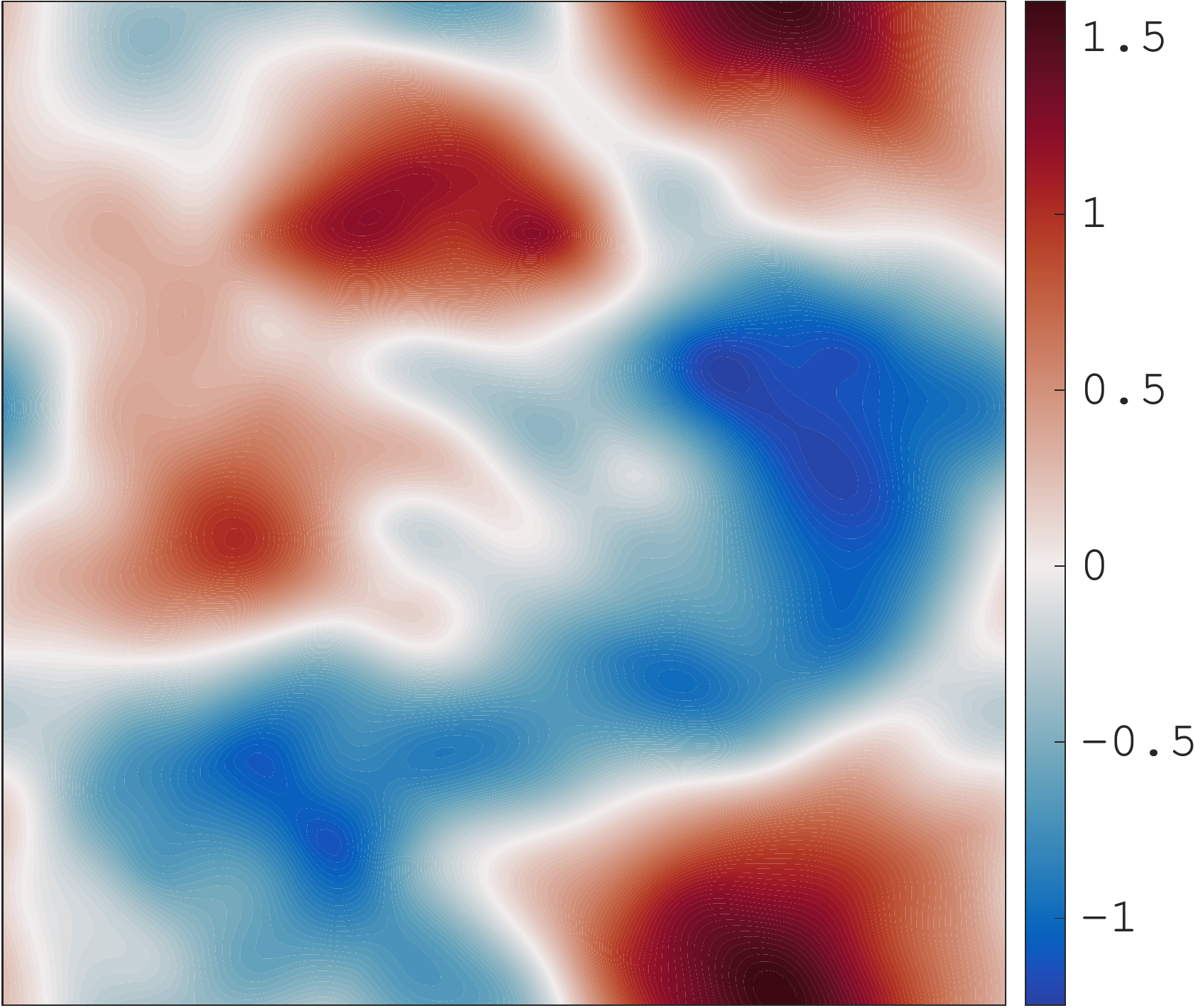}}\hfil
\subfloat{\includegraphics[width=1.02\tempwidth]{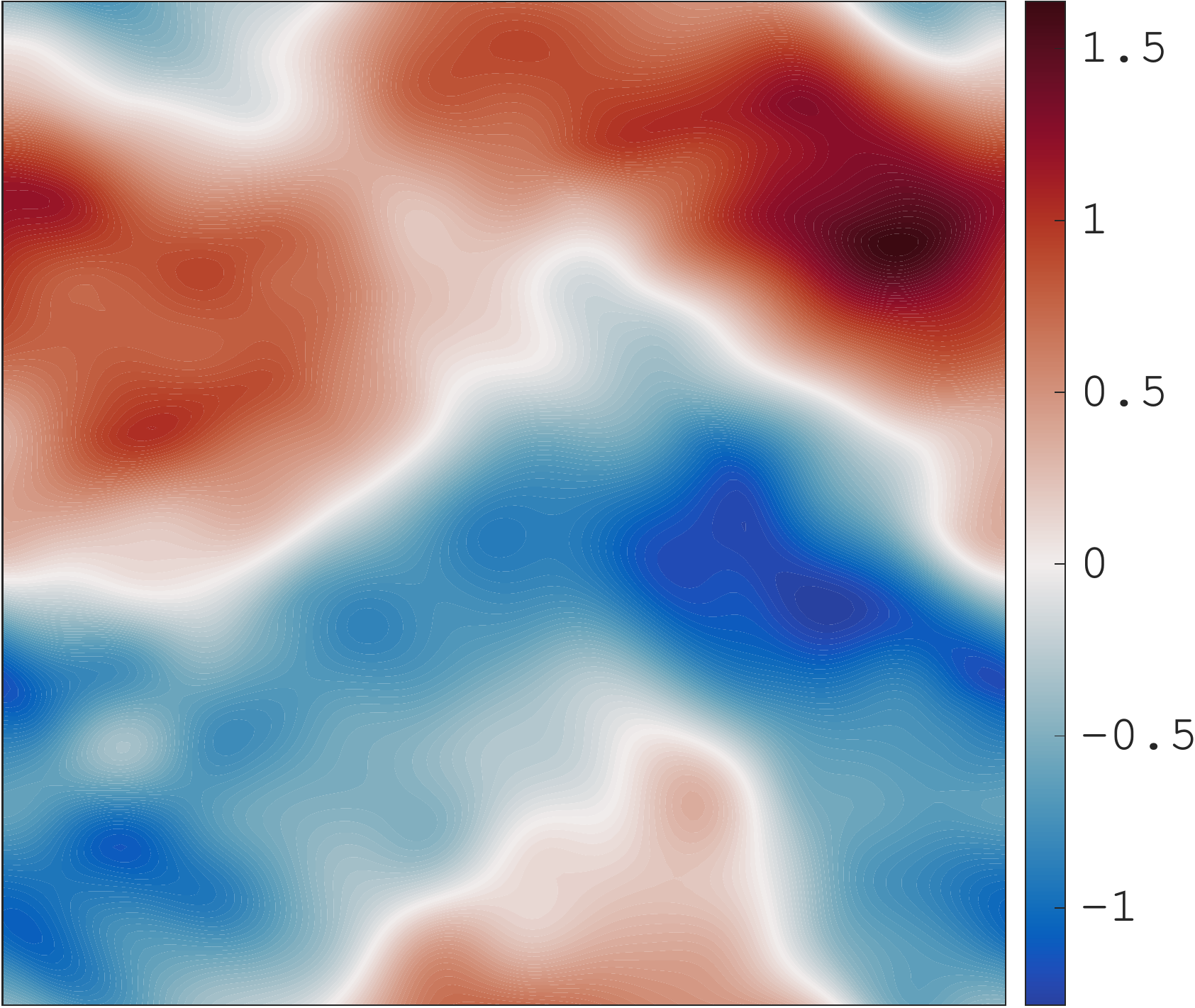}}
\\
\rowname{\hspace*{1cm}Pattern Corr  }
\subfloat{\includegraphics[width=4.06\tempwidth,height = 1\tempwidth]{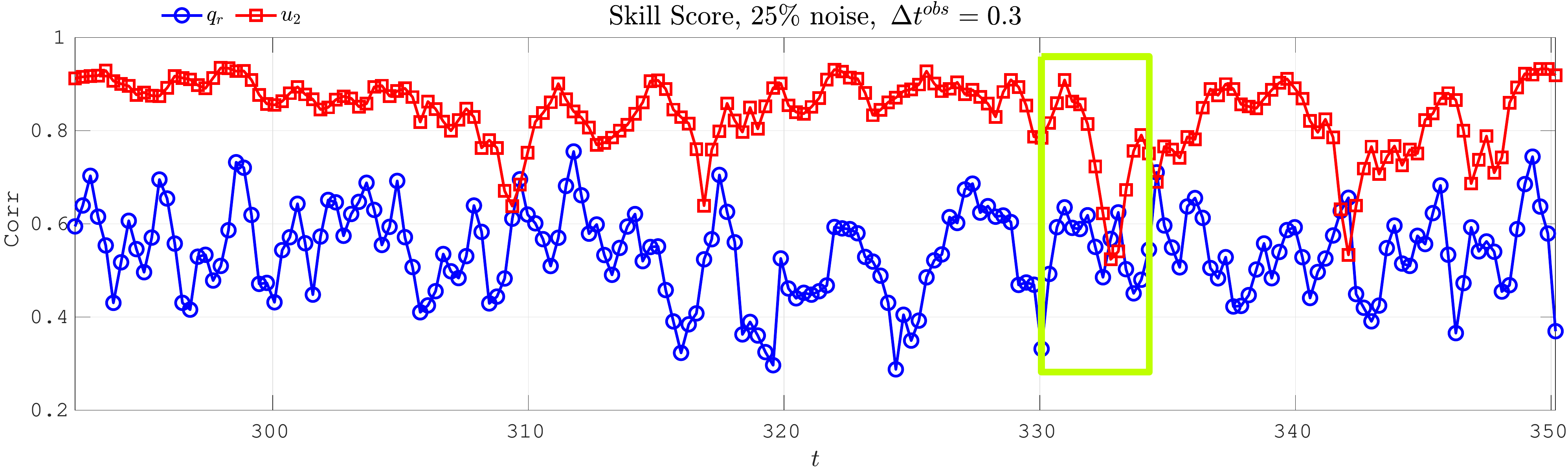}}
\caption{Recovered spatial fields and pattern correlation skill scores of 
rain water $q_r$ and zonal velocity at level 2 in the physical space at different time, with $\Delta t^{obs} =0.3$; the green box in the time series of pattern correlation denotes the time interval of recovered $q_r$ and $u_2$ fields. }
\label{figure5:recover-rain-vel}
\end{figure}

Figure \ref{figure5:recover-rain-vel} shows the rain water $q_r$ and zonal wind velocity field $u_2$ at the top level for both the truth and the filtered 
posterior mean result. Variation of $q_r$ from zero to positive values (shades of red) indicated the presence of phase changes, and there are regions of relatively large $q_r>0$
(darker red areas).
Here, $q_r$ and $u_2$ are calculated by solving the PV-and-M inversion equations \eqref{eqn:pv-m-inversion-1}--\eqref{eqn:pv-m-inversion-2}, and using the relations above \eqref{eqn:pv-m-inversion-1}--\eqref{eqn:pv-m-inversion-2}. It is worthwhile highlighting that, due to the sensitive behavior of the output with a small change in the input via the Heaviside nonlinearity in PV-and-M inversion, it is extremely challenging to reach a nearly perfect recovery of $q_r$. In particular, most of the significant features of $q_r$ are associated with small scales. Nevertheless, the reconstructed patterns of the $q_r$ field based on the data assimilation results reasonably resemble the truth. The pattern correlation is, on average, above the threshold Corr = $0.5$ within the entire period, indicating the skillful prediction of the timing and locations of the rainfall.
Furthermore, the variablilty (e.g meandering) of the large-scale, westerly zonal jet is successfully recovered.
This is consistent with the pattern correlation skill score in the bottom panel of Figure \ref{figure5:recover-rain-vel}, which is, on average, about $0.8$ across time.

\begin{figure}[H]
\includegraphics[width=\textwidth]{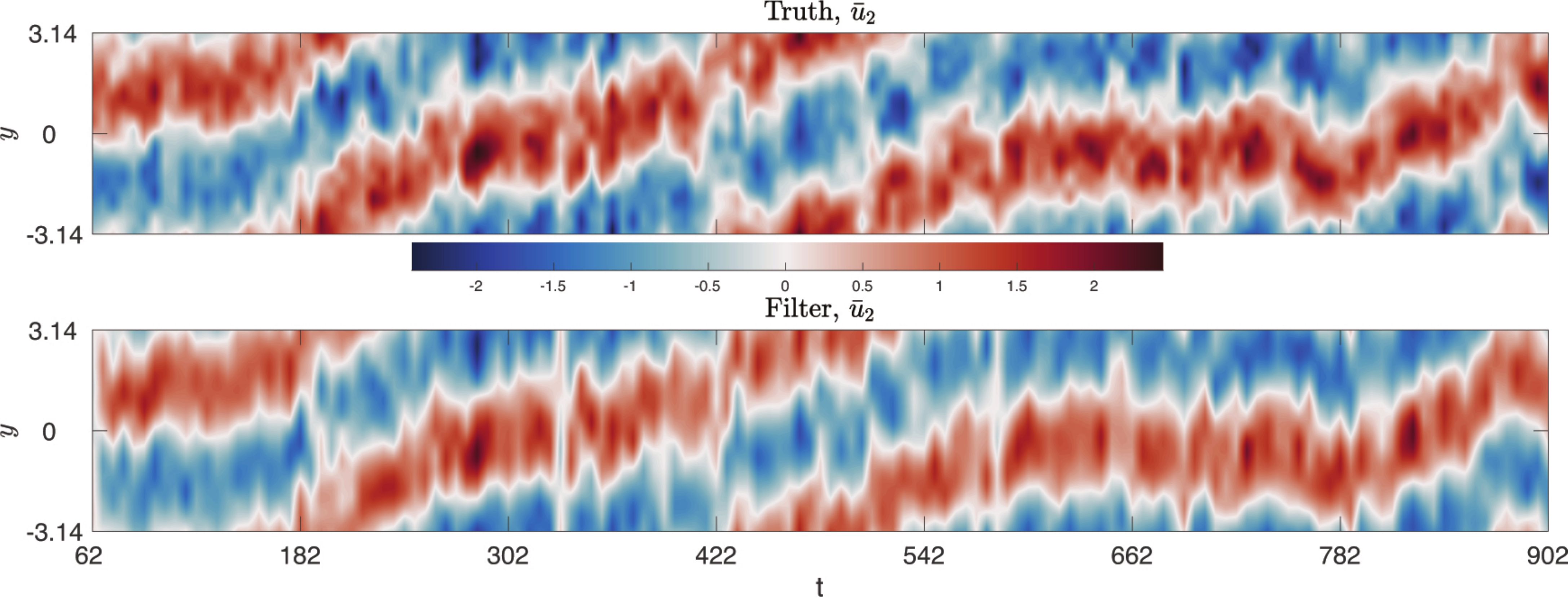}
\caption{Truth (top) and filter (bottom), recovered time evolution of $\bar{u}_2$, the zonally averaged zonal wind at the top level of atmosphere. 
}
\label{figure5-2:recover-zonal-vel-mean}
\end{figure}

Figure \ref{figure5-2:recover-zonal-vel-mean} displays the upper-level zonally averaged wind as a function of time for both the truth and the filter. The reconstructed velocity field successfully captures the location of zonal jets most of the time. Moreover, it recovers the northward propagation of the jet, which was first observed in \cite{hu2021initial}.

\subsubsection{Sensitivity tests}

\begin{figure}[H]
    \centering
    \includegraphics[width=.95\textwidth]{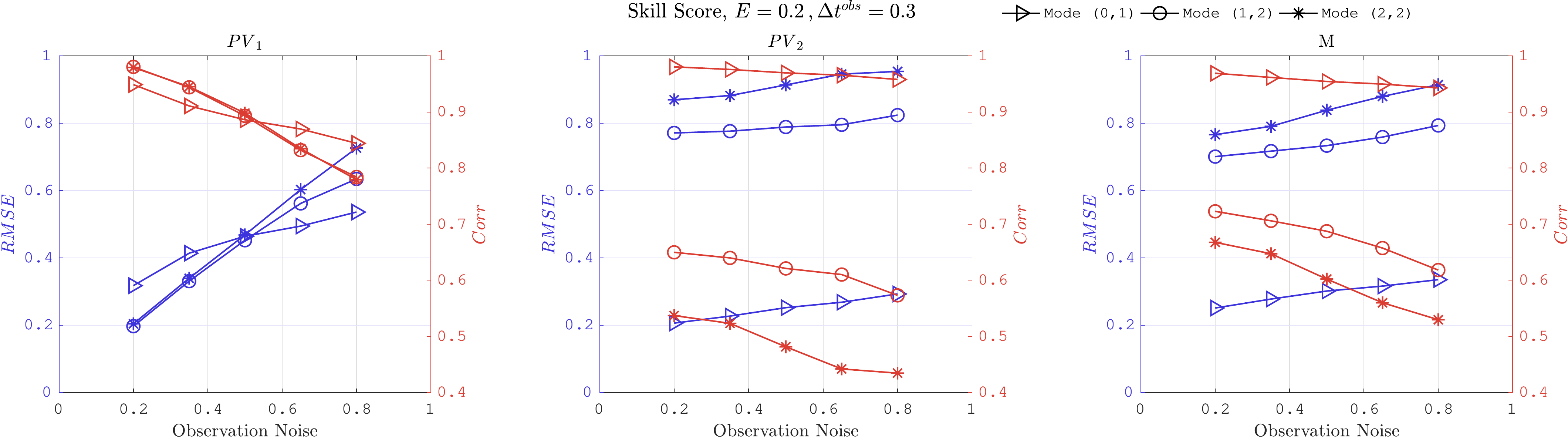}
    \caption{Skill scores, RMSE(in blue) and Corr (in red)  for different observational noise and fixed observational time $\Delta t^{obs} =0.3$ with three different Fourier modes. }
    \label{fig3-skill-score-fourier}
\end{figure}

Figure \ref{fig3-skill-score-fourier} shows skill scores for Fourier modes $(0,1),(1,2)$ and $(2,2)$ with different observational noise levels. For all the modes, increasing the observation noise leads to a larger RMSE and smaller pattern correlation, as expected. For $PV_1$, due to the availability of direct observations, the data assimilation remains skillful for all these Fourier modes. For the unobserved variables, namely $PV_2$ and $M$, the hybrid data assimilation results get worse from large- to small-scale variables. This is because smaller-scale modes are more intermittent and chaotic with shorter memory, and thus their recovery is more challenging. In addition to the observational noise, the sensitivity test on different observational time steps has also been carried out (not shown here). When the observational time step is reduced from $\Delta^{obs}t=0.3$ to $\Delta^{obs}t=0.1$, the recovery of the modes with $8\geq|k|>3$ is slightly more accurate as the observational time step becomes shorter than the decorrelation time. As a result, the accuracy of the recovered spatial reconstructed fields is slightly enhanced.

Two additional sensitivity tests have been implemented. First, if a simple linear regression replaces the RNN for detecting the relationship between the observed and unobserved state variables, then the data assimilation skill for the large-scale modes with $|k|\leq2$ becomes significantly worse. Consequently, the reconstructed spatial fields are more biased even for recovering the large-scale features. Second, suppose the SPEKF model is substituted by a simple linear Gaussian process, namely a complex OU process, as a cheap stochastic parameterization forecast model for the observed variables. In that case, the data assimilation result of $PV_1$ is less accurate in estimating intermittency and extreme events. This is consistent with the previous results shown in \cite{branicki2018accuracy}, which indicates the importance of adopting the additional stochastic processes to characterize the non-Gaussian features of these intermittent signals.

\subsection{Extrapolation \label{ss-sen}}


A robust data assimilation method means it remains skillful when carrying out extrapolation within a certain range. In other words, applying the current calibrated forecast model to different dynamical regimes (i.e., other geophysical scenarios) is expected to remain skillful. To this end, the SPEKF model and the LSTM trained in the regime with $E=0.2$ (the regime studied above) are tested on two other dynamical regimes with  $E=0.35$ and $E=0.02$, which correspond to $40\%$ and $2\%$ of the cloud fraction, respectively.

\begin{figure}[H]
\setlength{\tempwidth}{.3\linewidth}
\settoheight{\tempheight}{\includegraphics[width=\tempwidth]{example-image-a}}%
\centering
\rowname{Observation}
\subfloat{\includegraphics[width=\tempwidth]{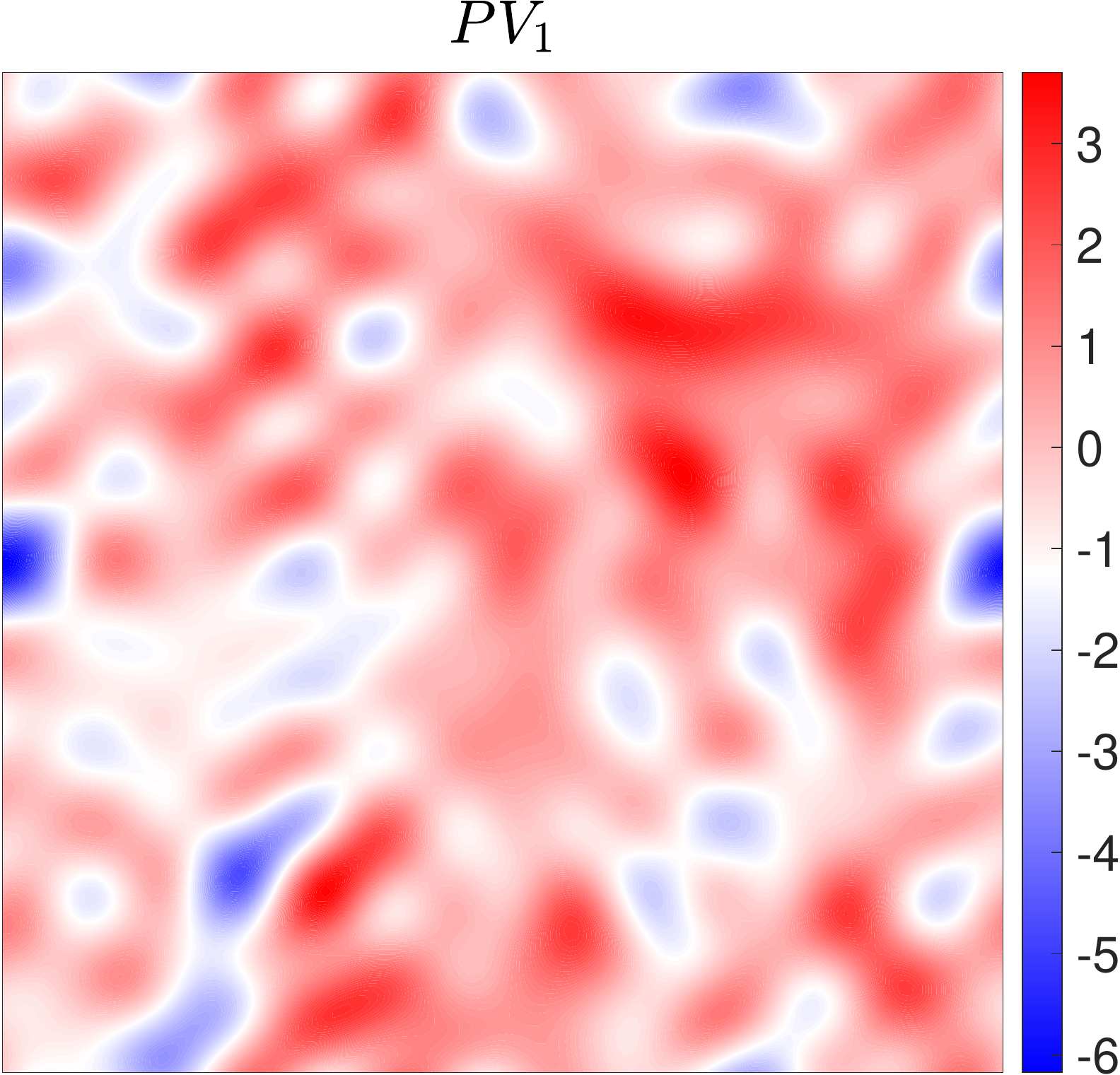}
}\hfil
\subfloat{\includegraphics[width=2\tempwidth]{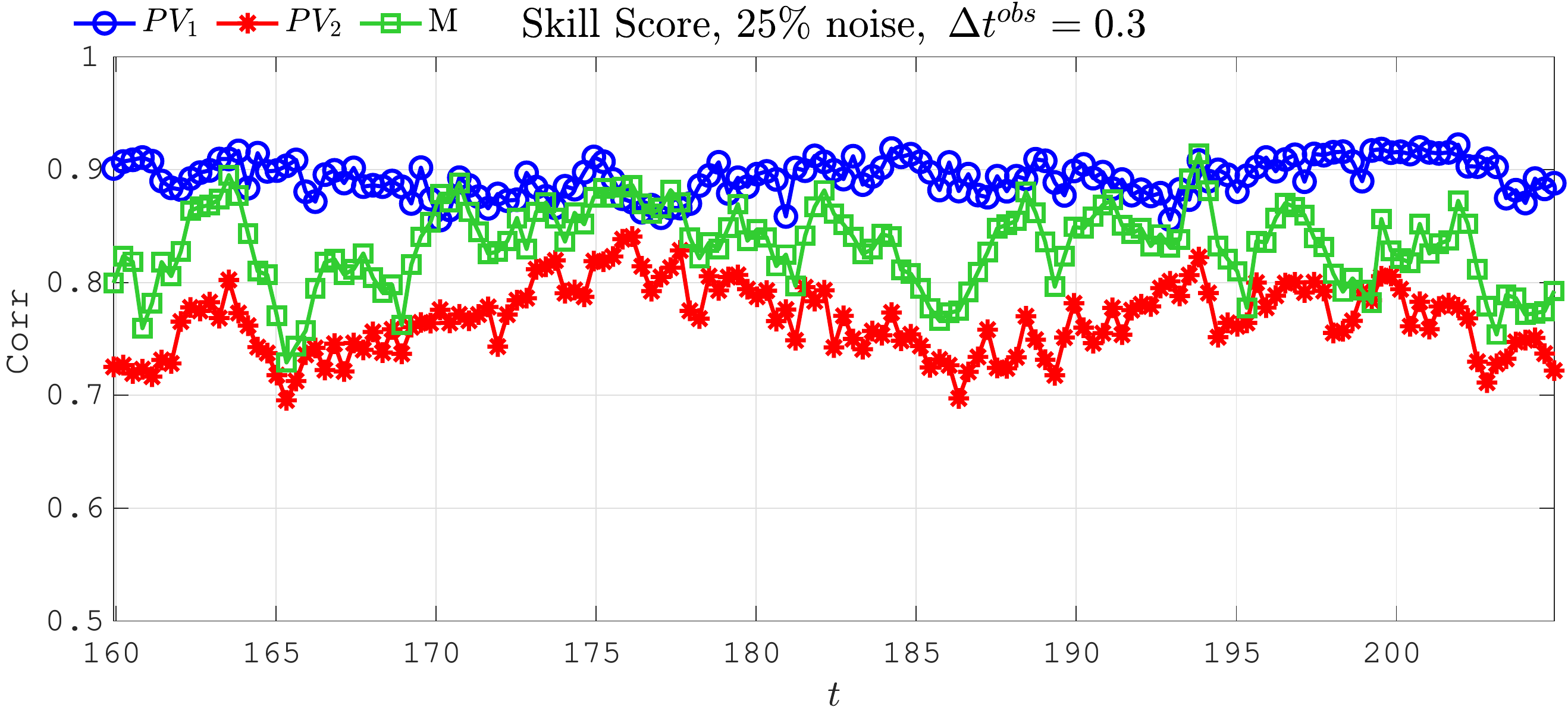}
}\\
\rowname{Truth}
\subfloat{\includegraphics[width=\tempwidth]{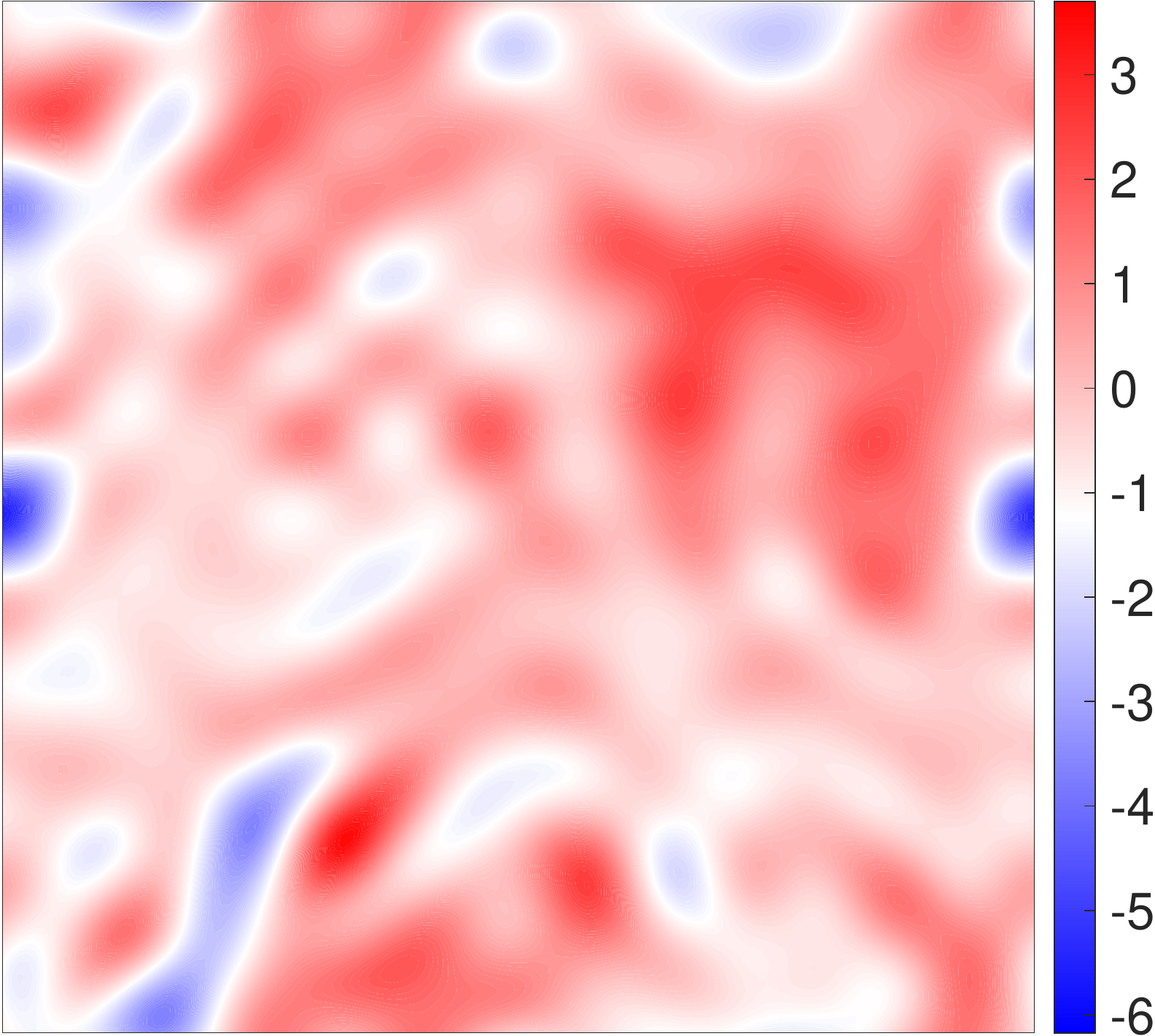}}\hfil
\subfloat{\includegraphics[width=\tempwidth]{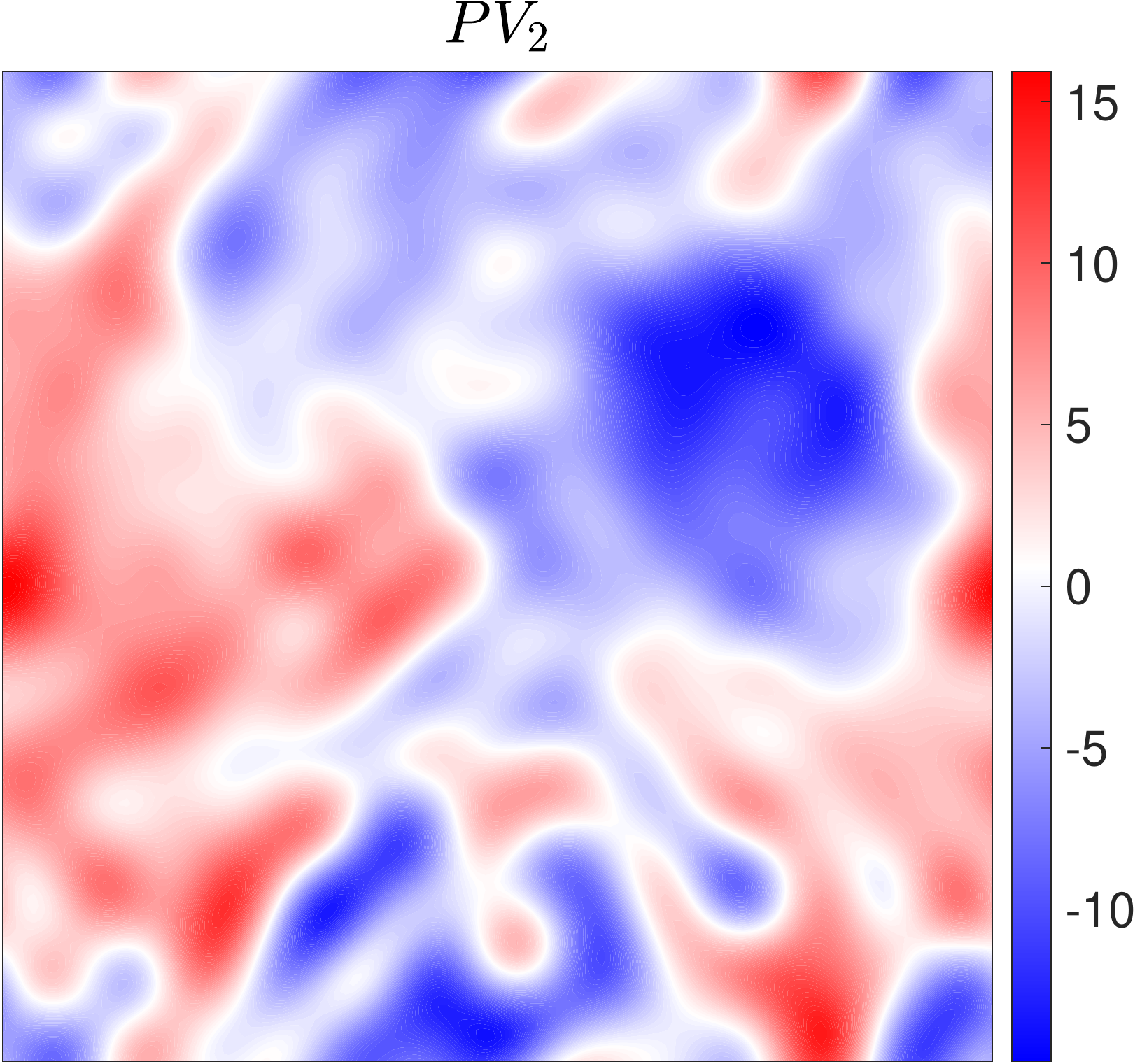}}\hfil
\subfloat{\includegraphics[width=\tempwidth]{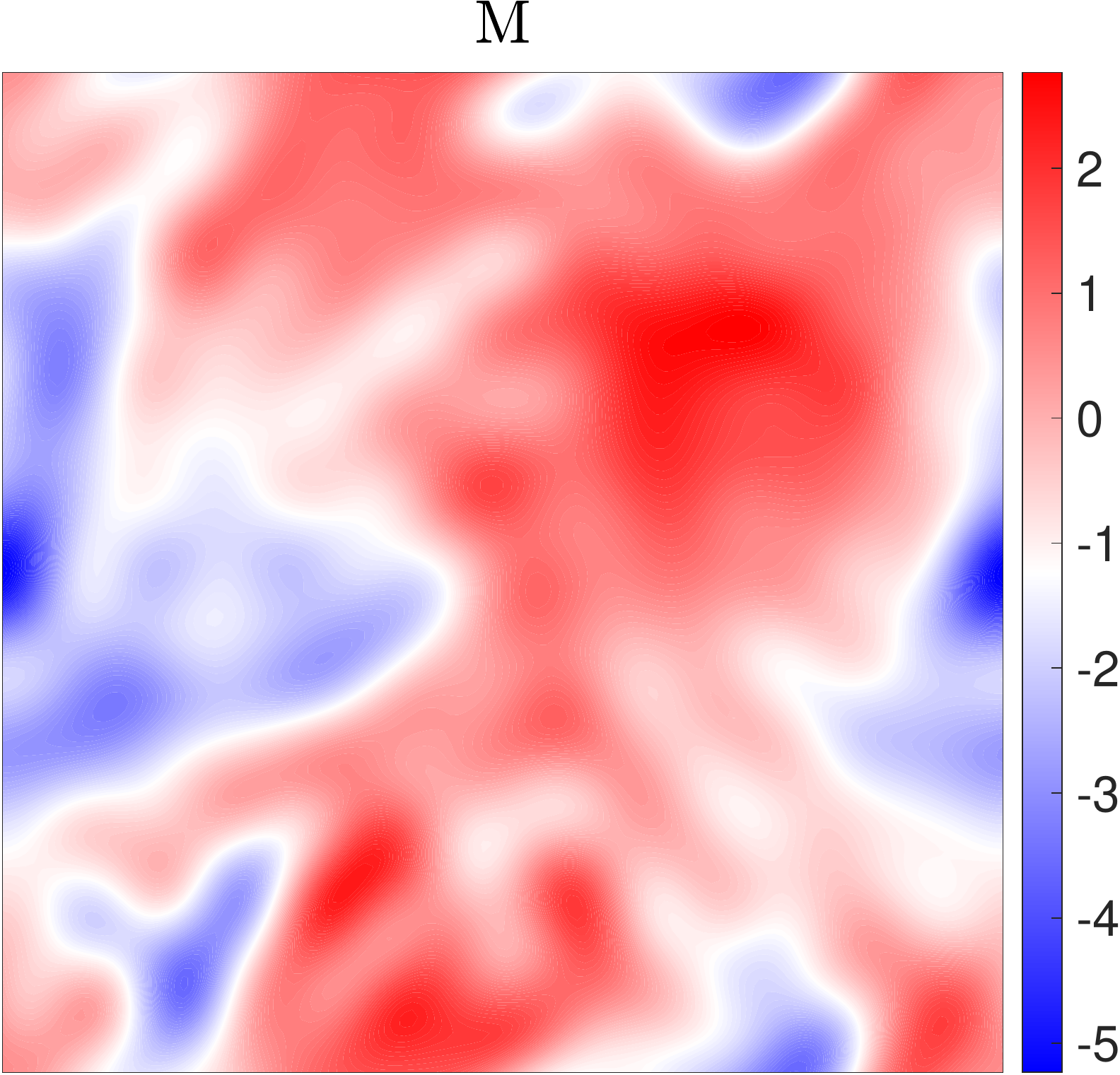}}
\\
\rowname{Filter}
\subfloat{\includegraphics[width=\tempwidth]{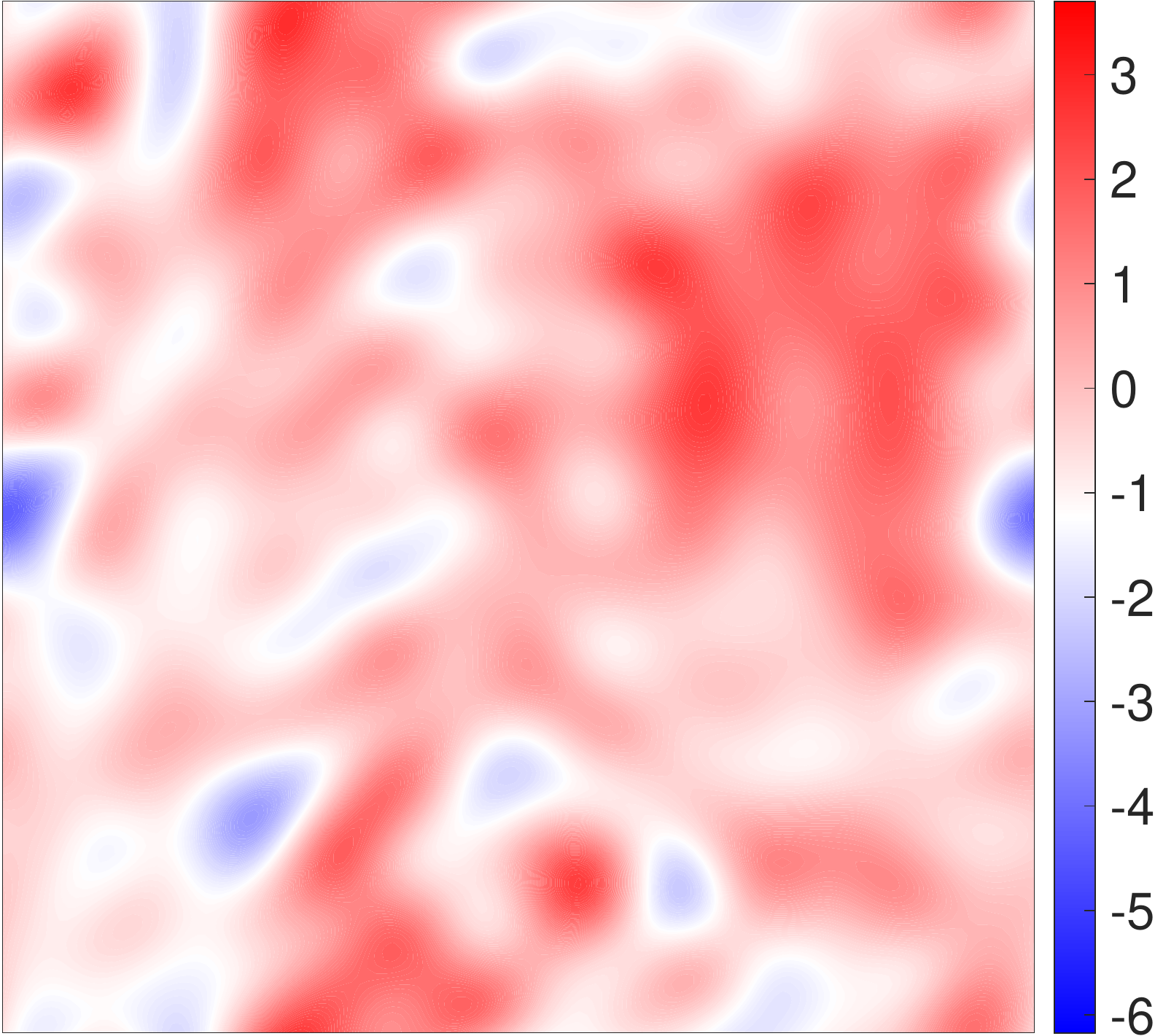}}\hfil
\subfloat{\includegraphics[width=\tempwidth]{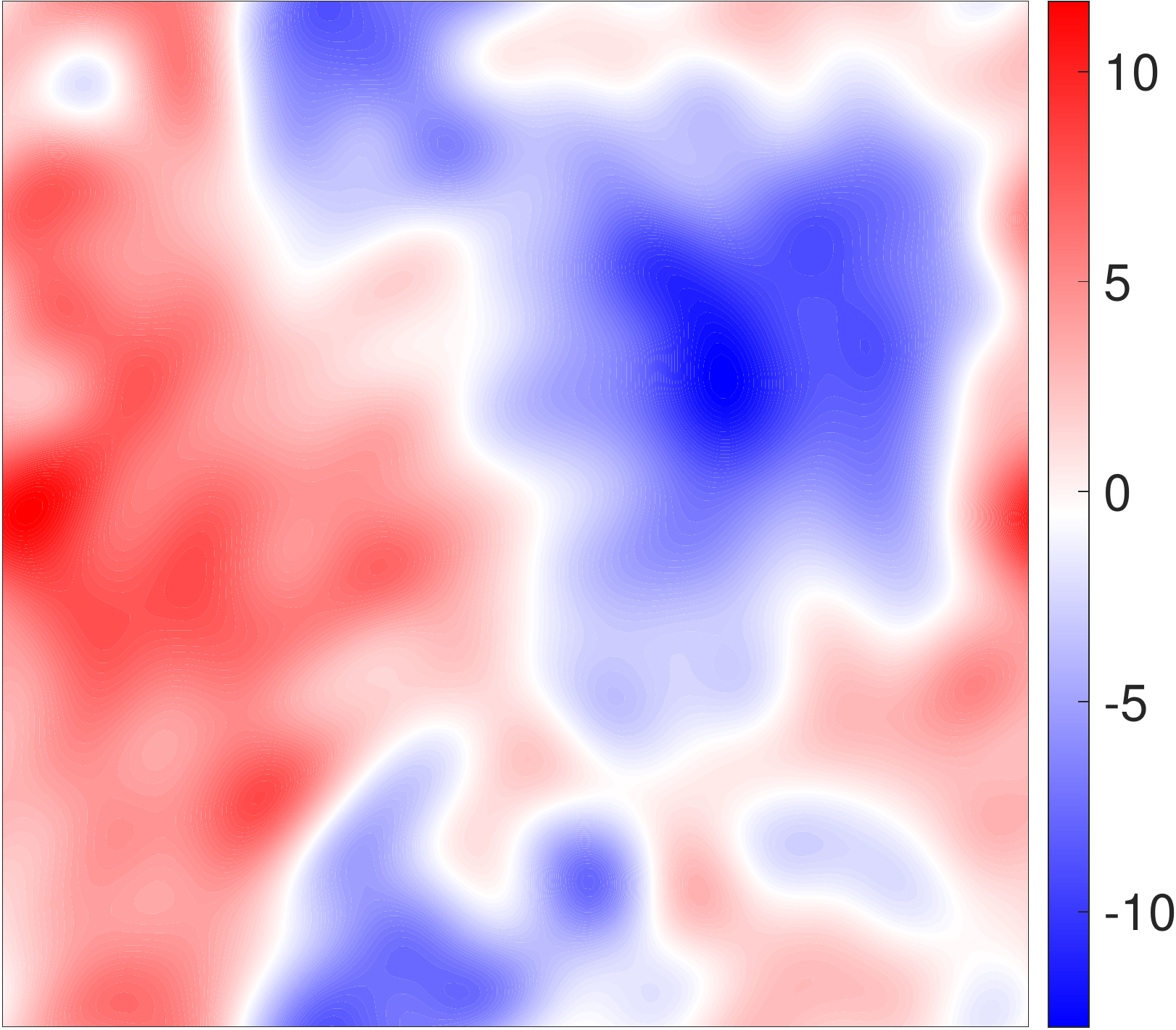}}\hfil
\subfloat{\includegraphics[width=.98\tempwidth]{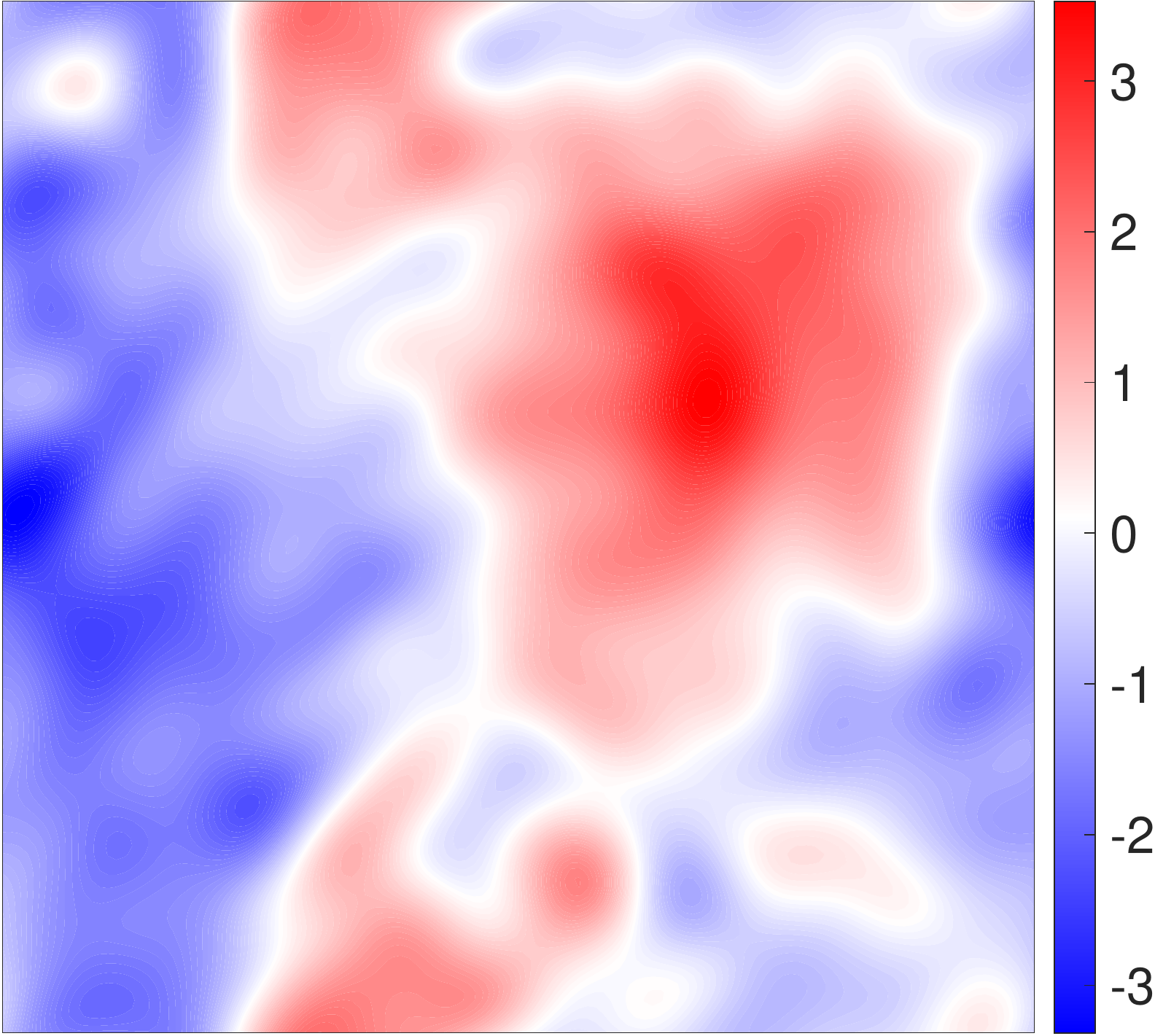}}
\caption{Reconstructed spatial fields and pattern correlation skill scores (upper right panel) of observed variable $PV_1$ (first column) and unobserved variables $PV_2$ (second column) and $M$ (third column) at $t = 190$, with observation time $\Delta t^{obs} =0.3$;
sensitivity test with $E=0.35$.}
\label{figure7:sen-variable-spatial}
\end{figure}

\begin{figure}[H]
\setlength{\tempwidth}{.3\linewidth}
\settoheight{\tempheight}{\includegraphics[width=\tempwidth]{example-image-a}}%
\centering
\rowname{Observation}
\subfloat{\includegraphics[width=\tempwidth]{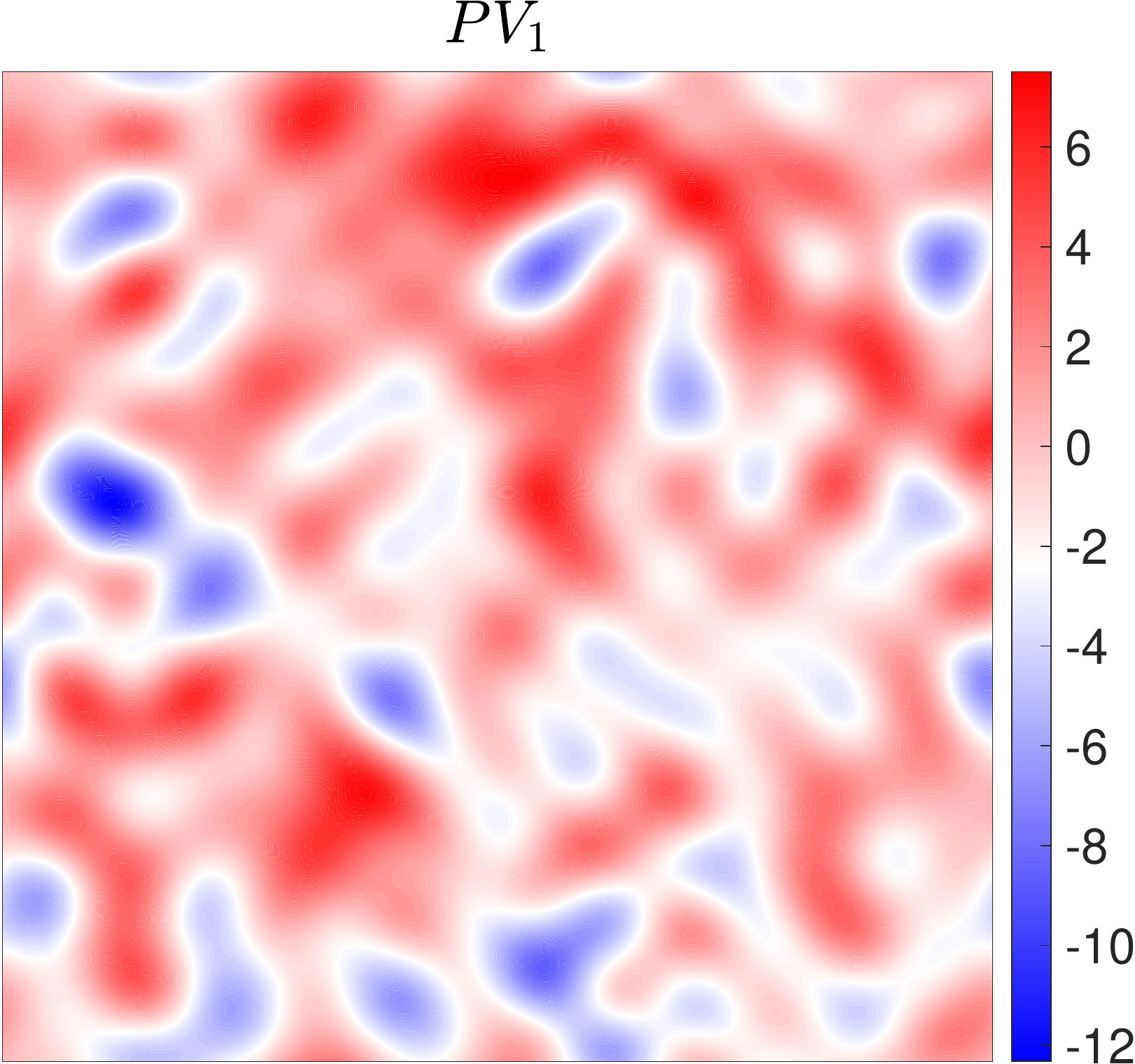}
}\hfil
\subfloat{\includegraphics[width=2\tempwidth]{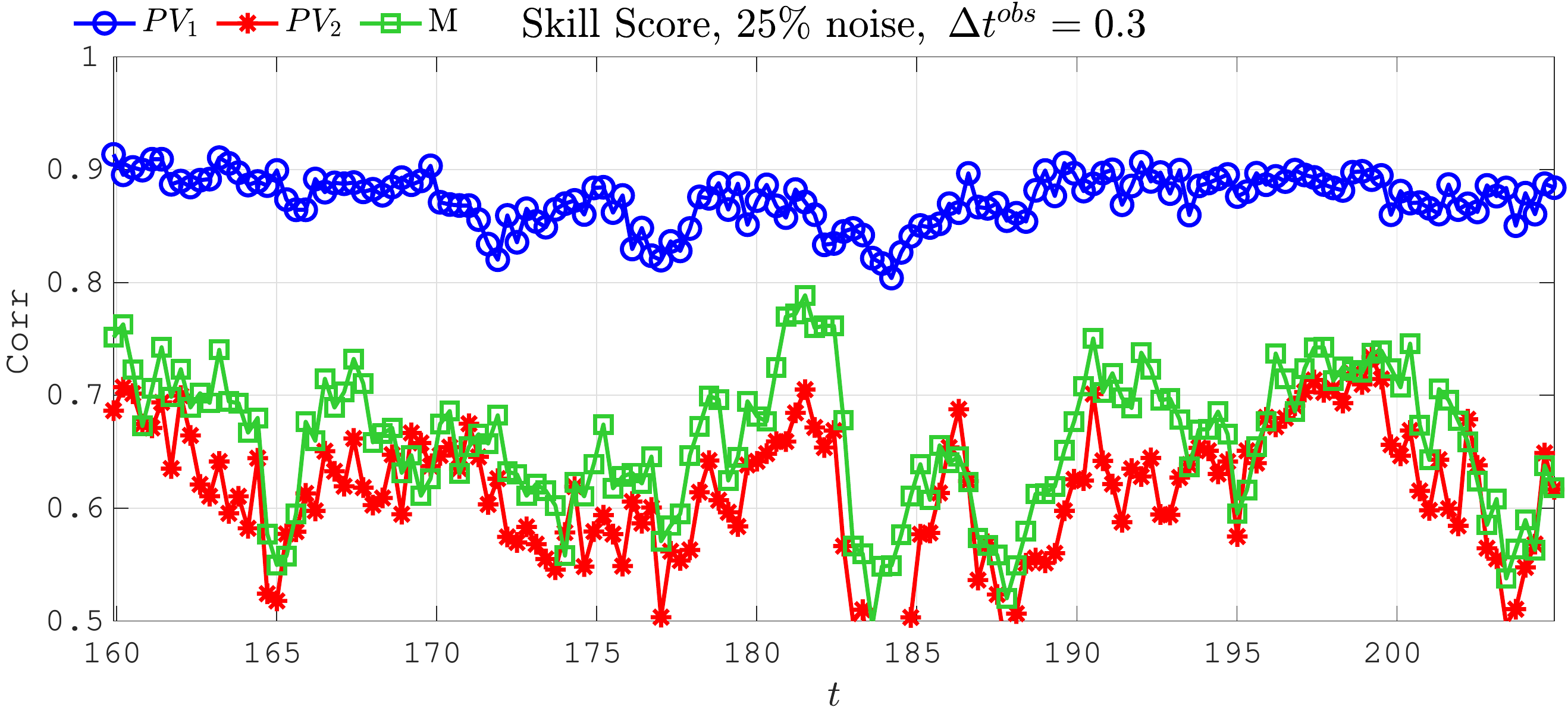}
}\\
\rowname{Truth}
\subfloat{\includegraphics[width=\tempwidth]{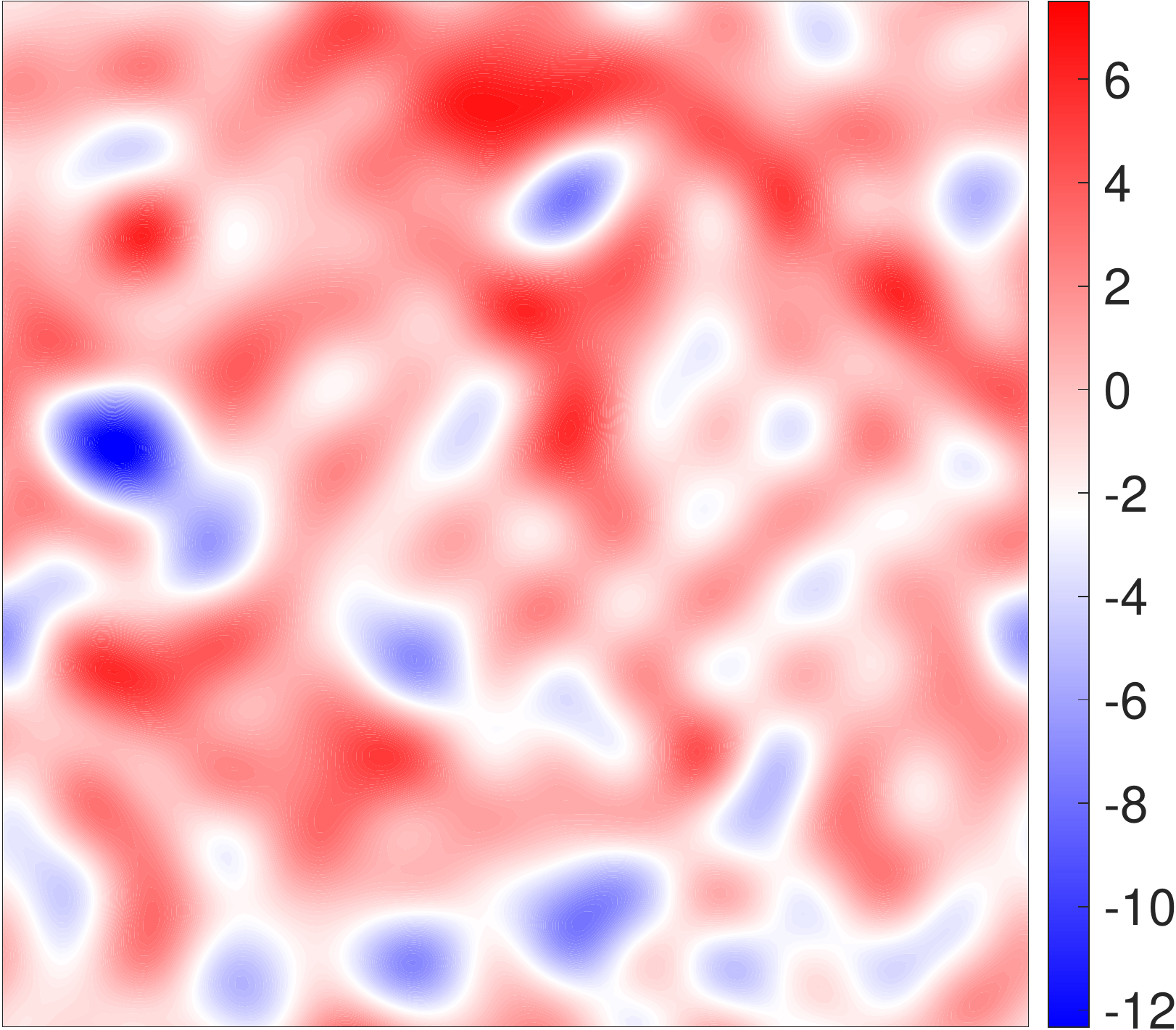}}\hfil
\subfloat{\includegraphics[width=\tempwidth]{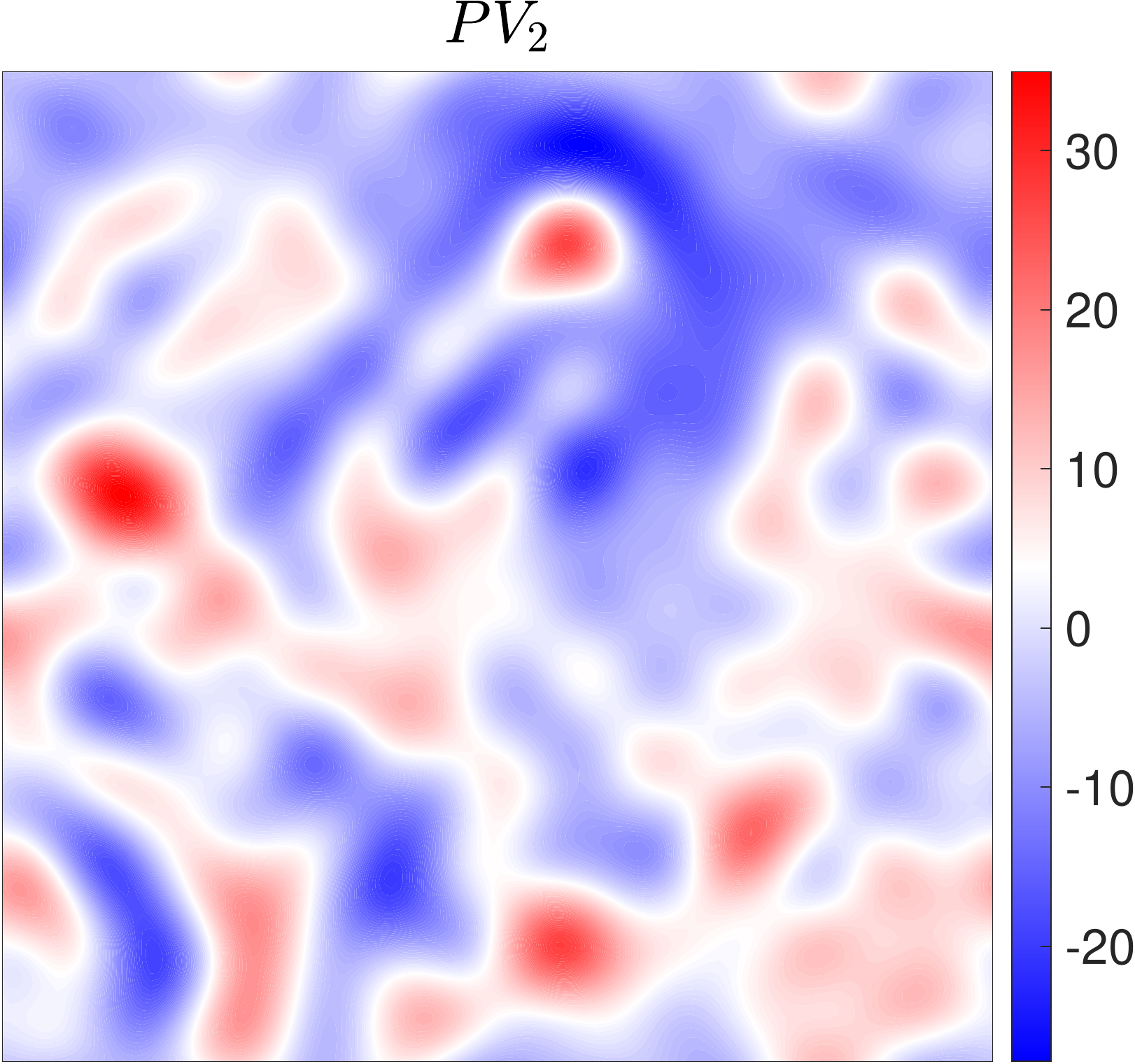}}\hfil
\subfloat{\includegraphics[width=\tempwidth]{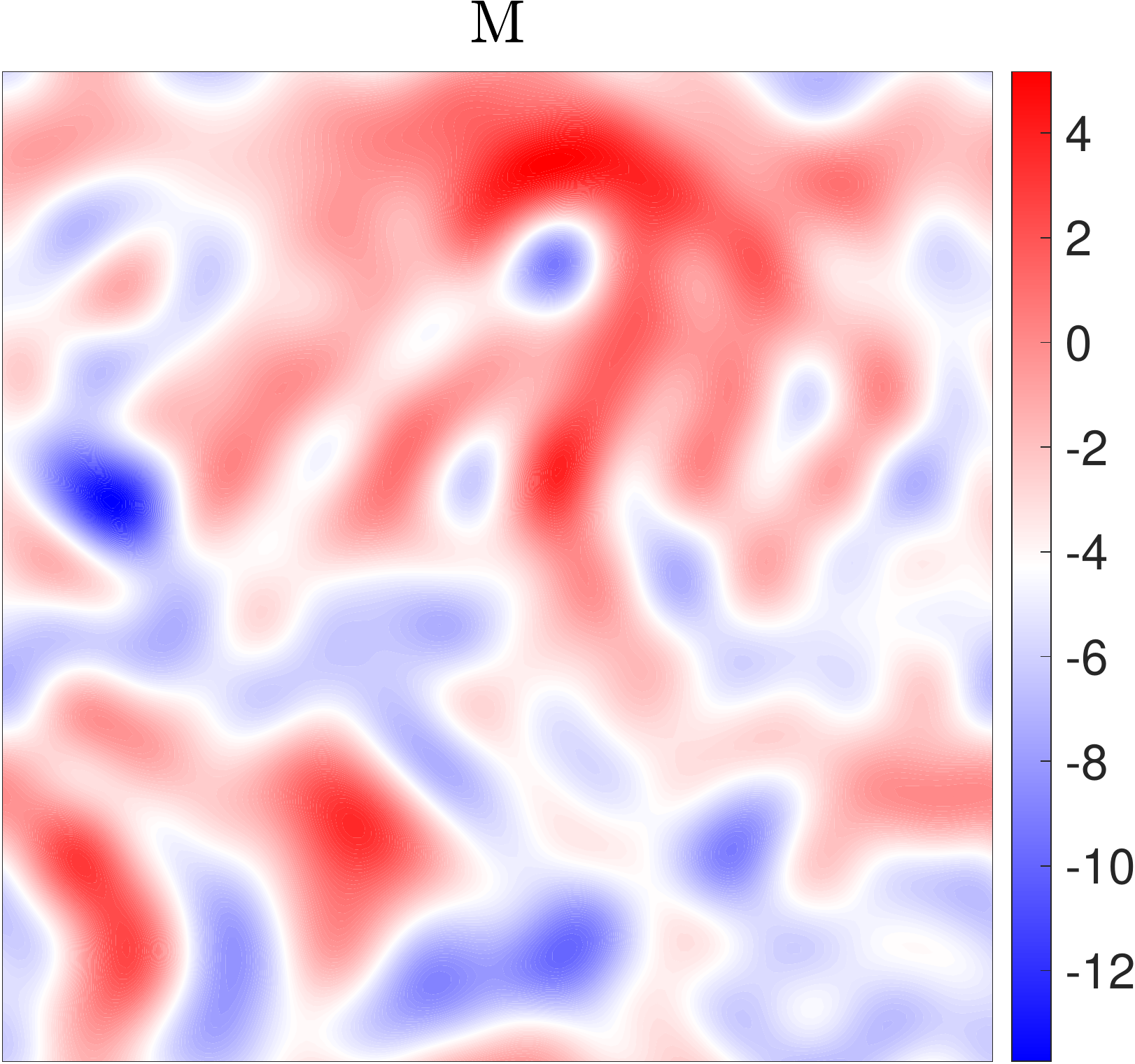}}
\\
\rowname{Filter}
\subfloat{\includegraphics[width=\tempwidth]{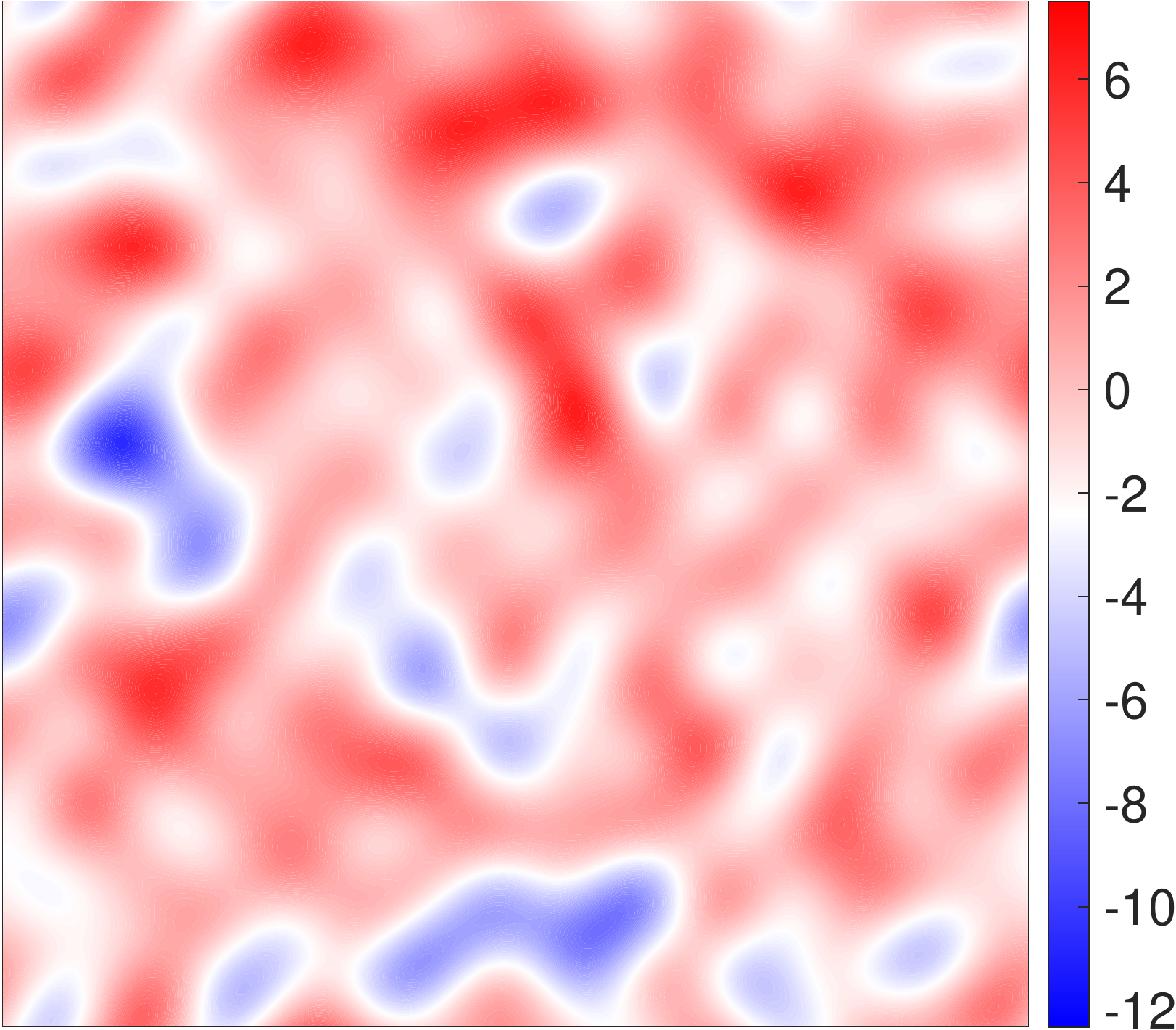}}\hfil
\subfloat{\includegraphics[width=\tempwidth]{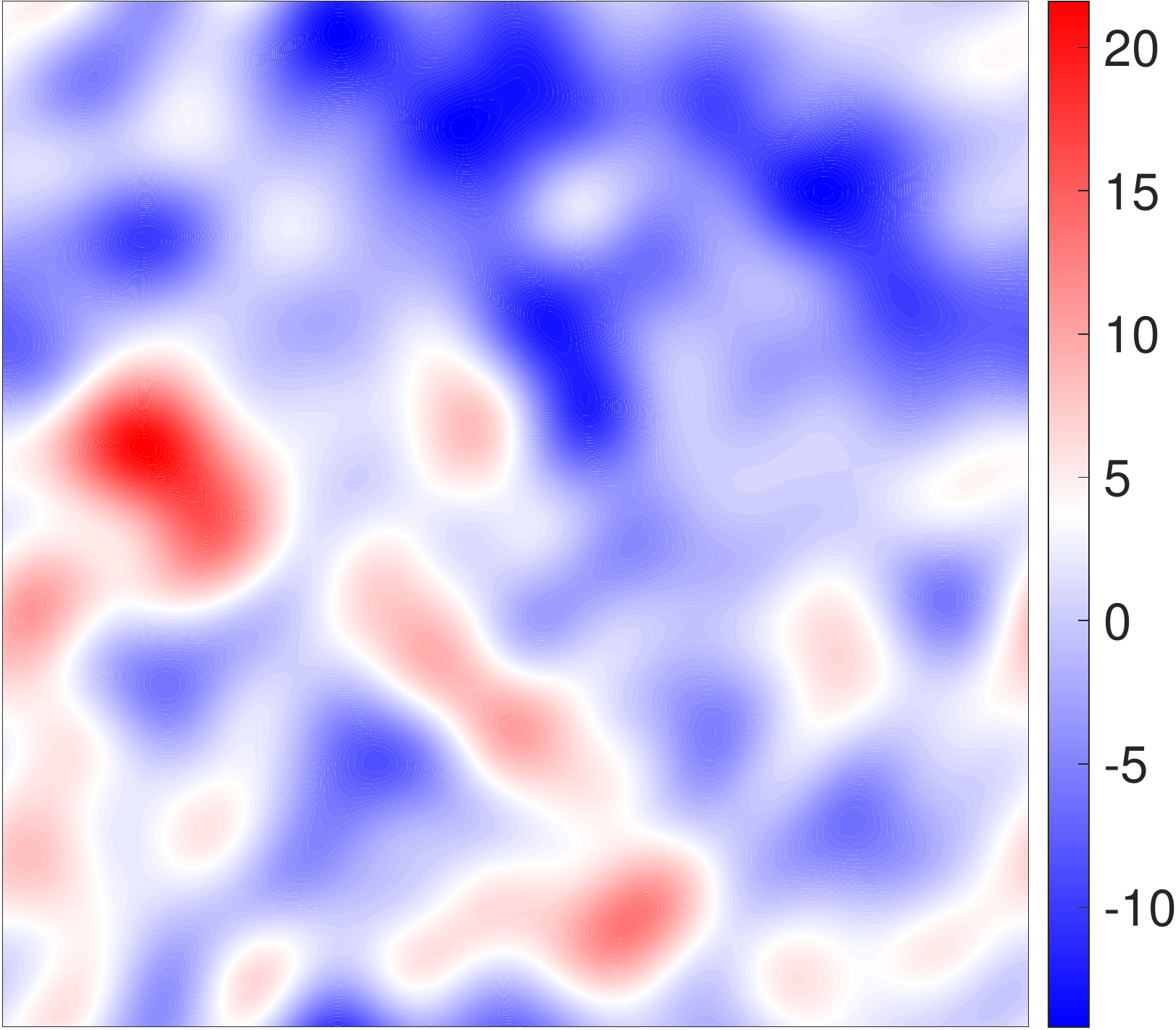}}\hfil
\subfloat{\includegraphics[width=.98\tempwidth]{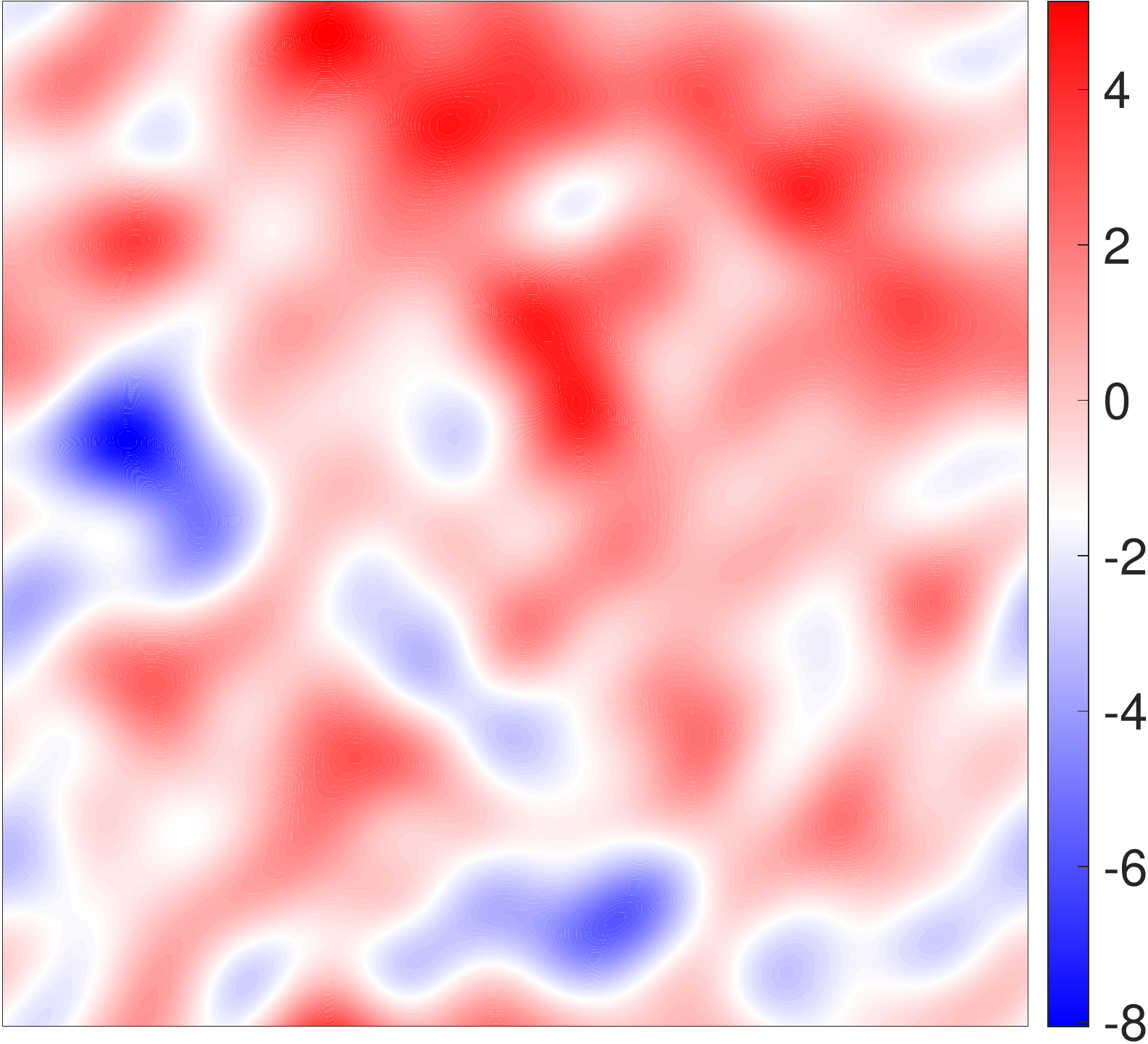}}
\caption{
Reconstructed spatial fields and pattern correlation skill scores (upper right panel) of observed variable $PV_1$ (first column) and unobserved variables $PV_2$ (second column) and $M$ (third column) at $t = 172$, with observation time $\Delta t^{obs} =0.3$;
sensitivity test with $E=0.02$.}
\label{figure6:sen-variable-spatial}
\end{figure}

Figures \ref{figure7:sen-variable-spatial}--\ref{figure6:sen-variable-spatial} show the spatially reconstructed fields of the observed and unobserved variables, $PV_1,PV_2$ and $M$, for the cases of $E=0.02$ and $E=0.35$, respectively. In both the extrapolation regimes, the data assimilation skill remains similar to that in Figure \ref{figure4-spacial-temporal-field}. The reconstructed observed variable $PV_1$ field is nearly perfect, and the reconstructed unobserved variables $PV_2$ and moisture $M$ accurately capture the large-scale spatial patterns. This is also confirmed from the time series of pattern correlations for $PV_1, PV_2$ and $M$ in Figures \ref{figure7:sen-variable-spatial}--\ref{figure6:sen-variable-spatial}. Among these two cases, the one with $E=0.02$ has slightly larger errors. Note that $E=0.02$ leads to almost a dry QG system, which is the extreme case of the PQG and is quite far from the typical features detected in the regime with $E=0.2$.

\begin{figure}[H]
\setlength{\tempwidth}{.22\linewidth}
\settoheight{\tempheight}{\includegraphics[width=\tempwidth]{example-image-a}}%
\centering
\hspace{\baselineskip}
\columnname{$q_r, E=0.02$}\hfil
\columnname{$u_2, E=0.02$}\hfil
\columnname{$q_r, E=0.35$}\hfil
\columnname{$u_2, E=0.35$}
\\
\rowname{Truth}
\subfloat{\includegraphics[width=1.02\tempwidth]{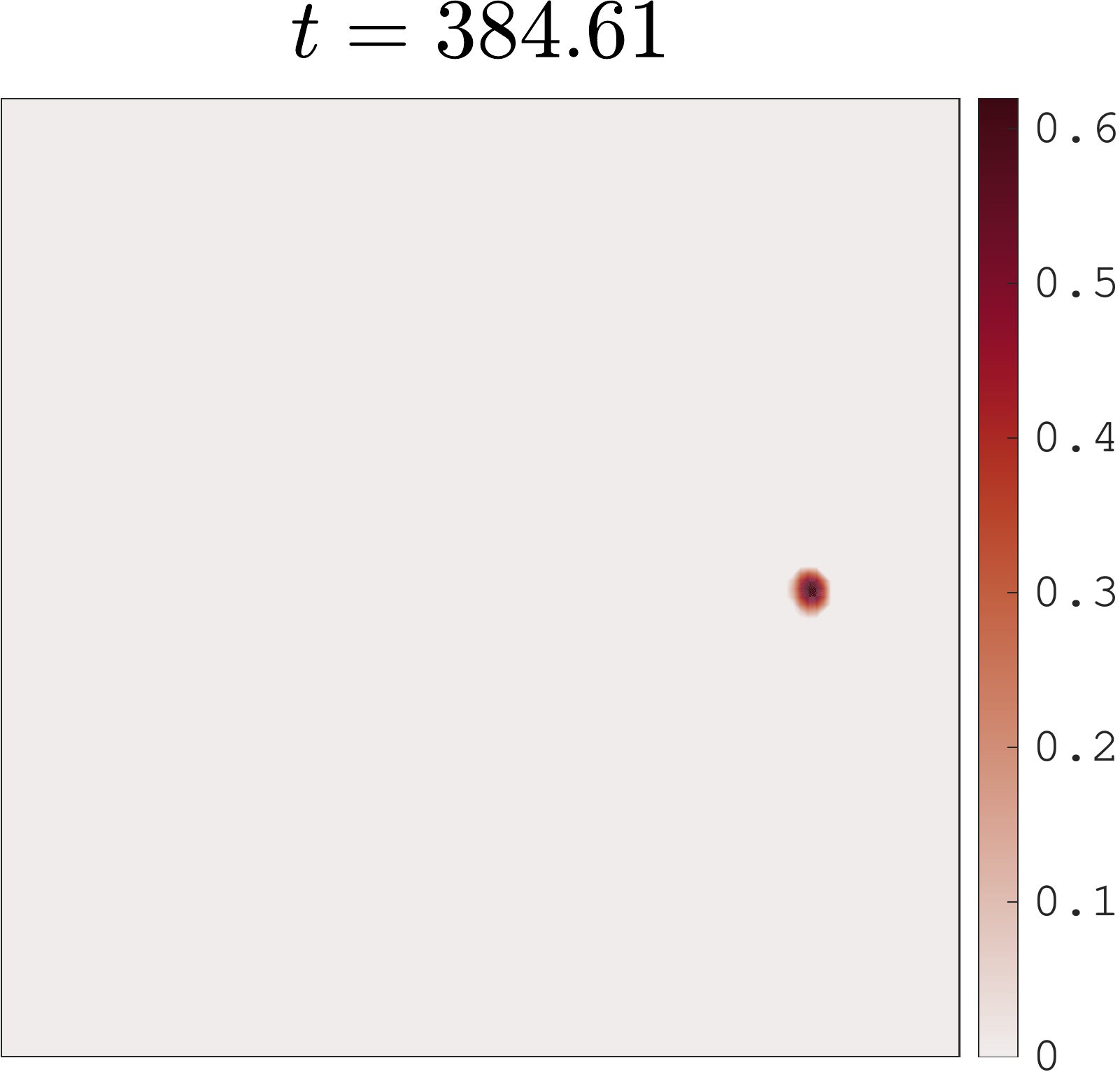}}\hfil
\subfloat{\includegraphics[width=1\tempwidth]{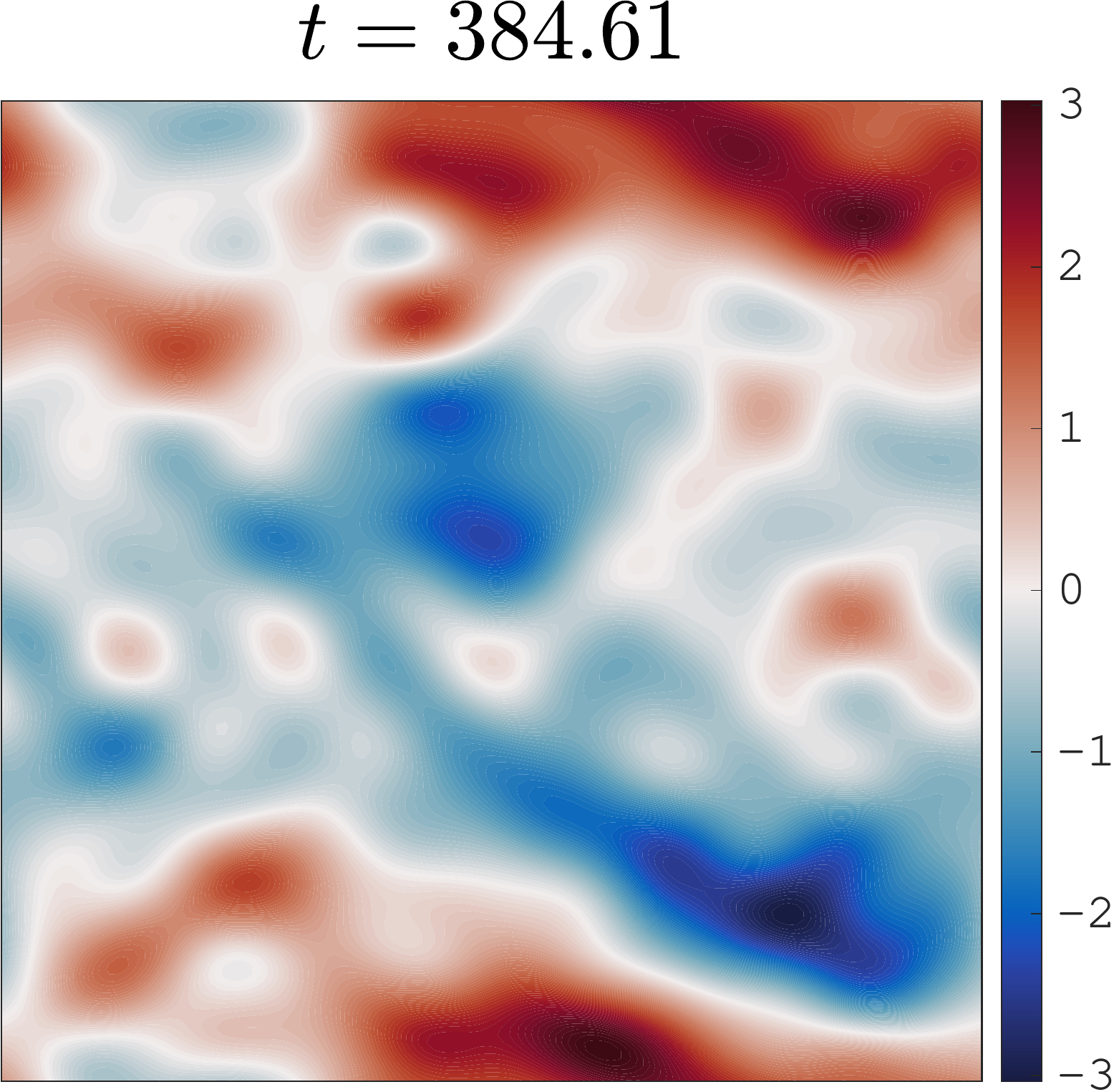}}\hfil
\subfloat{\includegraphics[width=1.02\tempwidth]{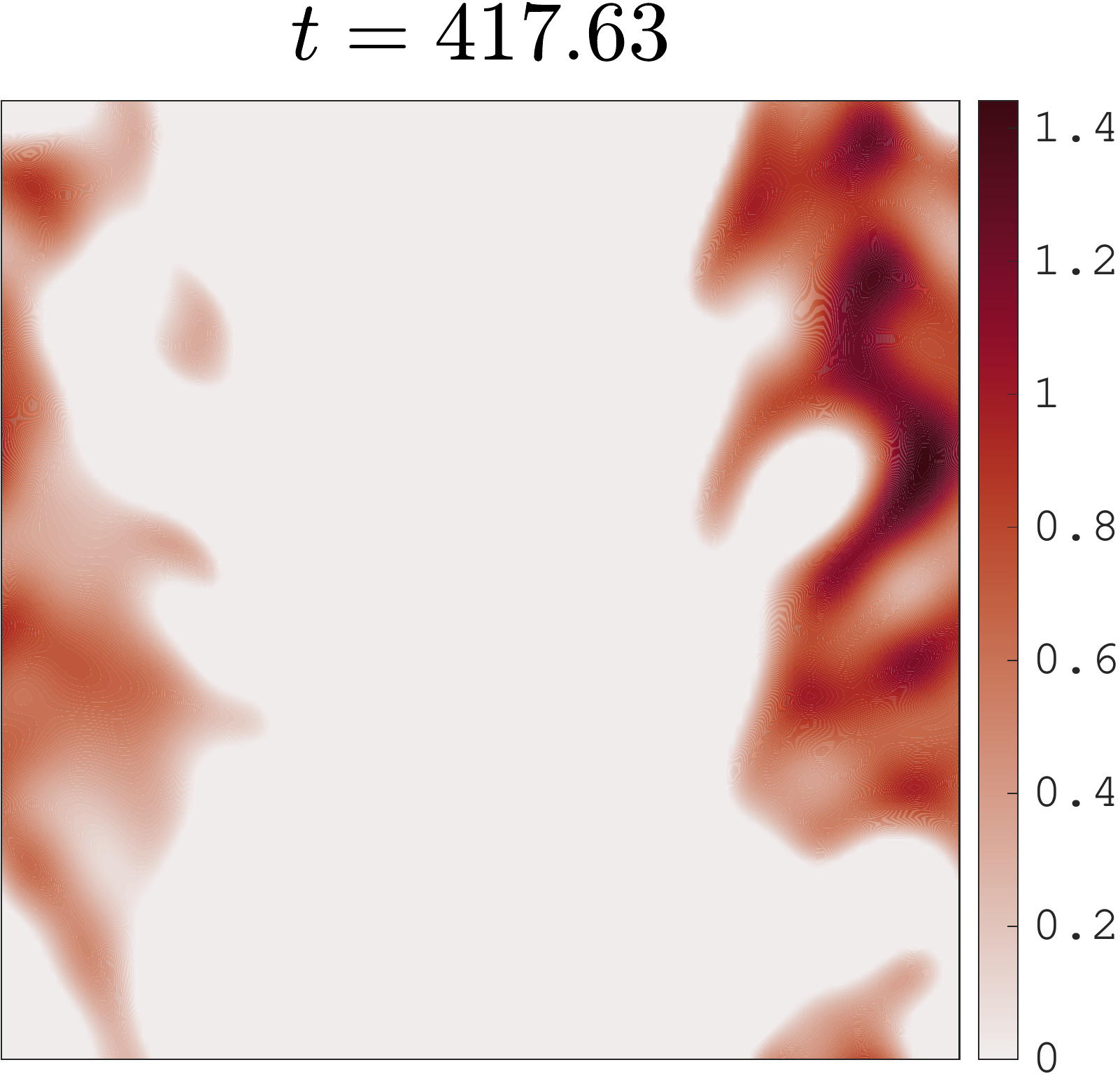}}\hfil
\subfloat{\includegraphics[width=1.05\tempwidth]{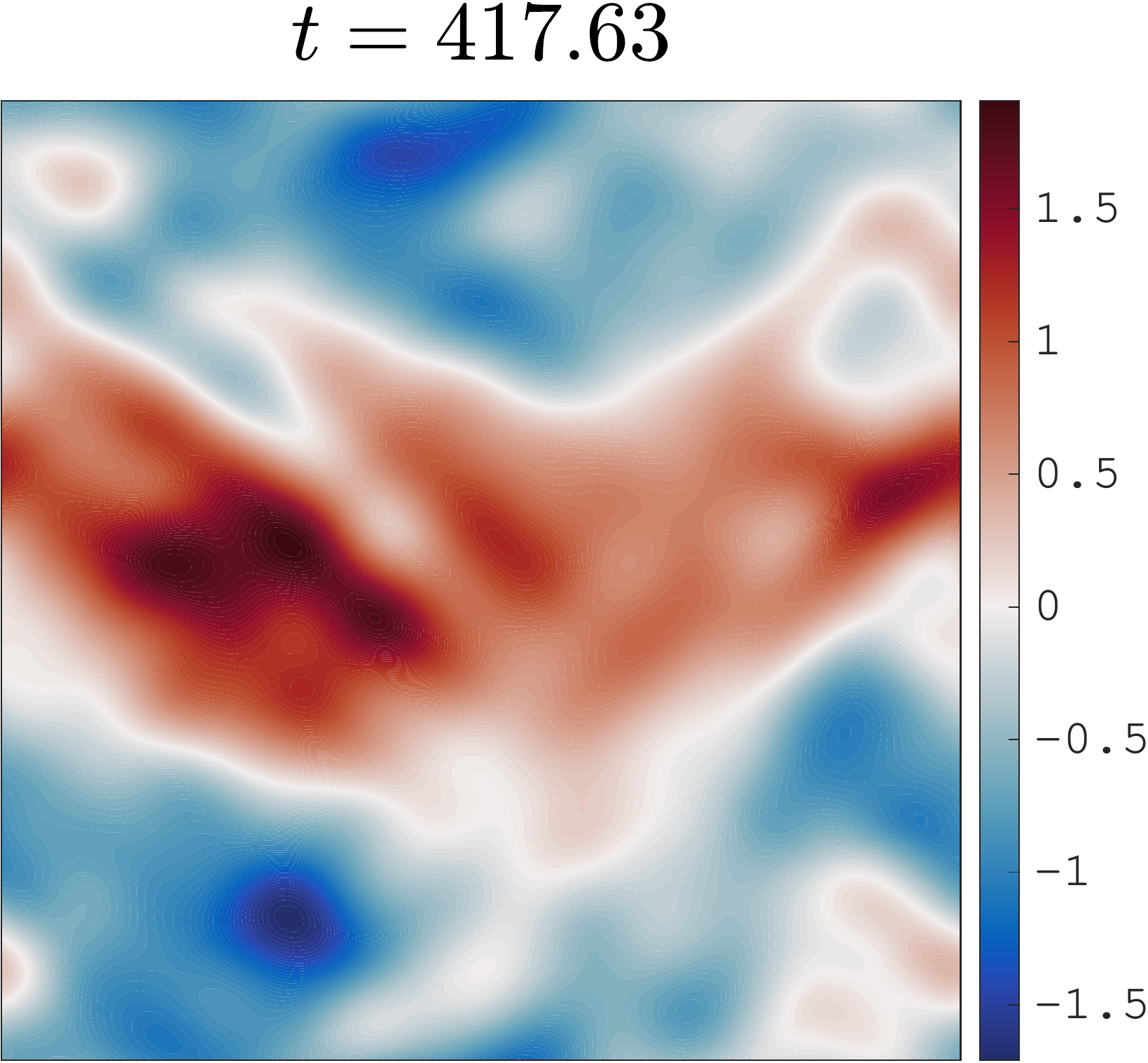}}\hfil
\\
\rowname{Filter}
\subfloat{\includegraphics[width=1.02\tempwidth]{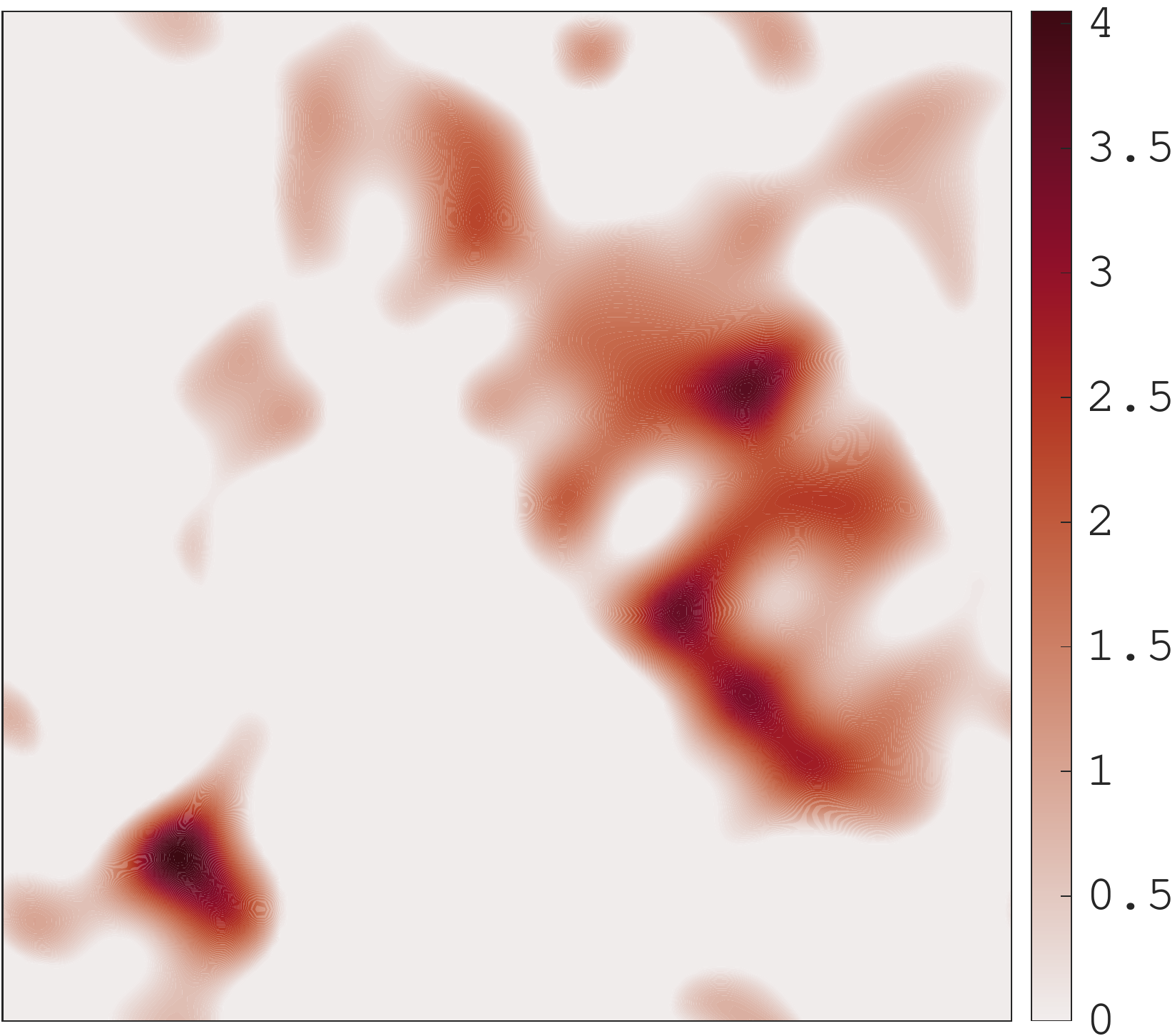}}\hfil
\subfloat{\includegraphics[width=1.05\tempwidth]{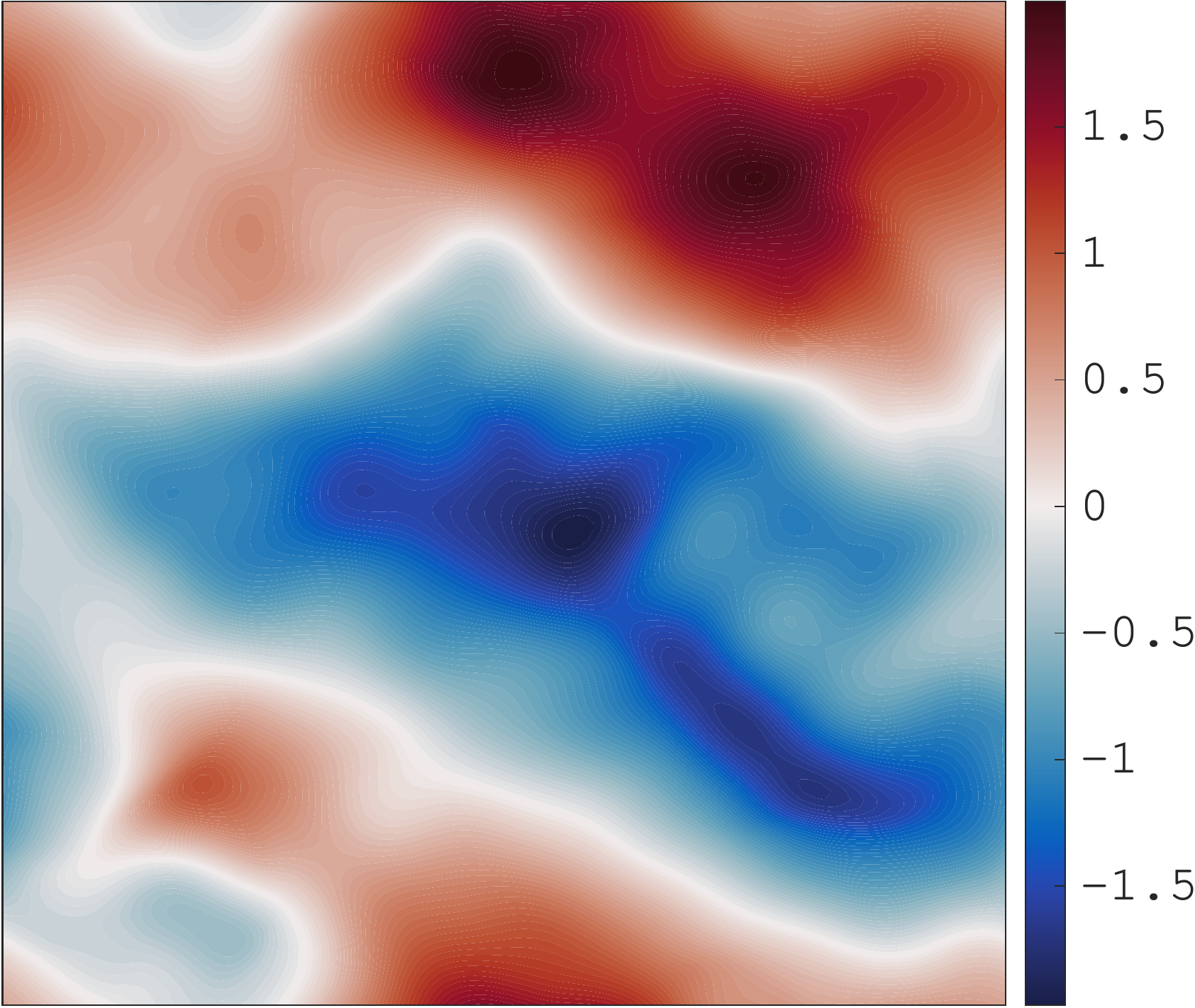}}\hfil
\subfloat{\includegraphics[width=1.02\tempwidth]{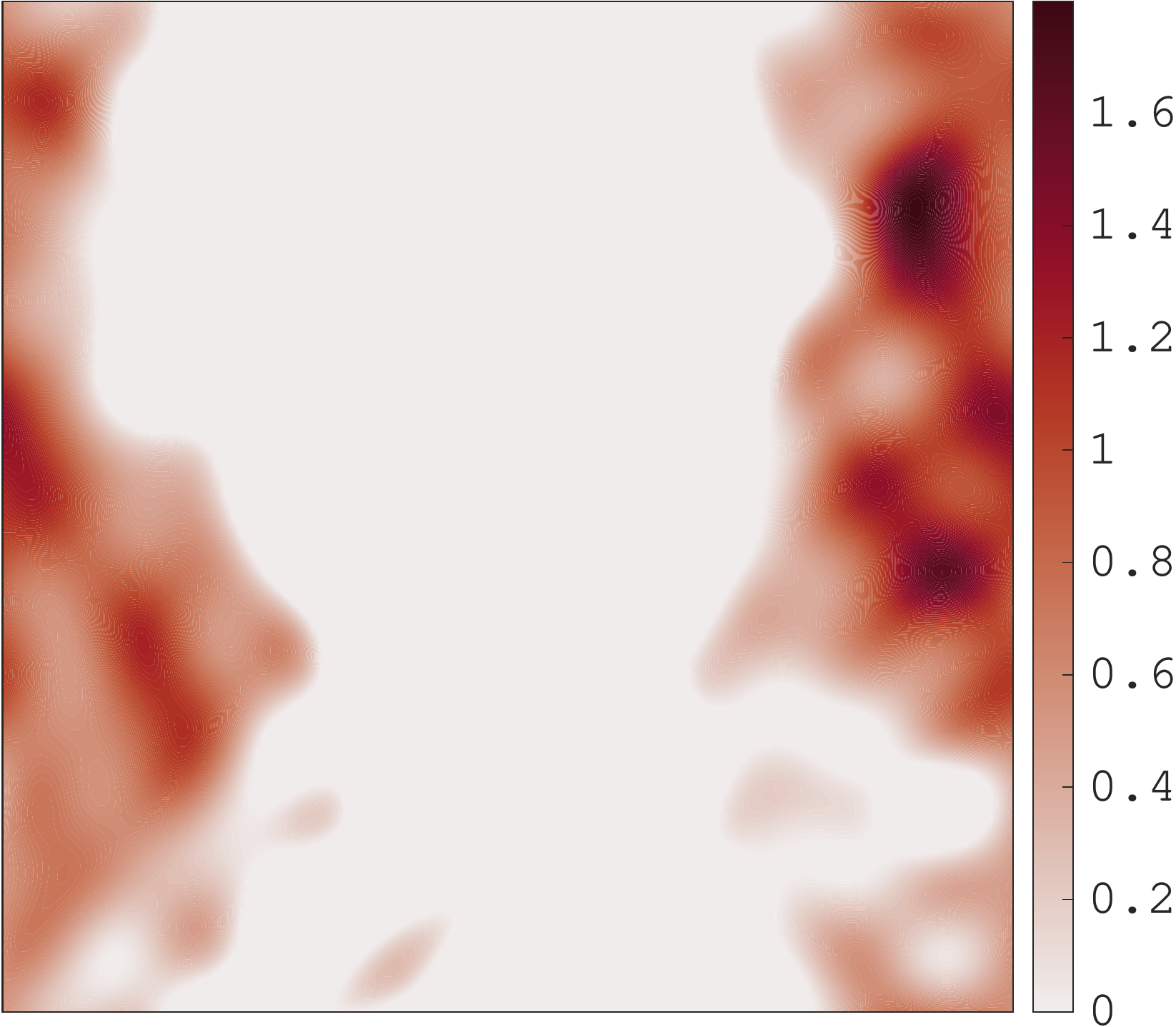}}\hfil
\subfloat{\includegraphics[width=1.05\tempwidth]{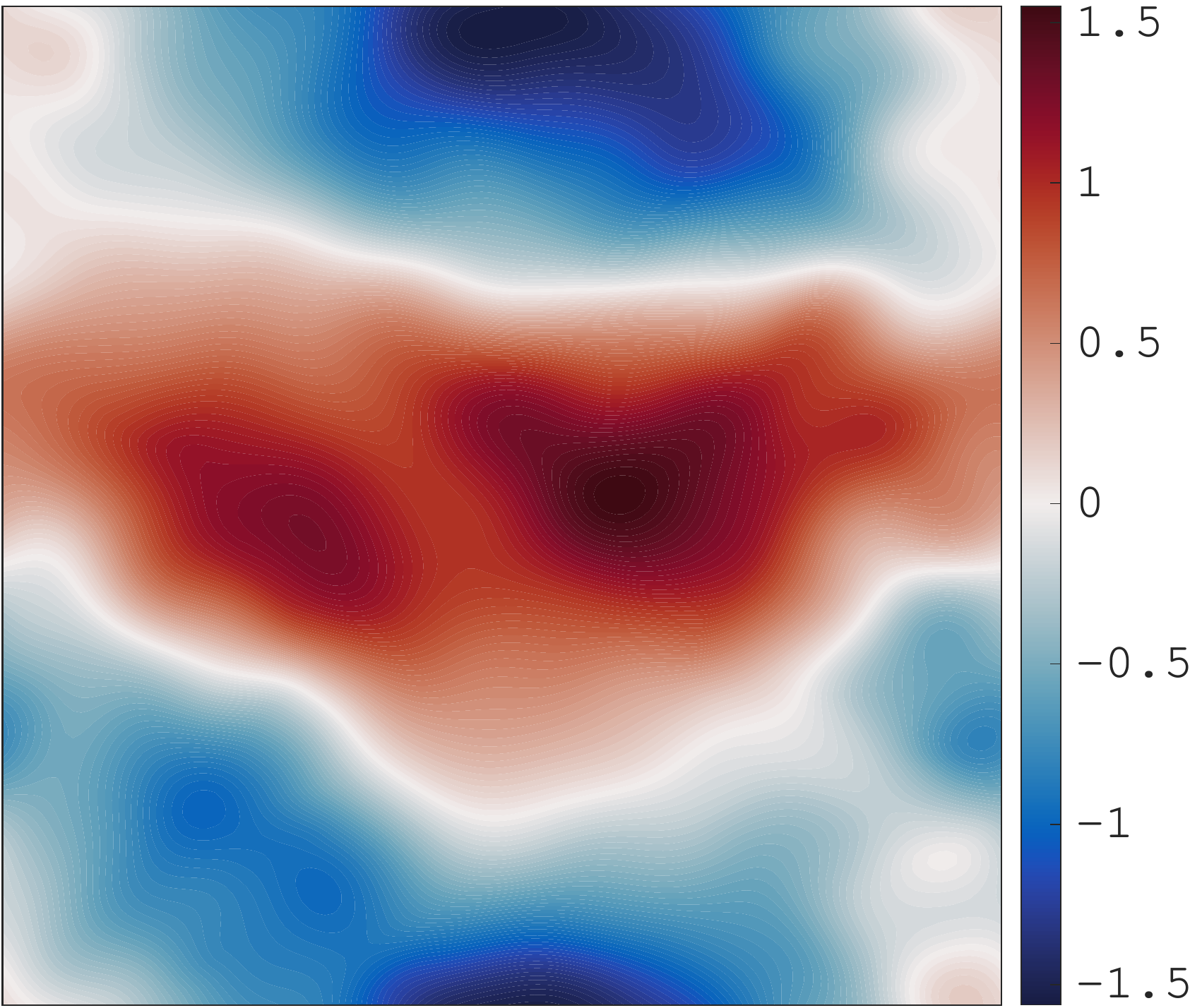}}
\\
\rowname{Corr}
\subfloat{\includegraphics[width=2.1\tempwidth]{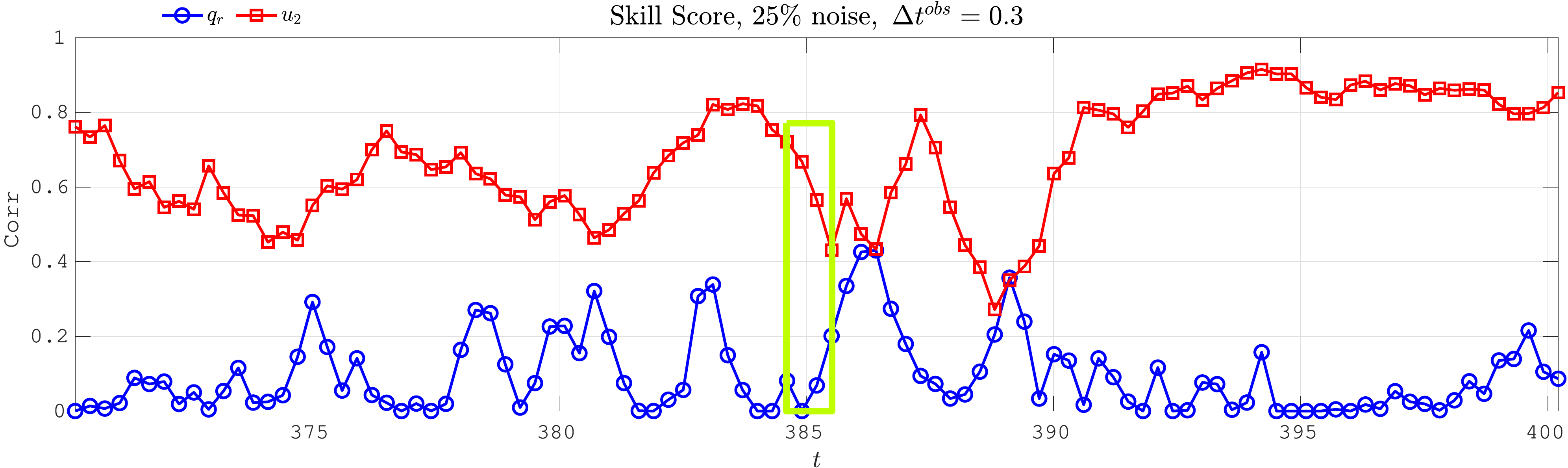}}
\hfil
\subfloat{\includegraphics[width=2.1\tempwidth]{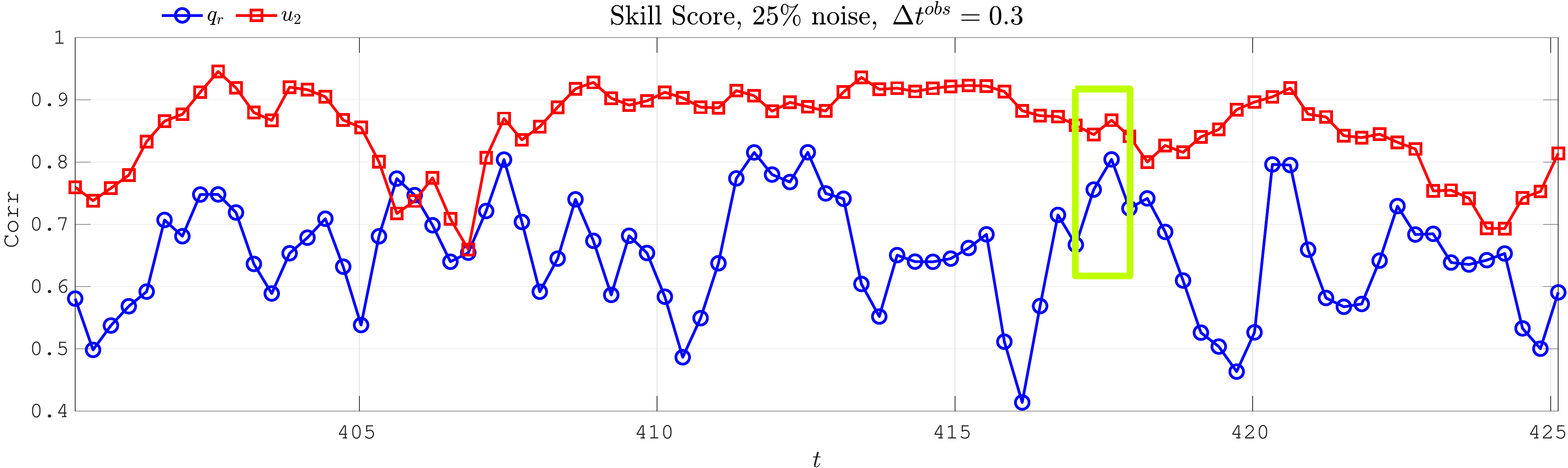}}
\caption{Recovered spatial fields and pattern correlation skill scores of rain water $q_r$ and zonal velocity $u_2$ (level 2) in physical space, at two different times, with $\Delta t^{obs} =0.3$; the green box in the time series of pattern correlation denotes the time interval of recovered $q_r$ and $u_2$ fields; Sensitivity test with $E=0.02$ (left panel) and $E=0.35$ (right panel).}
\label{figure8}
\end{figure}

Figure \ref{figure8} shows the reconstructed $q_r$ and $u_2$ in the two extrapolation regimes. In both regimes, the jet structure of $u_2$ is recovered quite accurately in terms of both the strength and the location. The rainwater area $q_r$ in the $E=0.35$ regime is also recovered reasonably well, with an averaged pattern correlation being $0.6$. Yet, the detected $q_r$ in the $E=0.02$ regime is far from the truth. This is not surprising as the recovered $PV_2$ and $M$ (see Figure \ref{figure6:sen-variable-spatial}) are already slightly less accurate. The Heaviside function magnifies such inaccuracies during the reconstruction of $q_r$. In addition, the training regime rarely has time instants that are almost clear skies. Thus, the features in such a case are not well-represented training phases for the SPEKF model and the LSTM network, which causes the deterioration of the data assimilation skill in such an extreme extrapolation regime.

\section{Conclusion and Discussions}\label{Sec:Conclusion}

In this paper, a new hybrid data assimilation framework is proposed. The new data assimilation combines partial observations with a machine learning model to recover the unobserved variables and quantify the uncertainties. This hybrid data assimilation framework exploits the cheap and low-order but effective SPEKF model as the forecast model in the EnKF for the observed variables. It constructs the RNN machine learning algorithm to recover the unobserved variables with ensembles. In addition to the ensemble estimation, posterior uncertainties are quantified using the mixture distribution with the machine learning residual. Numerical results with the PQG equations show that the proposed data assimilation framework accurately filters observed and unobserved variables and is effective in recovering cloud microphysics quantities and jet streams associated with PQG equations. The calibrated SPEKF and machine learning models are also effective and robust in different dynamical regimes of PQG equations as extrapolation.

There are several future directions that may also provide potential improvement of the current framework. First, recall that the SPEKF model is a data-driven forecast model independent of the exact physical formulation of the time evolution of each spectral mode. More specific feature-based decomposition can be used to improve the characterization of turbulent features for complex systems. For example, a nonlinear POD-type autoencoder \cite{cheng2022generalised} can be adopted to replace the Fourier decomposition in suitable applications. Second, only a simple LSTM is utilized for discovering the relationship between observed and unobserved variables. More advanced machine learning tools can be used to describe such a nonlinear relationship more accurately. In addition, some ideas from the rank histogram filter (RHF) \cite{anderson2020marginal} can also be helpful to accelerate the computation. Third, extrapolation studies from different viewpoints may have particular interest. For example, the three-dimensional Boussinesq system \cite{smith2002generation} includes the slowly-evolving QG component and the fast gravity waves. It is essential to explore the data assimilation skill of the slow part of the Boussinesq system for the model calibrated based on a relatively low-cost QG model.
\section*{Acknowledgements}
The research of L.S. is funded by NSF-DMS-1907667. The research of N.C. is funded by ONR N00014-21-1-2904 and ONR N00014-19-1-2421. C.M. is partially supported as a postdoc research associate under ONR N00014-19-1-2421.

\section{Appendix}

\subsection{Details of the Sampling Formula in the Parameter Estimation of the SPEKF Model}
Recall the parameter estimation of the SPEKF model \eqref{SPEKF} described in Section \ref{Subsec:SPEKF_ParameterEstimation}. The algorithm consists of an iterative procedure that alternates between updating the parameters and sampling the trajectories of the three stochastic processes $\gamma_\mathbf{k}(t),\omega_\mathbf{k}(t)$ and $b_\mathbf{k}(t)$. The update of the parameters is shown in \eqref{Parameter_Estimation}. The sampling of the trajectories can be carried out via a closed analytic formula by exploiting the following general conditional sampling framework. Denote by $\mathbf{X}$ the state variables with observations and by $\mathbf{Y}$ the stochastic parameterized processes. The SPEKF model fits into the following general model family with $\mathbf{X}=\hat{u}_\mathbf{k}$ and $\mathbf{Y}=(\gamma_\mathbf{k},\omega_\mathbf{k}, b_\mathbf{k})^\mathtt{T}$,
\begin{subequations}\label{CGNS}
\begin{align}
  \frac{\d\mathbf{X}}{\d t} &= \Big[\mathbf{A}_\mathbf{0}(\mathbf{X},t) + \mathbf{A}_\mathbf{1}(\mathbf{X},t) \mathbf{Y}\Big]   + \mathbf{B}_\mathbf{1}(\mathbf{X},t)\dot{\mathbf{W}}_\mathbf{1}(t),\label{CGNS_X}\\
  \frac{\d\mathbf{Y}}{\d t} &= \Big[\mathbf{a}_\mathbf{0}(\mathbf{X},t) + \mathbf{a}_\mathbf{1}(\mathbf{X},t) \mathbf{Y}\Big]   + \mathbf{b}_\mathbf{2}(\mathbf{X},t)\dot{\mathbf{W}}_\mathbf{2}(t).\label{CGNS_Y}
\end{align}
\end{subequations}
In \eqref{CGNS}, $\mathbf{A}_\mathbf{0}, \mathbf{a}_\mathbf{0}, \mathbf{A}_\mathbf{1}, \mathbf{a}_\mathbf{1}, \mathbf{B}_\mathbf{1}$ and $\mathbf{b}_\mathbf{2}$ are vectors or matrices that can depend nonlinearly on the state variables $\mathbf{X}$ and time $t$  while $\dot{\mathbf{W}}_\mathbf{1}$ and $\dot{\mathbf{W}}_\mathbf{2}$ are independent white noises.
For systems in \eqref{CGNS}, the conditional distribution $p(\mathbf{Y}(t)|\mathbf{X}(s),s\leq t)\sim\mathcal{N}(\boldsymbol\mu_{\mathbf{f}}, \mathbf{R}_{\mathbf{f}})$, which is also the posterior distribution of filtering $\mathbf{Y}$, is Gaussian. In addition, given one realization of $\mathbf{X}(t)$ up to the current time instant $t$, the conditional mean $\boldsymbol\mu_{\mathbf{f}}$ and the conditional covariance $\mathbf{R}_{\mathbf{f}}$ can be solved via the following closed analytic formulae \cite{liptser2013statistics},
\begin{subequations}\label{CGNS_Stat}
\begin{align}
  \d \boldsymbol{\mu}_{\mathbf{f}} &= (\mathbf{a}_\mathbf{0} + \mathbf{a}_\mathbf{1} \boldsymbol{\mu}_{\mathbf{f}}) + (\mathbf{R}_{\mathbf{f}}\mathbf{A}_\mathbf{1}^* ) (\mathbf{B}_\mathbf{1}\mathbf{B}_\mathbf{1}^*)^{-1} \left(\d\mathbf{X} - (\mathbf{A}_\mathbf{0} + \mathbf{A}_\mathbf{1}\boldsymbol{\mu}_{\mathbf{f}})\d t\right),\label{CGNS_Stat_Mean}\\
  \d\mathbf{R}_{\mathbf{f}} &= \left[\mathbf{a}_\mathbf{1} \mathbf{R}_{\mathbf{f}} + \mathbf{R}_{\mathbf{f}}\mathbf{a}_\mathbf{1}^* + \mathbf{b}_\mathbf{2}\mathbf{b}_\mathbf{2}^* - ( \mathbf{R}_{\mathbf{f}}\mathbf{A}_\mathbf{1}^*)(\mathbf{B}_\mathbf{1}\mathbf{B}_\mathbf{1}^*)^{-1}(\mathbf{A}_\mathbf{1}\mathbf{R}_{\mathbf{f}})\right]\d t,\label{CGNS_Stat_Cov}
\end{align}
\end{subequations}
with $\cdot^*$ being the complex conjugate transpose. Similarly, given one realization of  $\mathbf{X}(t)$  for $t\in[0,T]$, the optimal smoother estimate $p(\mathbf{Y}(t)|\mathbf{X}(s), s\in[0,T])\sim\mathcal{N}(\boldsymbol\mu_\mathbf{s}(t),\mathbf{R}_\mathbf{s}(t))$  is also Gaussian \cite{chen2020learning},
where the conditional mean $\boldsymbol\mu_\mathbf{s}(t)$ and conditional covariance $\mathbf{R}_\mathbf{s}(t)$ of the smoother at time $t$ satisfy the following backward equations\begin{subequations}\label{Smoother_Main}
\begin{align}
  \overleftarrow{\d \boldsymbol{\mu}_\mathbf{s}} &=  \left[-\mathbf{a}_\mathbf{0} - \mathbf{a}_\mathbf{1}\boldsymbol{\mu}_\mathbf{s}  + (\mathbf{b}_\mathbf{2}\mathbf{b}_\mathbf{2}^*)\mathbf{R}_{\mathbf{f}}^{-1}(\boldsymbol\mu_{\mathbf{f}} - \boldsymbol{\mu}_\mathbf{s})\right]\d t,\label{Smoother_Main_mu}\\
  \overleftarrow{\d \mathbf{R}_\mathbf{s}} &= \left[- (\mathbf{a}_\mathbf{1} + (\mathbf{b}_\mathbf{2}\mathbf{b}_\mathbf{2}^*) \mathbf{R}_{\mathbf{f}}^{-1})\mathbf{R}_\mathbf{s} - \mathbf{R}_\mathbf{s}(\mathbf{a}_\mathbf{1}^* + (\mathbf{b}_\mathbf{2}\mathbf{b}_\mathbf{2}^*)\mathbf{R}_{\mathbf{f}})  + \mathbf{b}_\mathbf{2}\mathbf{b}_\mathbf{2}^*\right]\d t,\label{Smoother_Main_R}
\end{align}
\end{subequations}
with $\boldsymbol\mu_{\mathbf{f}}$ and $\mathbf{R}_{\mathbf{f}}$ being given by~\eqref{CGNS_Stat}. Here, the subscript `$\mathbf{s}$' in the conditional mean $\boldsymbol{\mu}_{\mathbf{s}}$ and conditional covariance $\mathbf{R}_{\mathbf{s}}$ is an abbreviation for `smoother', which should not be confused with the time variable $s$ in $\mathbf{X}(s)$. The notation $\overleftarrow{\d \cdot}$ corresponds to the negative of the usual difference, which means that the system \eqref{Smoother_Main} is solved backward over $[0,T]$  with the starting value of the nonlinear smoother $(\boldsymbol\mu_\mathbf{s}(T), \mathbf{R}_\mathbf{s}(T))$ being the same as the filter estimate $(\boldsymbol\mu_{\mathbf{f}}(T), \mathbf{R}_{\mathbf{f}}(T))$. The backward equation takes into account future information. The forward run of \eqref{CGNS_Stat} and the backward run of \eqref{Smoother_Main} collect the past and future observational information, respectively, for the state estimation at time $t$.

Associated with the state estimation via the smoother in \eqref{Smoother_Main}, conditioned on one realization of $\mathbf{X}(s)$ for $s\in[0,T]$, the optimal strategy of sampling  the trajectories associated with the unobserved variable $\mathbf{Y}$ satisfies the following explicit formula \cite{chen2022conditional},
\begin{equation}\label{Sampling_Main}
  \overleftarrow{\d \mathbf{Y}} = \overleftarrow{\d \boldsymbol\mu_\mathbf{s}} - \big(\mathbf{a}_\mathbf{1} + (\mathbf{b}_\mathbf{2}\mathbf{b}_\mathbf{2}^*)\mathbf{R}_\mathbf{f}^{-1}\big)(\mathbf{Y} - \boldsymbol\mu_\mathbf{s})\d t + \mathbf{b}_\mathbf{2}\d{\mathbf{W}}_{\mathbf{Y}}(t),
\end{equation}
where ${\mathbf{W}}_{\mathbf{Y}}(t)$ is a Wiener process that is independent from ${\mathbf{W}}_\mathbf{2}(t)$ in \eqref{CGNS}. The optimality is in the Bayesian sense.

The closed analytic formula \eqref{Sampling_Main} is adopted to sample the trajectories of the three stochastic processes $\gamma_\mathbf{k}(t),\omega_\mathbf{k}(t)$ and $b_\mathbf{k}(t)$ in the SPEKF model \eqref{SPEKF}.

\subsection{The traditional ROM for data assimilation}

To construct an accurate traditional ROM with the LETKF data assimilation scheme, a truncation of the PQG equations to a coarse resolution grid is applied. Then the hyper-viscosity is increased to guarantee the stability of the solution, which is followed by adding stochastic noise to the truncated system that matches the equilibrium distribution of each mode with the truth. Furthermore, the complicated PV-and-M inversion is replaced by the linear inversion in the dry QG. The resulting system reads:
\begin{align}
    &\begin{aligned}
     \frac{\partial  { PV_{r,1}}}{\partial t}
     +
     J(\widetilde{\psi}_{r,1},{ PV_{r,1}})
   -
   U\frac{\partial { PV_{r,1}}}{\partial x}
   +
   &\beta \widetilde{v}_{r,1}
   +
    \widetilde{v}_{r,1}\frac{\partial { PV_{1,bg}}}{\partial y}
    =
    \\
    &
    -\kappa\Delta_h \widetilde{\psi}_{r,1}
   -\nu_{1}(-1)^{s}\Delta^{s}_h {PV_{r,1}} +\mathcal{F}_{r,1}
\end{aligned}
    \label{eqn:layer-pqg-rom-1}
    \\
    &\begin{aligned}
     \frac{\partial { PV_{r,2}}}{\partial t}
     +
    J(\widetilde{\psi}_{r,2}, {PV_{r,2}})
    +
    U\frac{\partial  { PV_{r,2}}}{\partial x}
    +
    &\beta \widetilde{v}_{r,2}
    +      \widetilde{v}_{r,2}\frac{\partial { PV_{2,bg}}}{\partial y}
    =
    \\
    &
    -\nu_{2}(-1)^{s}\Delta^{s}_h  { PV_{r,2}}+\mathcal{F}_{r,2},
\end{aligned}
    \label{eqn:layer-pqg-rom-2}
\\
    &\begin{aligned}
    {
     \frac{D_m M_{r,m}}{Dt} +
     \widetilde{v}_{r,m}\frac{\partial M_{bg}}{\partial y}
    =-\nu_M(-1)^{s}\Delta^{s}_h M_{r,m} +\mathcal{F}_{r,M} 
    }
    \label{eqn:layer-pqg-rom-3}
    \end{aligned}
\end{align}
where $\widetilde{\psi}_{r,j},\widetilde{u}_{r,j},\widetilde{v}_{r,j}$ are stream functions, zonal and meridional velocities using the dry QG relation and $\mathcal{F}_{r,1}, \mathcal{F}_{r,2}$ and $\mathcal{F}_{r,M}$ are stochastic forcings that mimic the interactions between the truncated small scales and retained scales.  In particular, the dry QG PV and streamfunction relation yields the following:
\begin{align}
    &PV_{\psi} = \Delta \psi \frac{PV_1+PV_2}{2} & \psi = \frac{\psi_1+\psi_2}{2}\\
    &PV_{\tau} = \Delta\tau - k_d^2 \tau= \frac{PV_1-PV_2}{2}, & \tau = \frac{\psi_1-\psi_2}{2},
\end{align}
where $PV_{\psi}$ and $PV_{\tau}$ are the potential vorticity in the barotrophic and baroclinic modes respectively while $\psi$ and $\tau$ are the corresponding barotrophic and baroclinic streamfunctions; $k_d$ is the baroclinic deformation wavenumber which is calculated as $k_d = 8(L/L_{du})^2$~\cite{qi2016low,hu2021initial}. Due to this dry QG PV and streamfunction relation, the moisture quantities, $E,V_r$ are removed from the PQG moisture evolution equation which are related to the phase changes of water in the original PQG system.
In addition, the stochastic forcing can be represented by Gaussian white noises for each individual Fourier mode:
\begin{align}
    \mathcal{F}_{r,1} & \approx  \sum_{l =-N_r/2}^{N_r/2} \sum_{k =-N_r/2}^{N_r/2}\sigma_{kl}^{(1)}\dot{W}  e^{i(kx+ly)}+c.c.
    \\
    \mathcal{F}_{r,2} & \approx  \sum_{l =-N_r/2}^{N_r/2} \sum_{k =-N_r/2}^{N_r/2}\sigma_{kl}^{(2)}\dot{W}  e^{i(kx+ly)}+c.c.
    \\
    \mathcal{F}_{r,M} & \approx \sum_{l =-N_r/2}^{N_r/2} \sum_{k =-N_r/2}^{N_r/2}\sigma_{kl}^{(M)}\dot{W} e^{i(kx+ly)}+c.c.
\end{align}
where, $\sigma_{kl}^{(1)},\sigma_{kl}^{(2)},\sigma_{kl}^{(M)}$, corresponds to the magnitudes of white noise of
 $(k,l)$th Fourier model of $PV_1,PV_2$ and moisture.

In addition, to improve the accuracy and robustness of the traditional ROM, the hyperviscosity and the white noise magnitude terms are tuned on an ad-hoc basis. More precisely, a wide range of different viscosity values $\nu_1,\nu_2,\nu_M$ as well as $\sigma_{kl}^{(1)},\sigma_{kl}^{(2)},\sigma_{kl}^{(M)}$ is run and the optimal set of parameters is chosen that generates the most accurate mean and variance compared with the truth time series.
The optimal results are presented as a comparison in Section~\ref{ss:result}.

\bibliographystyle{plain}
\bibliography{references}
\end{document}